\documentclass[useAMS,usenatbib]{mn2e}

\usepackage{graphicx}
\usepackage{bm}
\usepackage{epsfig} 
\usepackage{amssymb}
\usepackage{textcomp}

\begin{document}

\title{Gross-Pitaevskii model of pulsar glitches}
\author[L. Warszawski and A. Melatos]{L. Warszawski$^{1}$\thanks{lilaw@ph.unimelb.edu.au (LW)} and A. Melatos$^{1}$\\
$^{1}$School of Physics, University of Melbourne, Parkville, VIC 3010, Australia\\}

\date{\today}
\maketitle

\begin{abstract}
The first large-scale quantum mechanical simulations of pulsar glitches are presented, using a Gross-Pitaevskii model of the crust-superfluid system in the presence of pinning.  Power-law distributions of simulated glitch sizes are obtained, in accord with astronomical observations, with exponents ranging from $-0.55$ to $-1.26$.  Examples of large-scale simulations, containing $\sim 200$ vortices, reveal that these statistics persist in the many-vortex limit.  Waiting-time distributions are also constructed.  These and other statistics support the hypothesis that catastrophic unpinning mediated by collective vortex motion produces glitches; indeed, such collective events are seen in time-lapse movies of superfluid density.  Three principal trends are observed.  (1) The glitch rate scales proportional to the electromagnetic spin-down torque.   (2)  A strong positive correlation is found between the strength of vortex pinning and mean glitch size.  (3) The spin-down dynamics depend less on the pinning site abundance once the latter exceeds one site per vortex, suggesting that unpinned vortices travel a distance comparable to the inter-vortex spacing before repinning.   
\end{abstract}


\section{Introduction}\label{sec:intro}

Glitches are discrete, randomly timed jumps in the famously predictable angular velocity of a pulsar.  The statistical distributions of glitches in individual pulsars (power-law sizes and exponential waiting times) point to an underlying threshold-triggered collective process \citep{Melatos:2008p204}, akin to grain avalanches in sand piles \citep{Wiesenfeld:1989p44,Morley:1996p2128} or vortex avalanches in type-II superconductors \citep{Field:1995p155}.  In pulsars, the `grains' are quantised superfluid vortices. A comprehensive theory of why $\gtrsim10^7$ (out of $\sim 10^{18}$) vortices simultaneously unpin and self-reorganise  to create a glitch remains elusive.  A complete solution relies on connecting the microscopic (fm-scale) and macroscopic (km-scale) dynamics of the neutron superfluid in the pulsar interior. The challenge presented by this disparity of scales, especially in replicating collective phenomena, has proved formidable.  

In this paper, we present the first comprehensive quantum mechanical simulations of pulsar glitches, using the Gross-Pitaevskii equation (GPE) to model the temporal evolution of the superfluid inside a pulsar.  The GPE is a standard tool in condensed matter physics for investigating vortex dynamics in quantum condensates \citep{Pethick:2002p7935}.  Our simulations are unique in four ways:  (1) the system is much larger than those routinely simulated in condensed matter applications \citep[][for example]{Tsubota:2002p11,Tsubota:2010p10353,Penckwitt:2010p1036}, and therefore contains many more vortices; (2) in addition to the confining potential, we include a grid of pinning sites to model nuclear lattice defects in the pulsar's crust \citep{Jones:1998p34,Donati:2004p61,Donati:2006p32,Avogadro:2007p51,Avogadro:2008p29,Barranco:2010p10677}; (3) the container and superfluid are coupled via a feedback torque; and (4) we allow the angular velocity of the container to change non-adiabatically.  Thus, the simulations describe scales ranging from individual vortex cores to the collective behaviour of several hundred vortices in the presence of a large grid of pinning sites.  The results build on previous GPE studies of the general problem of spasmodic superfluid spin down \citep{Warszawski:2010lattice} and the microphysics of individual vortex unpinning, including knock-on processes involving acoustic radiation and the proximity effect \citep{Warszawski:2010individual}. 

The paper is structured as follows.  Section~\ref{sec:ch5:model} summarises the vortex unpinning paradigm of pulsar glitches, the main features of the  numerical model, and the feedback mechanism.  In Section~\ref{sec:define} we describe our algorithm for extracting glitches from simulated spin-down curves.  In Section~\ref{sec:canonical} this algorithm is employed in a canonical example.  In Section~\ref{sec:pinning} we construct glitch size and waiting-time distributions from relatively small-scale simulations ($\sim 30$ vortices) for different pinning parameters.  In Section~\ref{sec:structure}, we vary the inertia of the star and the electromagnetic spin-down torque.  Section~\ref{sec:ch5:big} reports results from larger-scale simulations involving $\sim 200$ vortices, in which we look for evidence of collective dynamics.  Parameter studies with the larger-scale simulation are too expensive computationally at this stage.  In Section~\ref{sec:ch5:conc_GPE} we synthesise the results and present our conclusions.

\begin{center}
\begin{table*}
\begin{tabular}{ | c| c| c| c| c|}
\hline
  Quantity			&Simulation		&Simulation		&Pulsar 	&Pulsar  \\
  	 			&(dimensionless) 	&(dimensional)  	&(dimensionless)&(dimensional)\\  \hline
  $R$ 				&$12.75$		&$10^{-16}\,\rm{m}$  	& $10^{21}$	&$10^{4}\,\rm{m}$\\
  $V_0$ 				&$16.6$			&$10^{4}\,\rm{MeV}$  	& $10^{-3}$	&$10^{0}\,\rm{MeV}$		\\
  $\Omega$ 			&$0.8$ 			&$10^{24}\,\rm{Hz}$  	& $10^{-24}-10^{-21}$	&$10^{-1}-10^{2}\,\rm{Hz}$\\
  $N_{\rm{EM}}/I_{\rm{c}}$		&$10^{-3}$		&$10^{45}\,\rm{Hz~s}^{-1}$  & $10^{-63}-10^{-59}$	&$10^{-15}-10^{-10}\,\rm{Hz~s}^{-1}$	\\
  $t_{\rm{total}}$		&$10^{3}$ 		&$10^{-21}\,\rm{s}$  	& $10^{32}-10^{33}$	&$10^{8}-10^{9}\,\rm{s}$	\\
  $n_{\rm{F}}$			&$0.13$ &$2.8\times10^{-16}\,\rm{m}^{-2}$& $10^{8}-10^{12}$&$10^{-8}-10^{-4}\,\rm{m}^{-2}$\\
  $n_{\rm{pin}}$		&$0.19$ 	&$2.6\times10^{-16}\,\rm{m}^{-2}$& $10^{0}-10^{4}$ 	&$10^{-14}-10^{-10}\,\rm{m}^{-2}$\\
  $N_{\rm{v}}$			&$30-200$  		&$30-200$		&$10^{16}-10^{19}$&$10^{16}-10^{19}$\\
  $\eta$			&$1$		&$1$	&$10^{-6}$	&$10^{-6}$\\
  $\Delta x$		&$0.15$ 	&$1.5\times10^{-16}\,\rm{m}$& $---$ 	&$---$\\
  $\Delta t$		&$5\times 10^{-4}$ 	&$5\times10^{-28}\,\rm{m}$& $---$ 	&$---$\\\hline
\end{tabular}
\caption{Dimensionless and dimensional parameters for a typical simulation (first and second columns) and pulsar (third and fourth columns).  Dimensionless quantities are scaled so as to recover Eq.~(\ref{eq:GPE_dimensionless}) from the dimensional form of the GPE, making the replacements described in Section~2 of \citet{Warszawski:2010lattice}.  Pulsar quantities are quoted as a range representative of the pulsar population.}
\label{tab:ch5:pulsar}
\end{table*}
\end{center}

\begin{figure}
\includegraphics[scale=0.365,angle=90]{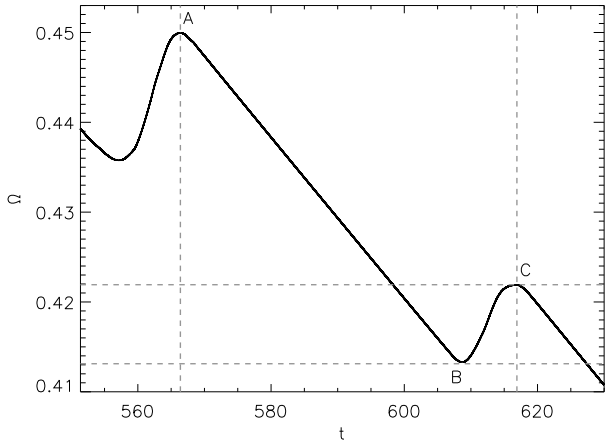}
\caption{Schematic of how the glitch-finding algorithm operates.  A and B mark the positions of the two glitches in the interval $552<t<630$ on the plot of angular velocity  as a function of time $\Omega(t)$.  The size of glitch C is the difference in $\Omega$ between B (the local minimum between A and C), and C (the vertical distance between the horizontal \emph{dashed} lines). Simulation parameters:  $V_0 = 16.6$, $\eta=1$, $\Delta V_i/V_0=0.0$, $R=12.5$, $\Delta x=0.15$, $\Delta t=5\times 10^{-4}$, $N_{\rm{EM}}/I_{\rm{c}}=10^{-3}$, $\Omega(t=0) = 0.8$, $n_{\rm{pin}}/n_{\rm{F}}=0.97$.}
\label{fig:ch5:defineglitch}
\end{figure}

\begin{figure*}
\includegraphics[scale=0.75,angle=90]{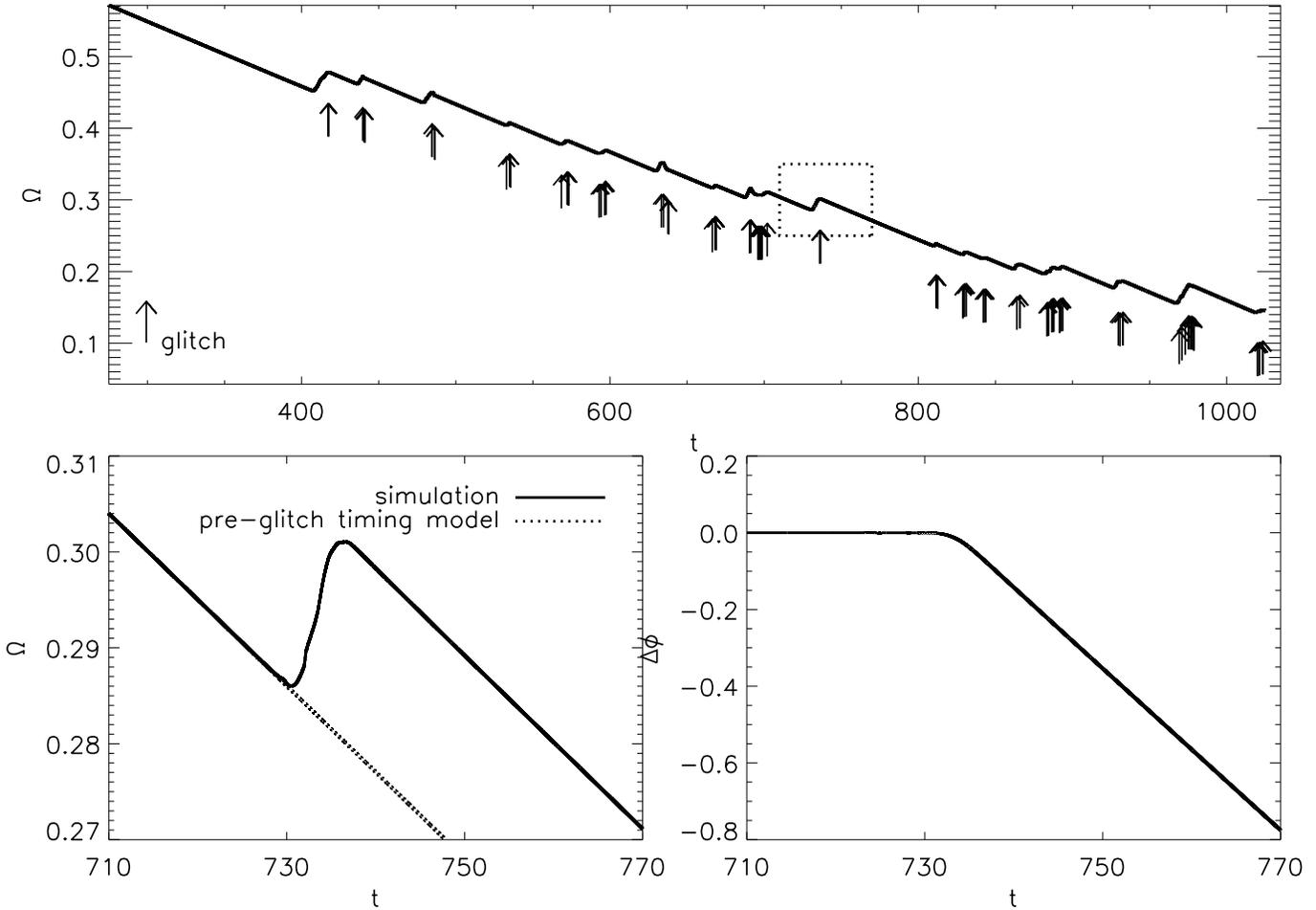}
\caption{Timing signature of a simulated glitch.  \emph{Top}: Angular velocity of the crust as a function of time $\Omega(t)$ for $\Omega(t=0)=0.8$.  $\Omega(t)$ is smoothed with a top-hat function of width $\Delta t_{\rm{sm}}=1.0$.  The \emph{arrows} mark the positions of glitches identified using the automated glitch finder described in Section~\ref{sec:define}.  \emph{Bottom left}:  Close-up of a glitch at $t=736$, bracketed by the \emph{dotted rectangle} in the \emph{top} panel.  The \emph{dotted} curve represents the linear model fitted to the pre-glitch spin-down curve.  \emph{Bottom right}: Phase residual after removing a pre-glitch linear fit to the interval  $710<t<730$ [$\Omega(t=710) =  0.942$, $\dot{\Omega} = -8.98\times 10^{-4}$] from the interval $710<t<770$ shown in the \emph{bottom left} panel.  Simulation parameters:  $V_0 = 16.6$, $\eta=1$, $\Delta V_i/V_0=0.0$, $R=12.5$, $\Delta x=0.15$, $\Delta t=5\times 10^{-4}$, $N_{\rm{EM}}/I_{\rm{c}}=10^{-3}$, $n_{\rm{pin}}/n_{\rm{F}}=0.97$.}
\label{fig:ch5:residual}
\end{figure*} 

\begin{figure*}
\includegraphics[scale=1.0]{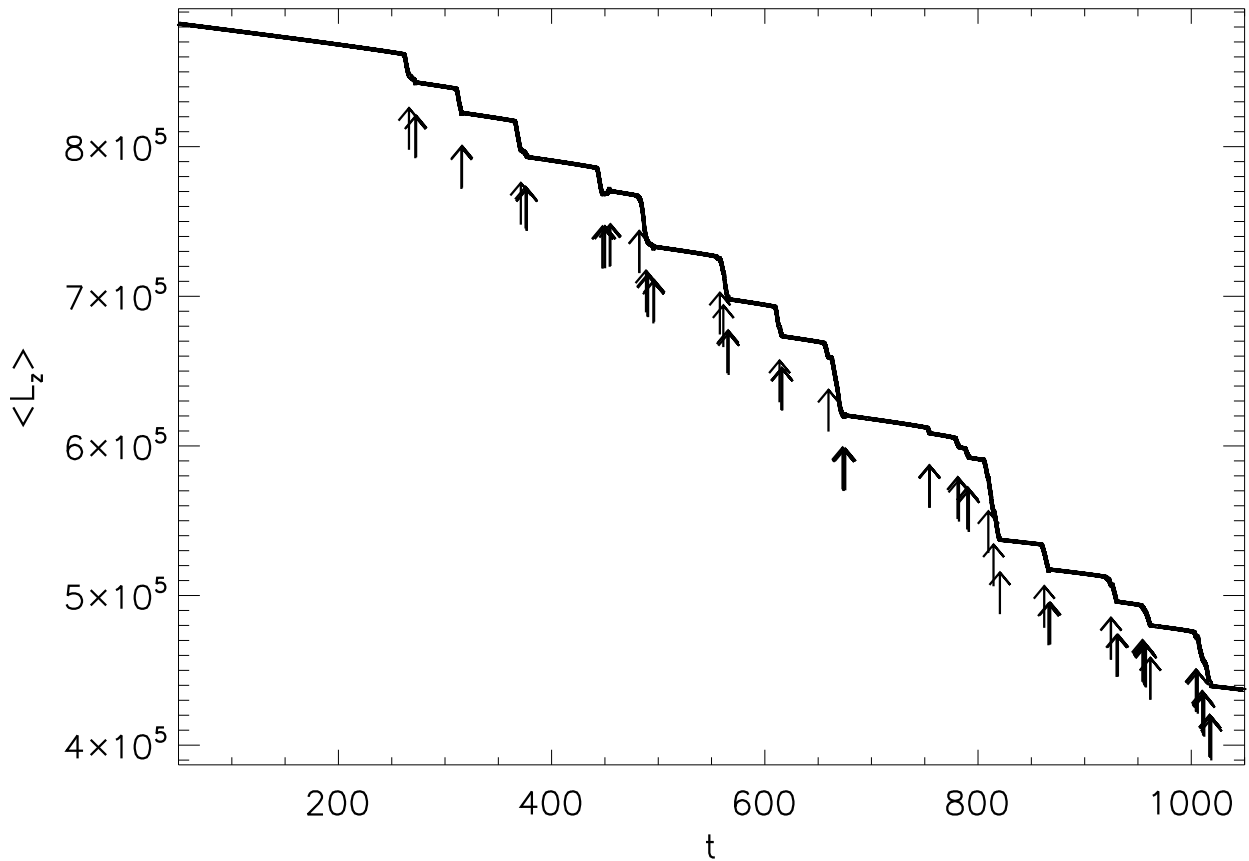}
\caption{Superfluid angular momentum as a function of time $\langle L_{\rm{z}}\rangle(t)$, in units of $\hbar/2$, for the canonical spin-down experiment described in Section~\ref{sec:canonical}. The \emph{arrows} indicate the position of glitches identified using the glitch-finding algorithm defined in Section~\ref{sec:define}.  Simulation parameters:  $\eta=1$, $\Delta V_i/V_0=0.0$, $R=12.5$, $\Delta x=0.15$, $\Delta t=5\times 10^{-4}$, $N_{\rm{EM}}/I_{\rm{c}}=10^{-3}$, $\Omega(t=0) = 0.8$, $n_{\rm{pin}}/n_{\rm{F}}=0.97$.}
\label{fig:ch5:L_define}
\end{figure*}

\begin{figure*}
\includegraphics[scale=0.7,angle=90]{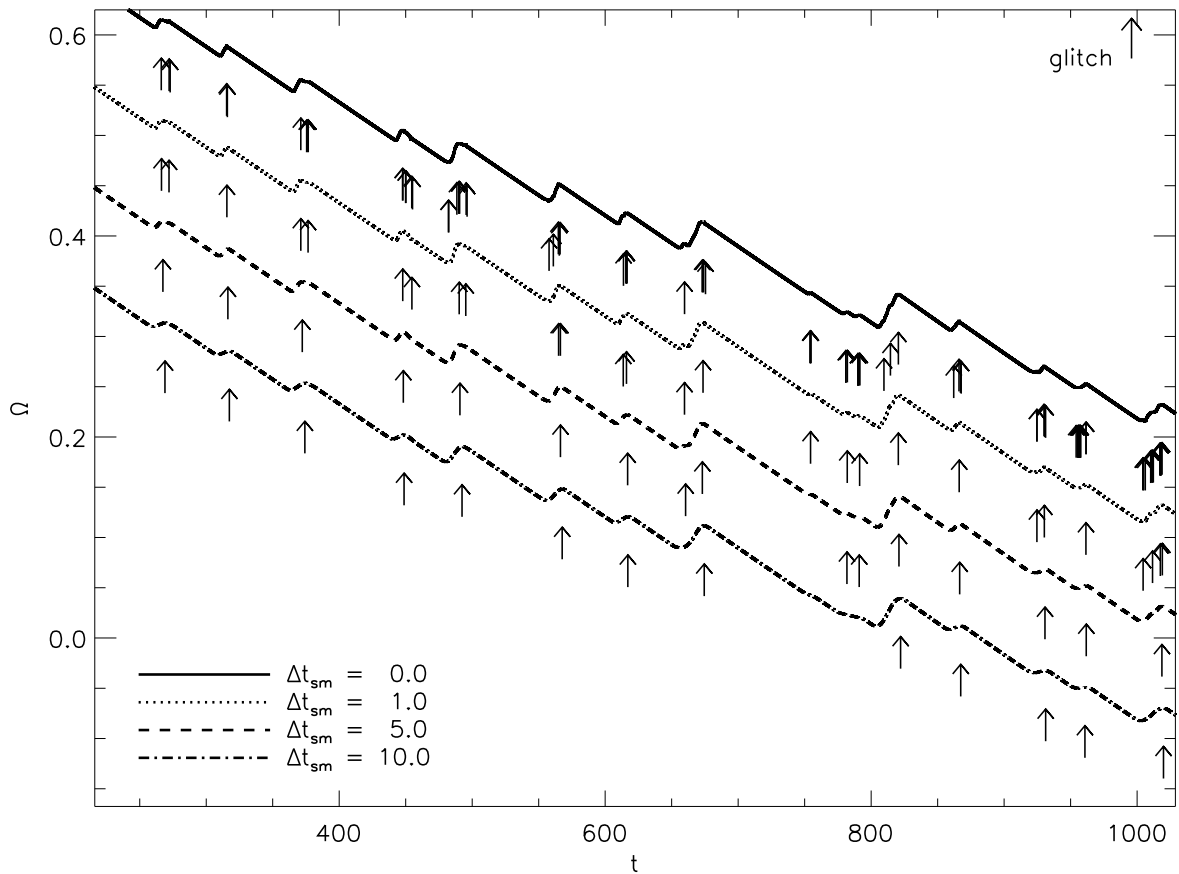}
\caption{Angular velocity as a function of time $\Omega (t)$ for the canonical example described in Section~\ref{sec:canonical}, smoothed with a top-hat window function of width $\Delta t_{\rm{sm}}=0.0$, 1.0, 5.0 and 10 (\emph{solid}, \emph{dotted}, \emph{dashed}, \emph{dot-dashed} curves respectively; the curves are shifted vertically down by 0.0, 0.1, 0.2 and 0.3 respectively to improve visibility).  Arrows below each curve mark the positions of glitches found using the glitch-finding algorithm described in Section~\ref{sec:define}; as $\Delta t_{\rm{sm}}$ increases, the number of found glitches decreases.  Simulation parameters:  $V_0 = 16.6$, $\eta=1$, $\Delta V_i/V_0=0.0$, $R=12.5$, $\Delta x=0.15$, $\Delta t=5\times 10^{-4}$, $N_{\rm{EM}}/I_{\rm{c}}=10^{-3}$, $\Omega(t=0) = 0.8$, $n_{\rm{pin}}/n_{\rm{F}}=0.97$.}
\label{fig:ch5:smooth_full}
\end{figure*}

\begin{figure*}
\includegraphics[scale=0.365,angle=90]{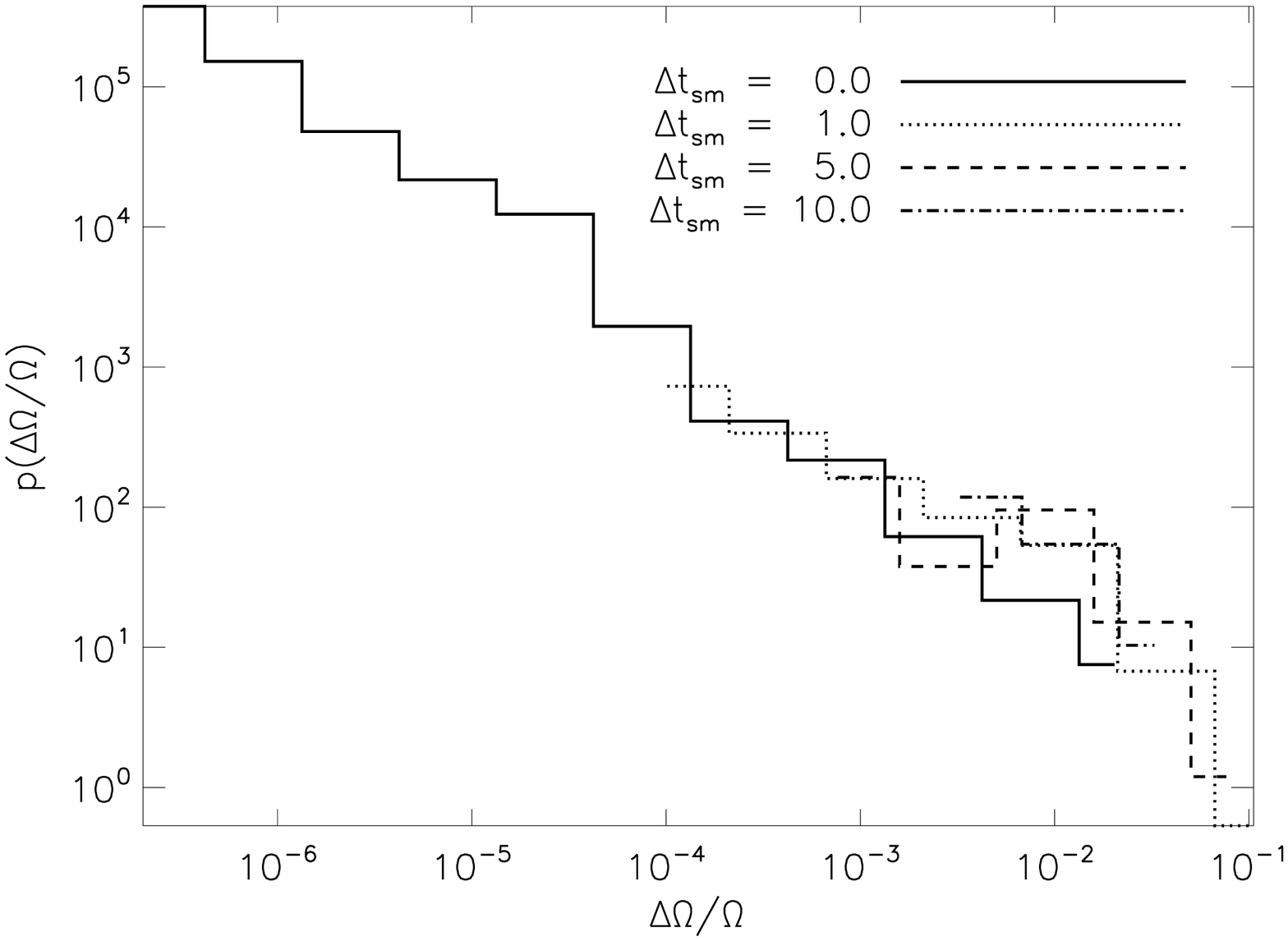}
\includegraphics[scale=0.365,angle=90]{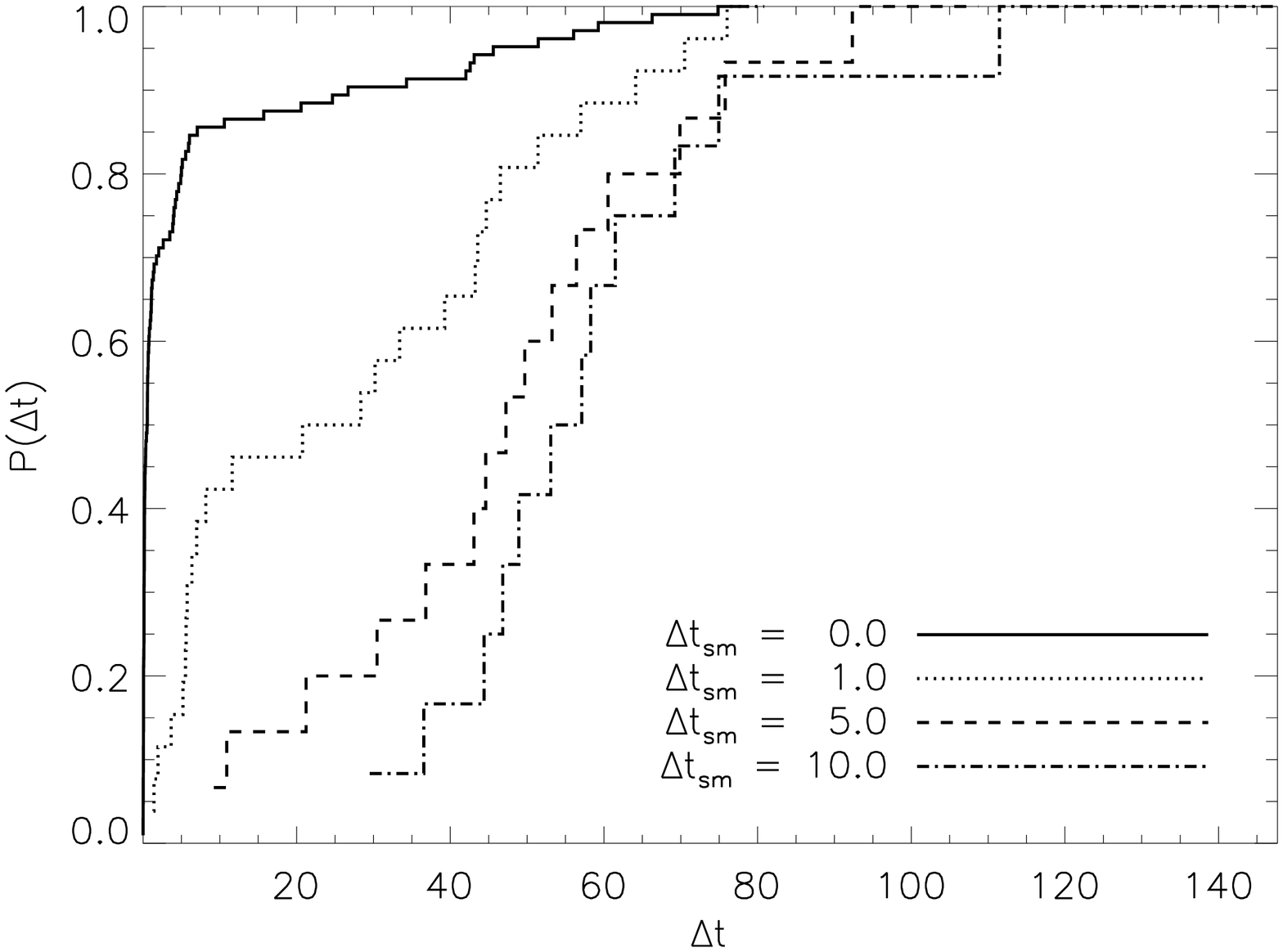}
\caption{Glitch statistics as a function of the degree of smoothing for the spin-down experiment in Fig.~\ref{fig:ch5:smooth_full}.  \emph{Left}:  Probability density function (pdf) of fractional glitch sizes $p(\Delta\Omega/\Omega)$, binned logarithmically and graphed on log-log axes, for top-hat window functions of width $\Delta t_{\rm{sm}}=0.0$, 1.0, 5.0 and 10 (\emph{solid}, \emph{dotted}, \emph{dashed}, \emph{dot-dashed} curves respectively).  \emph{Right}:  Cumulative distribution of glitch waiting times $P(\Delta t)$ for top-hat window functions of width $\Delta t_{\rm{sm}}=0.0$, 1.0, 5.0 and 10 (\emph{solid}, \emph{dotted}, \emph{dashed}, \emph{dot-dashed} curves respectively), graphed on linear axes. Simulation parameters:  $V_0 = 16.6$, $\eta=1$, $\Delta V_i/V_0=0.0$,  $R=12.5$, $\Delta x=0.15$, $\Delta t=5\times 10^{-4}$, $N_{\rm{EM}}/I_{\rm{c}}=10^{-3}$, $\Omega(t=0) = 0.8$, $n_{\rm{pin}}/n_{\rm{F}}=0.97$.}
\label{fig:ch5:smooth_stats}
\end{figure*}

\section{Vortex unpinning paradigm}\label{sec:ch5:model}

\subsection{Overview}\label{sec:ch5:overview}

The superfluid in a pulsar is composed of neutron Cooper pairs, which interpenetrate the nuclear lattice in the crust \citep{Migdal:1959p10771}.  The superfluid attempts to match the rotation of the crust by forming quantised vortices, each carrying a quantum of circulation $\kappa=h/(2m)$, whose areal density is proportional to the angular velocity of the superfluid.  In the absence of pinning, vortices form a hexagonal Abrikosov lattice, causing the superfluid to rotate as a rigid body.  In equilibrium, the mean areal density of vortices is $n_{\rm{F}}=2\Omega/\kappa$.  In reality, however, vortices pin at or between nuclear lattice sites and/or defects, thus distorting the equilibrium vortex configuration.  The extent of distortion depends on the configuration and abundance of pinning centres supplied by the crustal lattice of nuclei.  Pinning prevents the superfluid from decelerating smoothly with the crust.  Glitches are believed to occur when a large number of vortices simultaneously unpin and move outward to new pinned positions.  Indeed, GPE simulations reported in \cite{Warszawski:2010lattice} exhibit stick-slip vortex dynamics and resulting glitch-like interruptions to the smooth spin down of the rotating container.  Similar spin-up events were observed in laboratory experiments using rotating containers filled with helium II \citep{Tsakadze:1980p2131}.

We construct a simple, two-dimensional pulsar model comprising a superfluid inside a rotating, decelerating crust, in the presence of a pinning grid that co-rotates with the crust.  We include self-consistent coupling between the superfluid and crust by ensuring that angular momentum is conserved overall.

A popular explanation for pulsar glitches is that angular momentum is transferred from the superfluid to the crust in response to sudden unpinning, outward motion, and repinning of vortices \citep{Packard:1972p5868,ANDERSON:1975p84}.  Crust-superfluid differential rotation produces a fictitious Magnus force, $\mathbf{F}_{\rm{M}}$, that unpins vortices when it exceeds some threshold.  $\mathbf{F}_{\rm{M}}$ is transverse to the relative velocity between the vortex line and superfluid flow; it is similar in origin to the Bernoulli lift force in conventional hydrodynamics.  For a vortex line moving with velocity $\mathbf{v}$ through a superfluid of density $\rho$ and velocity $\mathbf{v}_{\rm{s}}$, the Magnus force is given by
\begin{equation}
 \mathbf{F}_{\rm{M}} = \rho\bm{\kappa}\times (\mathbf{v}-\mathbf{v}_{\rm{s}})~,
\end{equation}
where $\bm{\kappa}$ is directed along the vortex axis.  
For a pinned vortex, $\mathbf{v}$ is the crust velocity; for a vortex at $r=b$, one has $\mathbf{v}=\Omega b\hat{\mathbf{z}}$.  For $N_{\rm{v}}$ vortices in an Abrikosov lattice, the Feynman relation approximates (the result is exact in the infinite-vortex limit) the azimuthal component of the superfluid velocity at radius $r$ as $v_{\phi}\approx\kappa N_{\rm{v}}/(2\pi r)$.

The vortex unpinning paradigm presumes that a catalyst exists to spontaneously unpin between $10^7$ and $10^{15}$ vortices during a single glitch, with the range deduced from measurements of glitch sizes.  To date, the catalyst has not been identified securely.  If vortices unpin independently, the number of unpinning events in any time interval follows Poisson statistics \citep{Hakonen:1998p10787}.  On its own, a Poisson process does not account for the observed power-law distribution of glitch sizes.  Hence, a mechanism is needed which correlates vortices, so that the unpinning rate of a particular vortex increases when another vortex (either nearby or far away) unpins; collective unpinning avalanches are essential to explaining pulsar glitches.  In recent GPE simulations, \cite{Warszawski:2010individual} demonstrated two mechanisms -- in addition to the growing global velocity shear as the crust spins down -- for triggering and sustaining vortex avalanches:  (i) acoustic radiation from moving vortices, and (ii) vortex proximity knock-on, as unpinned vortices move past their pinned neighbours.  

Acoustic radiation results from the propagation of Kelvin waves on vortices \citep{Vinen:2001p134520}, from vortex reconnection \citep{Leadbeater:2001p10352} and due to dissipation by accelerating vortices \citep{Parker:2004p7936,Barenghi:2005p6878}.  The disturbance in the density field as the sound waves propagate is capable of unpinning weakly pinned vortices. The amplitude of the disturbance is a function of vortex acceleration.  Therefore, the strongest acoustic radiation results from the repinning phase, as a vortex spirals in to the new pinning site \citep{Warszawski:2010individual}.  When the level of dissipation in the system is high, the acoustic disturbance generated by a repinning vortex is weaker than for a low-dissipation system, and hence it is harder to unpin vortices acoustically.

The second effect, proximity knock-on, occurs when two vortices approach sufficiently closely that the Magnus force on at least one of them exceeds the unpinning threshold.  It was found that a ballistic unpinned vortex can unpin a pinned vortex at a larger radius via the proximity effect.  The distance of closest approach between the two vortices is inversely proportional to pinning strength \citep{Warszawski:2010individual}.

\subsection{Gross-Pitaevskii model}\label{subsec:ch5:GPE}
In this section we summarise the quantum mechanical pulsar model first introduced in \cite{Warszawski:2010lattice}.  The interested reader who wishes to reproduce our results is referred to the latter paper for technical details.  In the model, a two-dimensional, zero-temperature Bose-Einstein condensate represents an equatorial projection of the stellar superfluid.  The time evolution of the superfluid density $|\psi|^2$, defined in terms of the order parameter $\psi(\mathbf{x},t)$, in a potential $V$, with chemical potential $\mu$, observed in a reference frame rotating with angular velocity $\Omega$, is described by the dimensionless GPE
\begin{equation}
\label{eq:GPE_dimensionless}
(i-\gamma)\frac{\partial\psi}{\partial t}= 
	-\nabla^2\psi +(\mu-V-|\psi|^2)\psi-
	\Omega \hat{L}_z\psi~.
\end{equation}
In arriving at Eq.~(\ref{eq:GPE_dimensionless}), we define the sound speed $c_s = (n_0 g/m)^{1/2}$, the healing length (also the characteristic length-scale) $\xi = \hbar/(m n_0 g)^{1/2}$, the characteristic time-scale $\hbar/(n_0 g)$, and the energy scale $n_0 g$,  where $g$ is the self-interaction strength, $m$ is the mass of each boson (twice the neutron mass), and $n_0$ is the mean particle density.  The angular momentum operator is given by $ \hat{L}_{\rm z} = -i \mathbf{\hat{z}}\partial/\partial \phi$ (with units $\hbar/2$), where $\phi$ is the azimuthal angle. $V$ is the sum of the confining potential representing the crust and the pinning potential.

In line with standard practice \citep{Jin:1996p420,Choi:1998p9985,Tsubota:2002p11,Penckwitt:2002p1045,Kasamatsu:2003p1051}, we include in Eq.~(\ref{eq:GPE_dimensionless}) a phenomenological treatment of dissipation, modelled by the term $-\gamma\partial\psi/\partial t$.  Its effect is to suppress sound waves, thus reducing the relaxation time-scale of the superfluid and ensuring that the wave function can adjust to non-adiabatic changes in $\Omega$ due to glitches.  On the other hand, as reported in \cite{Warszawski:2010individual}, sound waves play an important role in vortex unpinning avalanches, by facilitating knock on.  Therefore, one must be careful not to raise $\gamma$ to the point where it quells the latter effect.  An accurate estimate of $\gamma$ in a pulsar is not available at present.

The dissipationless GPE ($\gamma=0$) models a superfluid at $T=0$, which does not contain viscous excited-state components.  In contrast, solutions to Eq.~(\ref{eq:GPE_dimensionless}) for $\gamma \neq 0$ are by definition at non-zero temperature.  \cite{Penckwitt:2002p1045} attributed dissipation in experiments to atom transfer between the viscous thermal cloud and the condensate.  Experiments with Bose-Einstein condensates \citep{Haljan:2001p210403} showed that the thermal cloud also influences vortex nucleation.  A self-consistent description of this phenomenon requires additional terms in the GPE \citep{Wouters:2007}, which are not included in most studies, including ours.  Laboratory experiments claim $\gamma=0.03$ in helium II \citep{AboShaeer:2002p8832}; we use $\gamma=0.05$.  The dissipative term drives particle loss from the system, which we correct by adjusting the chemical potential, such that the number of particles remains constant \citep{Kasamatsu:2003p1051,Warszawski:2010lattice}.  

\subsection{Container and pinning potential}\label{subsec:ch5:pinning}
Using a fourth-order Runge-Kutta algorithm in time, and a fourth-order finite difference scheme for the spatial derivatives, we solve Eq.~(\ref{eq:GPE_dimensionless}) for a superfluid in a rotating potential $V$, in the frame co-rotating with the potential.  

$V$ is the sum of a confining potential, representing the edge of the pulsar inner crust, and the pinning potential, $V_{\rm{pin}}$, representing the nuclear lattice in the crust, which co-rotates with the confining potential. $V_{\rm{pin}}$ comprises a square grid of equidistant spikes, each one described by
\begin{equation}
V_{\rm{pin},i}(r) = V_i\left[ 1+\tanh\left(\Delta|\mathbf{r}-\mathbf{R}_i|\right)\right]~,
\end{equation}
where $\mathbf{R}_i$ is the position of the centre of the pinning site, and $V_i$ and $\Delta^{-1}$ parametrise its strength and width respectively.  In all cases, the spikes in the pinning grid are equally spaced, with a mean areal density $n_{\rm{pin}}$.  In Section~\ref{subsec:pinning:sites}, we use the ratio $n_{\rm{pin}}/n_{\rm{F}}$ as a measure of the relative abundance of pinning sites and vortices. In Section~\ref{subsec:pinning:delta}, we draw $V_i$ from a top-hat distribution of pinning strengths with mean $V_0$ and width  $\Delta V_i$; in all other simulations, $V_i$ is constant ($V_i=V_0$).  A non-zero range of pinning strengths ($\Delta V_i/V_0\sim 1$) is likely to play an important role in glitch dynamics.  Each small unpinning event biases the distribution of pinning strengths of \emph{occupied} pinning sites towards large $V_i$ \citep{Newman:1996p1484,Melatos:2009p4511}, until a large global velocity shear or strong acoustic perturbation comes along and unpins many strongly pinned vortices simultaneously.
 
The parameters of vortex pinning within a pulsar have been studied theoretically in great detail
\citep{DeBlasio:1998p127,Jones:1998p34,Hirasawa:2001p145,Jones:2003p10,Avogadro:2008p29,Barranco:2010p10677}.  Of particular interest is the pinning strength, the density of pinning sites, and the region within the star where pinning is strongest (the superfluid pairing state changes with depth).  The strength of pinning is usually calculated from the energy difference between a vortex sitting atop a column of spherical defects and a vortex positioned interstitially.  In general, vortices pin to nuclear sites to minimise the region from which superfluid is excluded (the vortex core is devoid of superfluid).  Depending on the superfluid density, however, pinning can also occur interstitially \citep{Pizzochero:1997p45,Donati:2004p61,Donati:2006p32,Avogadro:2007p51,Pizzochero:2007p144}.  For a pulsar, the characteristic size of the pinning site (the width of a nucleus) is approximately equal to the superfluid healing length ($\sim 1~\rm{fm}$), which is significantly smaller than the lattice ($\sim 10~\rm{fm}$) and inter-vortex ($\sim 10^{12}~\rm{fm}$) spacings.

\subsection{Feedback torque}\label{subsec:ch5:feedback}

Vortex motion changes the superfluid angular momentum $\langle \hat{L}_z\rangle$, the expectation value of the operator $\hat{L}_z$.  The response of the crust to these changes is an essential element of any pulsar glitch theory.  We employ a simple conservation argument, wherein changes in $\langle \hat{L}_z\rangle$ are communicated instantaneously to the crust.  In reality, changes in the superfluid flow are communicated to the crust by sound and Kelvin waves along vortex lines \citep{Jones:1998p33}, but these processes are fast.  Given an electromagnetic (EM) spin-down torque, $N_{\rm{EM}}$, the angular velocity of the crust, $\Omega$, evolves according to
\begin{eqnarray}
 I_{\rm{c}}\frac{d\Omega}{dt} &=& -\frac{d\langle \hat{L}_z\rangle}{dt}+N_{\rm{EM}}~,
\label{eq:ch5:feedback}
\end{eqnarray}
where $I_{\rm{c}}$ is the moment of inertia of the crust, and $\Omega$ is the same angular velocity that appears in the GPE.  We emphasise that, even for smooth deceleration, $N_{\rm{EM}}/I_{\rm{c}}$ does not equal the observed pulsar spin-down rate, as this rate is the combination of two effects:  the external EM torque, and the spin-up response to gradual (non-glitch) spin down of the superfluid \citep{Warszawski:2010lattice}.  We also note that if the vortex lattice is significantly distorted by pinning, the superfluid does not rotate rigidly.  Hence it is not meaningful to assign to it a unique angular velocity or moment of inertia.   We solve Eq.~(\ref{eq:GPE_dimensionless}) simultaneously with Eq.~(\ref{eq:ch5:feedback}) in the following sections.

\begin{figure}
\includegraphics[scale=0.365,angle=90]{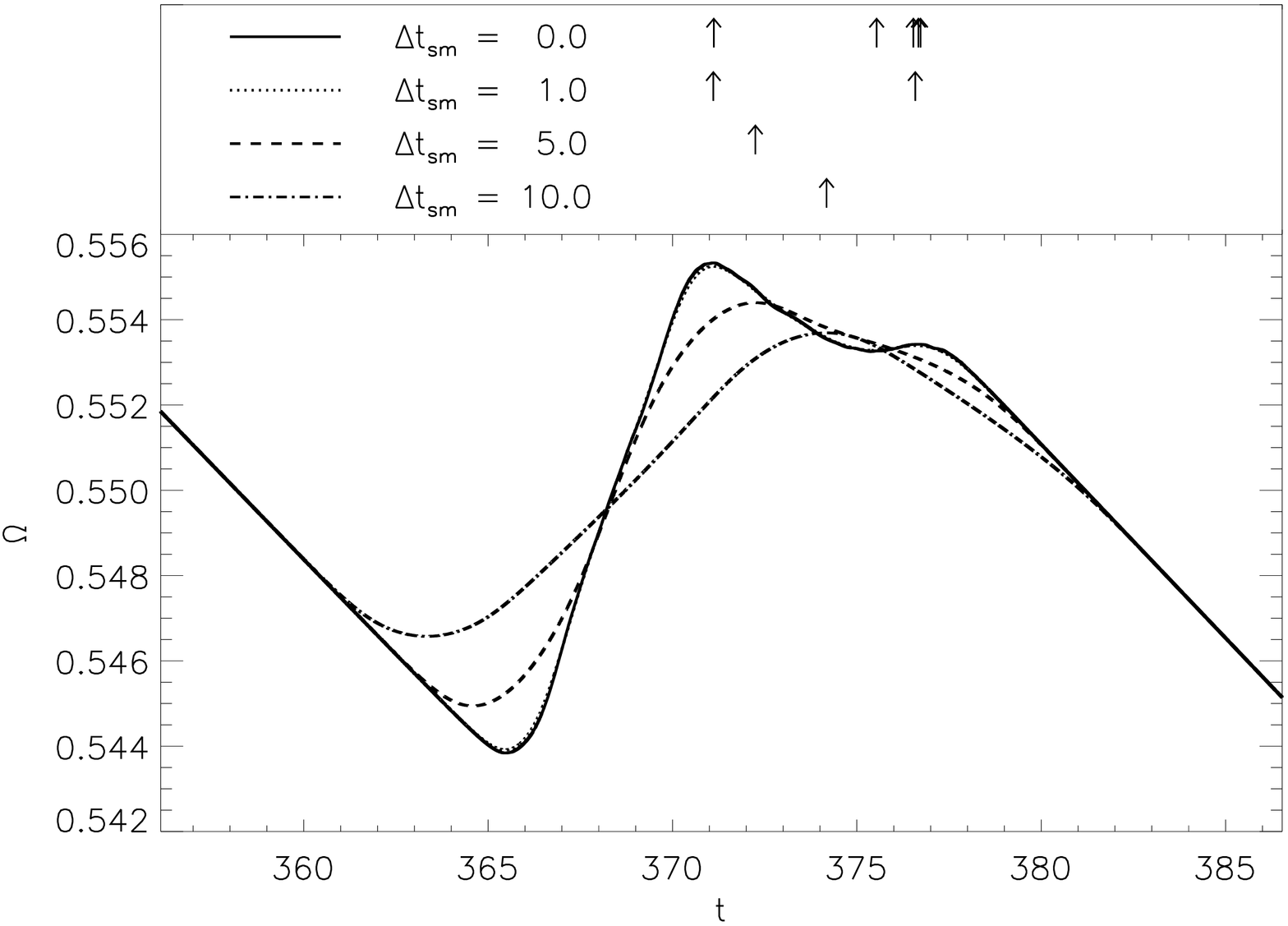}
\caption{Close-up of glitches identified using the glitch-finding algorithm described in Section~\ref{sec:define}, as a function of the degree of smoothing.  The plots cover the interval $356<t<386$ (taken from the $\Omega$ curve graphed in Fig.~\ref{fig:ch5:smooth_full}).  The curves are smoothed with a top-hat window function of width $\Delta t_{\rm{sm}}=0.0$, 1.0, 5.0 and 10 (\emph{solid}, \emph{dotted}, \emph{dashed}, \emph{dot-dashed} curves respectively).  The number and position of glitches, shown as arrows in the \emph{top} panel, changes with $\Delta t_{\rm{sm}}$; for $\Delta t_{\rm{sm}}\geq 5$, one glitch is detected.  Simulation parameters:  $V_0 = 16.6$, $\eta=1$, $\Delta V_i/V_0=0.0$, $R=12.5$, $\Delta x=0.15$, $\Delta t=5\times 10^{-4}$, $N_{\rm{EM}}/I_{\rm{c}}=10^{-3}$, $\Omega(t=0) = 0.8$, $n_{\rm{pin}}/n_{\rm{F}}=0.97$.}
\label{fig:ch5:define_close}
\end{figure}

\begin{figure*}
\hspace{1.cm} \includegraphics[scale=0.45]{fig7a.epsi}
\includegraphics[scale=0.45]{fig7b.epsi}
\includegraphics[scale=0.45]{fig7c.epsi}
\includegraphics[scale=0.45]{fig7d.epsi}
\hspace{-1.5cm} \includegraphics[scale=1.0]{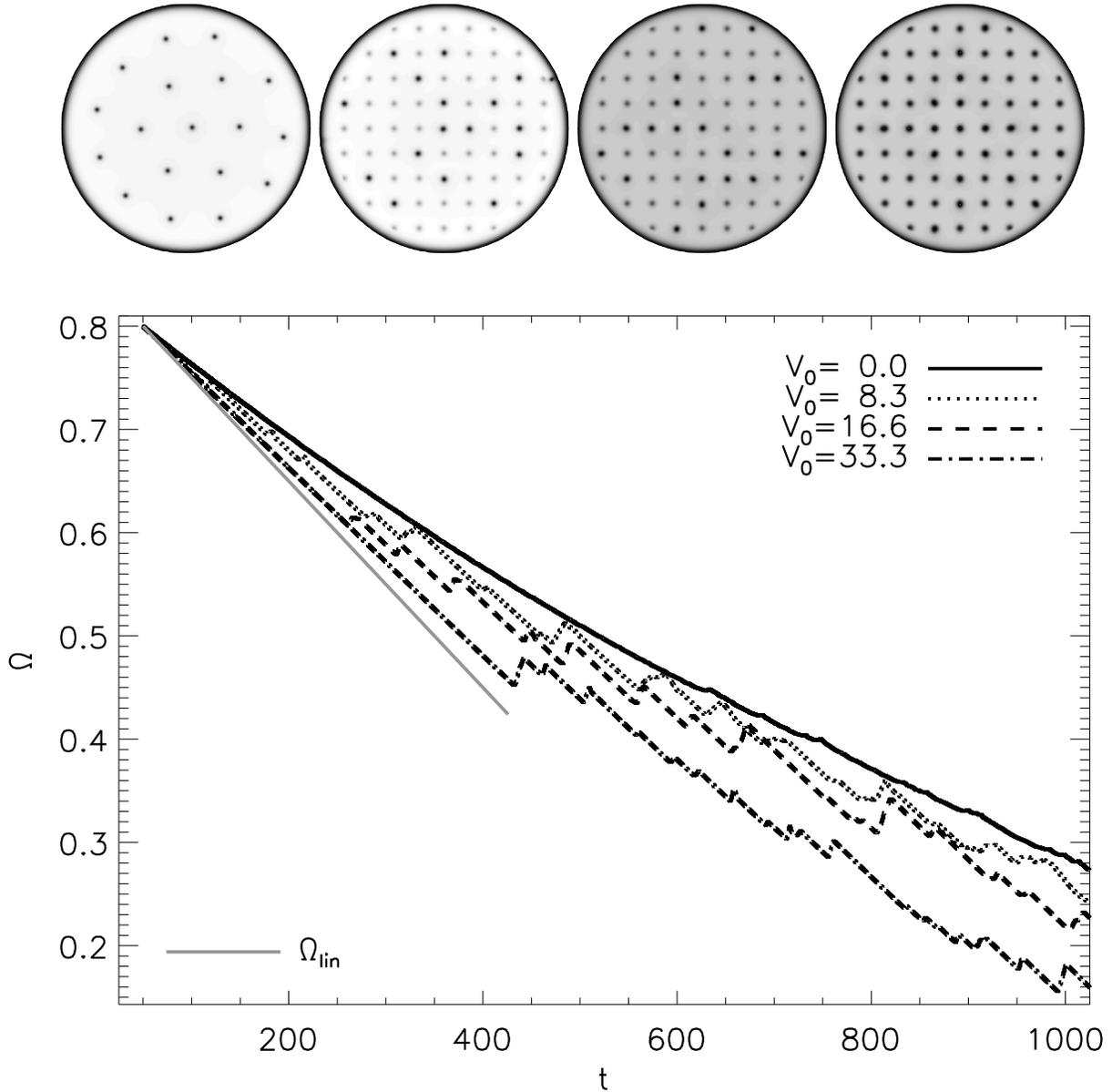}
\caption{Vortex configurations and spin-down curves for different pinning strengths $V_0$.  \emph{Top}: Greyscale plots of the final ($t=1000$) superfluid density for $V_i=0$, 8.3, 16.6 and 33.7 (\emph{left} to \emph{right} respectively).  The density greyscale runs from dark (low density) to light (high density).  \emph{Bottom}:  Angular velocity of the crust as a function of time $\Omega(t)$ for $V_i=0$, 8.3, 16.6 and 33.7 (\emph{solid}, \emph{dotted}, \emph{dashed} and \emph{dot-dashed} respectively).  Simulation parameters: $\eta=1$, $\Delta V_i/V_0=0.0$, $R=12.5$, $\Delta x=0.15$, $\Delta t=5\times 10^{-4}$, $N_{\rm{EM}}/I_{\rm{c}}=10^{-3}$, $\Omega(t=0) = 0.8$, $n_{\rm{pin}}/n_{\rm{F}}=0.97$, $\Delta t_{\rm{sm}}=1.0$.}
\label{fig:ch5:om_V}
\end{figure*}

\begin{figure*}
\includegraphics[scale=0.365,angle=90]{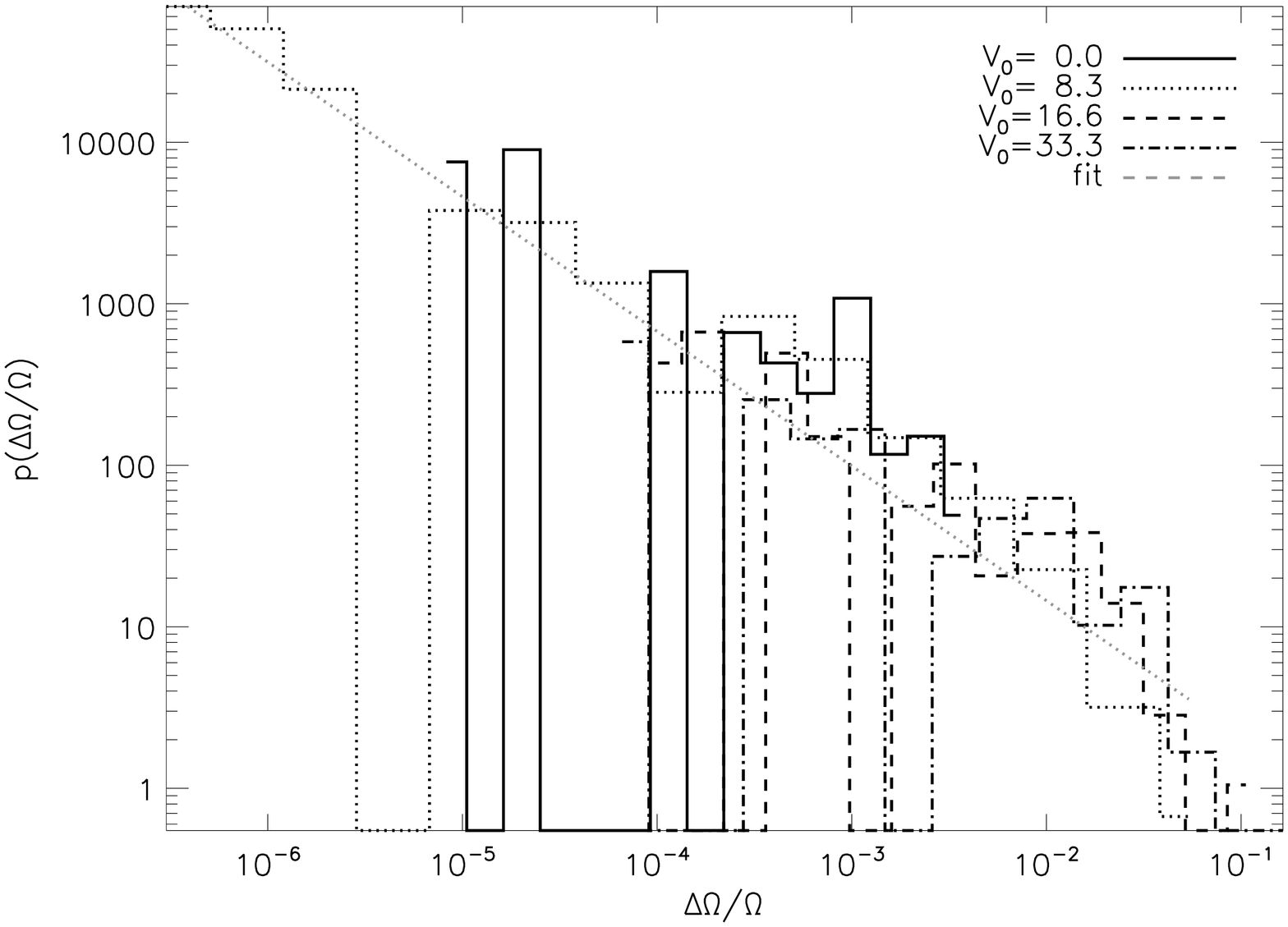}
\includegraphics[scale=0.365,angle=90]{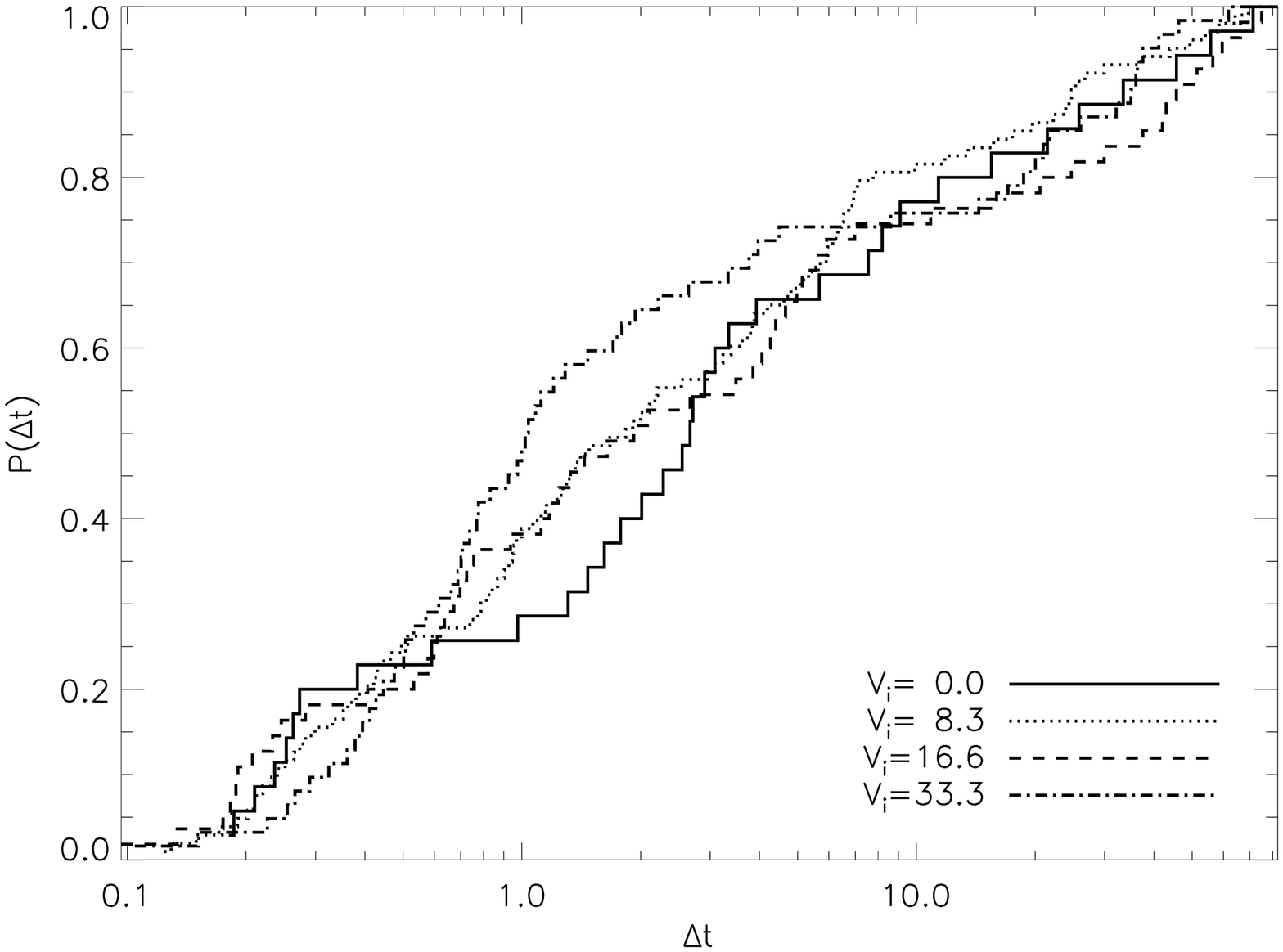}
\caption{Glitch statistics for different pinning strengths corresponding to the $\Omega$ curves graphed in Fig.~\ref{fig:ch5:om_V}.  \emph{Left}:  Probability density function (pdf) of fractional glitch sizes $p(\Delta\Omega/\Omega)$ for $V_0=0$, 8.3, 16.6 and 33.7 (\emph{solid}, \emph{dotted}, \emph{dashed} and \emph{dot-dashed} respectively) logarithmically binned and graphed on log-log axes.  The \emph{dotted} curve is a power-law least squares fit to the \emph{dotted} histogram with index $-0.818$.  \emph{Right}:  Cumulative probability of glitch waiting times $P(\Delta t)$ graphed on log-linear axes for $V_0=0$, 8.3, 16.6 and 33.7 (\emph{solid}, \emph{dotted}, \emph{dashed} and \emph{dot-dashed} respectively). Simulation parameters:  $\eta=1$, $\Delta V_i/V_0=0.0$, $R=12.5$, $\Delta x=0.15$, $\Delta t=5\times 10^{-4}$, $N_{\rm{EM}}/I_{\rm{c}}=10^{-3}$, $\Omega(t=0) = 0.8$, $n_{\rm{pin}}/n_{\rm{F}}=0.97$, $\Delta t_{\rm{sm}}=1.0$.}
\label{fig:ch5:V_stats}
\end{figure*} 

\section{Automated glitch finder}\label{sec:define}

Our first task is to define exactly what we mean by a glitch.  In practice, glitches in pulsar data are extracted `by hand', by looking for discontinuities in the slope and/or curvature of the pulse phase.  Recently, \citet{Chukwude:2010} employed an automated glitch-finding algorithm, based on a least-squares goodness-of-fit parameter, to identify micro-glitches.  Since our simulations provide extremely high time resolution ($\Omega \Delta t\sim 10^{-4}$~revolutions/time step compared to $\Omega \Delta t\sim 10^{5}$~revolutions/time step for the observations), we resolve the glitch spin-up phase; we can identify glitches directly by looking for spin-up events, rather than resorting to polynomial fitting of the phase residuals.  The difference between simulation and astrophysical time scales is also reflected in which physical processes can be resolved.  In our simulations, we track the motion of individual vortices between pinning sites, whereas observed glitches appear as unresolved spin-up events, which likely result from a multi-generation vortex cascade.

For the purposes of this investigation, we employ the following simple algorithm for identifying glitches in $\Omega(t)$ curves.  
\begin{enumerate}
 \item Smooth $\Omega(t)$ with a top-hat window function $w(t)$ of width $\Delta t_{\rm{sm}}$, viz. $\Omega_{\rm{sm}}(t)=\int_0^t dt^{\prime}w(t-t^{\prime})\Omega(t^{\prime})$.
 \item Calculate $\Delta\Omega_{\rm{sm}}(t_i)=\Omega_{\rm{sm}}(t_i)-\Omega_{\rm{sm}}(t_{i-1})$ for all discrete time points $t_i$ in the simulated range.
 \item\label{gt0} Find all $t_s$ for which  $\Delta\Omega_{\rm{sm}}(t_s)>0$.  For each $t_s$, find the smallest $t_g(>t_s)$ such that $\Delta\Omega_{\rm{sm}}(t_g)\geq0$ and $\Delta\Omega_{\rm{sm}}(t_{g+1})<0$ (points A and C in Fig.~\ref{fig:ch5:defineglitch} meet this criterion), and record $\Omega_{\rm{sm}}(t_g)$.
 \item Locate the global $\Omega_{\rm{sm}}(t)$ minimum between $t_{g-1}$ and $t_{g}$, occurring at $t_{g,\rm{min}}$ (point B in Fig.~\ref{fig:ch5:defineglitch}).  Define glitch size as $\Delta\Omega/\Omega=[\Omega_{\rm{sm}}(t_g)-\Omega_{\rm{sm}}(t_{g,\rm{min}})]/\Omega(t_g)$ (the vertical distance between the \emph{dashed} lines in Fig.~\ref{fig:ch5:defineglitch}), and waiting time as $\Delta t=t_g-t_{g-1}$ (the horizontal distance between points A and C in Fig.~\ref{fig:ch5:defineglitch}).
\end{enumerate}
The principal weakness of this algorithm is the stipulation in step~\ref{gt0} that $\Delta\Omega_{\rm{sm}}>0$, which does not capture decreases in $\langle \hat{L}_z\rangle$ that are smaller than $N_{\rm{EM}}\Delta t$.  Depending on the moment of inertia of the crust, $\Delta\Omega_{\rm{g}}$ due to real glitches may not be greater than the negative $\Delta\Omega_{\rm{sm}}$ due to the spin-down torque [$(N_{\rm{EM}}/I_{\rm{c}})\Delta t$], and hence this algorithm is not suitable for observational data. 

The \emph{arrows} under the spin-down curve in the top panel of Fig.~\ref{fig:ch5:residual} indicate the positions of glitches found using the above algorithm for a curve smoothed with a top-hat window function of width $\Delta t_{\rm{sm}}=0.1$.  The glitch sizes and waiting times are then used to construct probability density functions.  In Section~\ref{sec:canonical} we discuss the two bottom panels of Fig.~\ref{fig:ch5:residual}, which demonstrate how we subtract a linear fit to the pre-glitch spin-down curve (\emph{bottom left} panel), and the resulting phase residual curve (\emph{bottom right} panel). 

\begin{center}
\begin{table*}
\begin{tabular}{ | c | c | c | c | c | c | c | c |}
\hline
 $\Delta t_{\rm{sm}}$ 	&$N_{\rm{g}}$ 	&$10^4\langle\Delta\Omega/\Omega\rangle$&$\langle\Delta t\rangle$ &$r_{\rm{p}}$	&$\lambda$ ($p_{\rm{KS}}$) & $\gamma$ ($p_{\rm{KS}}$)\\
 \hline
 0.0	 &105	&51.97	&7.176	&$-0.057$&0.865(0.000)	&$-0.994$(0.745)\\
 1.0	 &27	&188.0	&27.93	&0.050&	0.035(0.449)	&$-0.847$(0.613)\\
 5.0	 &16	&264.9	&50.51	&0.428&	0.016(0.312)	&$-0.710$(0.709)\\
 10.0	 &13	&253.5	&62.31	&0.334&	0.011(0.130)	&$-1.104$(0.948)\\\hline
\end{tabular}
\caption{Summary of simulation results for different values of the smoothing width $\Delta t_{\rm{sm}}$ [corresponding $\Omega(t)$ curves shown in Fig.~\ref{fig:ch5:smooth_full}]. $N_{\rm{g}}$ is the number of glitches detected, $\langle \Delta\Omega/\Omega\rangle$ is the mean glitch size, $\langle\Delta t\rangle$ is the mean waiting time, and $r_{\rm{p}}$ is the Pearson correlation coefficient; $r_{\rm{p}}=1$, $-1$ and 0 indicate positive, negative and no correlation respectively.  $\lambda$ is the mean glitch rate that parametrises the cumulative distribution of waiting times, $\gamma$ is the power-law index that parametrises the pdf of glitch sizes.  The K-S probability $p_{\rm{KS}}$ is given in parenthesis;  the fit can be rejected at the $1-p_{\rm{KS}}$ confidence level.  Simulation parameters:  $V_0 = 16.6$, $\eta = 1$, $\Delta V_i/V_0=0.0$, $R=12.5$, $\Delta x=0.15$, $\Delta t=5\times 10^{-4}$, $N_{\rm{EM}}/I_{\rm{c}}=10^{-3}$, $\Omega(t=0) = 0.8$, $n_{\rm{pin}}/n_{\rm{F}}=0.97$.}
\label{tab:ch5:define}
\end{table*}
\end{center}

\section{Canonical glitch}\label{sec:canonical}
Let us now examine a canonical example, which demonstrates the ability of the model in Section~\ref{sec:ch5:model} to generate astrophysically relevant results.  Table~\ref{tab:ch5:pulsar} lists the set of canonical parameters for this investigation.  In the first column, the simulation parameters are given in dimensionless units; their dimensional counterparts are provided in the second column.  The third and fourth columns quote the range of values associated with a typical pulsar in dimensionless and dimensional units respectively.

\subsection{Set up} 

The nuclear lattice of the neutron star inner crust is represented by an $11\times11$ square grid of equal-strength pinning sites ($V_0=16.6$, $\Delta V_i/V_0=0.0$), within a box of side length 15, with crust radius $R=12.5$ (in dimensionless units).  The simulation grid is a $200\times200$ rectangular grid; the greyscale plots appear circular because $|\psi|^2=0$ outside the crust.  The time step and grid spacing are $\Delta t = 5\times 10^{-4}$ and $\Delta x=0.15$ respectively.  

The initial conditions are established by accelerating the crust from rest to $\Omega = 0.8$ instantaneously and allowing the system to reach a steady state (defined as when the fractional change in total energy during one time step drops below $10^{-5}$).  Once a steady-state solution is obtained, the external spin-down torque $N_{\rm{EM}}(=10^{-3}I_{\rm{c}})$ is switched on, and the superfluid-crust system is allowed to evolve.  

\subsection{Spin-down curve}
The spin down observed after $N_{\rm{EM}}$ switches on is \emph{spasmodic}.  By watching movies of how the density field $|\psi(\bm{x},t)|^2$ evolves, we can easily see why:  pinning prevents vortices from moving smoothly towards the boundary as the crust decelerates; instead, vortices hop between pinning sites, resulting in step-like decreases in $\langle L_{\rm{z}}\rangle$.  A graph of $\langle \hat{L}_z\rangle$ versus time is presented in Fig.~\ref{fig:ch5:L_define}.   The positions of glitches identified using the automated glitch-finding algorithm described in Section~\ref{sec:define} are marked with \emph{arrows}.  Equation~(\ref{eq:ch5:feedback}) tells us that each step down in $\langle \hat{L}_z\rangle$ is accompanied by a step up in $\Omega$ (so long as the change in $\langle \hat{L}_z\rangle$ exceeds $N_{\rm{EM}}\Delta t$).  We see this clearly by comparing the \emph{top} panel of Fig.~\ref{fig:ch5:residual} and Fig.~\ref{fig:ch5:L_define}.  

A close-up of one of the glitches (bracketed by the \emph{dotted} rectangle in the \emph{top} panel of Fig.~\ref{fig:ch5:residual}) is shown in the \emph{bottom left} panel of Fig.~\ref{fig:ch5:residual}.  Does this look like a pulsar glitch?  An important difference is that the spin-down rate before and after the glitch does not change; the $\Omega(t)$ curve after the glitch parallels the linear timing model describing the pre-glitch spin down.  This contrasts with some (not all) observed glitches, which frequently produce a permanent increase in the spin-down rate \citep{Wong:2001p192}.  The simulation output also lacks a post-glitch recovery phase, during which spin down is faster than the global average.  Given that time scales associated with the post-glitch recovery phase, \emph{i.e.} $\sim 10^{29}\hbar/(n_0g)$, [typically lasting days to weeks \citep{Wong:2001p192}] are associated with coupling of the superfluid to a viscous fluid component not included in these simulations, this result is unsurprising \citep{VanEysden:2010}.  In our simulations, the time for $\Omega(t)$ to return to its pre-glitch value is proportional to the size of the glitch and is comparable to the sound-crossing of the system, $R/c_{\rm{s}}$ \citep{Warszawski:2010lattice}. 

Practical pulsar glitch detection involves comparing measured pulse times of arrival (TOAs) against TOAs predicted by a spin-down model.  When a glitch occurs, pulses arrive more frequently than the model predicts, and the phase residuals (the difference between the number of revolutions predicted by the timing model and the number of revolutions inferred from the TOAs) turn negative and grow linearly with time.  In the \emph{bottom right} panel of Fig.~\ref{fig:ch5:residual}, we plot the difference between the simulation output and a timing model defined by
\begin{equation}
 \Omega_{\rm{model}}(t)=\Omega(t=0)+\dot{\Omega}t~,
 \label{eq:timingmodel}
\end{equation}
for the interval bracketed by the \emph{dotted} rectangle in the \emph{top} panel.  The timing model, with slope $\dot{\Omega} = -8.98\times 10^{-4}$, has been extracted by performing a least squares fit of Eq.~(\ref{eq:timingmodel}) to $\Omega(t)$ ($710\leq t\leq 730$) and is graphed as a \emph{dotted} curve in the \emph{bottom left} panel of Fig.~\ref{fig:ch5:residual}.  The phase residual $\Delta\phi$, in units of radians, is therefore
\begin{equation}
 \Delta\phi(t) = \int_0^t\Omega_{\rm{model}}(t)dt-\int_0^t\Omega(t)dt~.
\end{equation}
The shape of the residual curve, graphed in the \emph{bottom right} panel of Fig.~\ref{fig:ch5:residual}, qualitatively resembles observations; see Fig.~1 in \cite{Shemar:1996p677}, for example.

\subsection{Statistics}\label{subsec:canon_stats}

We now construct probability density functions (pdfs) of sizes and waiting times using the automated glitch finder.  Motivated by the hypothesis that glitches are produced by avalanche dynamics, we look for evidence that the pdfs of sizes and waiting times are distributed as power-laws [$p(\Delta\Omega_{\rm{g}}/\Omega)\propto (\Delta\Omega_{\rm{g}}/\Omega)^{\gamma}$] and exponentials [$p(\Delta t)= \lambda \exp (-\lambda\Delta t)$] respectively.  Evidence is building that pdfs of this form yield good fits to observational data \citep{Melatos:2008p204}.  To fit these functional forms, we find the parameter that minimises the $D$ statistic (the maximum distance between the cumulative distribution of the model and that of the data).  The fit can be rejected with $1-p_{\rm{KS}}$ confidence, where $p_{\rm{KS}}$ is the Kolmogorov-Smirnov probability \citep{Press:1992}.

To begin with, we note that the smoothing width $\Delta t_{\rm{sm}}$ changes the glitch statistics, as summarised in Table~\ref{tab:ch5:define}.  Glitch positions are marked on the spin-down curve in Fig.~\ref{fig:ch5:smooth_full}  using \emph{arrows}  for $\Delta t_{\rm{sm}}=0.0$, 1.0, 5.0 and 10.0 (\emph{solid}, \emph{dotted}, \emph{dashed} and \emph{dot-dashed} curves respectively).  The curves have been separated vertically for clarity.  As $\Delta t_{\rm{sm}}$ increases, the number of glitches decreases from $N_{\rm{g}}=105$ for $\Delta t_{\rm{sm}}=0.0$ to $N_{\rm{g}}=13$ for $\Delta t_{\rm{sm}}=10.0$.  The mean waiting time $\langle\Delta t\rangle$ and glitch size $\langle\Delta \Omega/\Omega\rangle$ increase five- and nine-fold respectively.  The trend in $\langle\Delta \Omega/\Omega\rangle$ versus $\Delta t_{\rm{sm}}$ is evident in the size pdfs graphed on log-log axes in the \emph{left} panel of Fig.~\ref{fig:ch5:smooth_stats}:  the pdfs for longer $\Delta t_{\rm{sm}}$ are weighted further towards the large-($\Delta \Omega/\Omega$) end.  Notably, the best-fit power-law index, $\gamma$ (we warn the reader that this is \emph{not} the same quantity as the dissipation parameter $\gamma$ that appears in Eq.~\ref{eq:GPE_dimensionless}), does not change dramatically with $\Delta t_{\rm{sm}}$; we find $\gamma = -0.994$ for  $\Delta t_{\rm{sm}}=0$ ($\gamma$ can be rejected at the $25.5\%$ confidence level), and $\gamma = -1.104$ for $\Delta t_{\rm{sm}}=10.0$ (rejected at the $5.3\%$ confidence level).   

On the other hand, increasing $\Delta t_{\rm{sm}}$ changes markedly the shape of the cumulative waiting-time distribution (\emph{right} panel of Fig.~\ref{fig:ch5:smooth_stats}, graphed on linear axes). For $\Delta t_{\rm{sm}}\leq 1.0$, the cumulative probability is dominated by short waiting times (\emph{i.e.} convex down).  For $\Delta t_{\rm{sm}}\geq 5.0$, waiting times are more evenly distributed.  Exponential fits to the cumulative probability return mean glitch rates between $\lambda =0.011$ (for $\Delta t_{\rm{sm}}=10.0$; rejected at the $87\%$ confidence level) and $\lambda = 0.87$ (for $\Delta t_{\rm{sm}}=0.0$; rejected at the with near certainty).  An exponential functional form for the cumulative distribution for $\Delta t_{\rm{sm}}=1.0$ is rejected with the least confidence ($55.1\%$).

Glitch positions and distributions depend on $\Delta t_{\rm{sm}}$ because smoothing $\Omega(t)$ affects the morphology of individual glitches.  Understanding this tendency is essential to interpreting observational data properly using our algorithm.  For example, increasing $\Delta t_{\rm{sm}}$ reduces the size of a glitch, and in some cases erases the glitch altogether.  To visualise this, in Fig.~\ref{fig:ch5:define_close} we graph an interval of $\Omega(t)$ containing an obvious bump (when viewed on the scale graphed in Fig.~\ref{fig:ch5:smooth_full}) for different smoothing widths.  The \emph{arrows} in the \emph{top} panel mark the positions of glitches identified by the automated algorithm for different $\Delta t_{\rm{sm}}$.  In the unsmoothed ($\Delta t_{\rm{sm}}=0.0$, \emph{solid}) and $\Delta t_{\rm{sm}}=1.0$ (\emph{dotted}) curves, we see two obvious bumps at $t=371$ and $367.5$ (for $\Delta t_{\rm{sm}}=0.0$ the algorithm finds $8$ glitches that are not visible by eye on the scale plotted).  However, for $\Delta t_{\rm{sm}}=5.0$ and 10.0, the two bumps merge, and the algorithm finds only one glitch.  For $\Delta t_{\rm{sm}}=10.0$, the glitch is $\sim 16\%$ smaller than for $\Delta t_{\rm{sm}}=5.0$.  Increased waiting times are a direct corollary of glitches being erased by smoothing.

After watching movies of $|\psi|^2(\bm{x},t)$, we conclude that glitches too small to be discerned by eye on the scale depicted in Fig.~\ref{fig:ch5:define_close} are associated with `jiggling', not unpinning.  By `jiggling' we mean vortices wobble around their pinning site without unpinning.  Multi-peaked glitches arise from multi-stage vortex motion (\emph{i.e.} one vortex moves, causing a second vortex to move a short time later), precisely the collective avalanche behaviour we are seeking. Observationally, we can only detect pulsar glitches resulting from the motion of many vortices (the smallest observed glitches correspond to $\sim10^7$ unpinnings).  For this reason, we henceforth adopt $\Delta t_{\rm{sm}}=1.0$, which also happens to produce the most exponential $P(\Delta t)$.  Thus we retain multi-peaked glitches like the one in Fig.~\ref{fig:ch5:define_close}, whilst excluding glitches not associated with unpinning.  

An important recent result from pulsar data analysis is the absence of a reservoir effect:  glitch size is uncorrelated with waiting time \citep{Wong:2001p192,Melatos:2008p204}.  In Table~\ref{tab:ch5:define} we list the Pearson correlation coefficient, $r_{\rm{p}}$, relating $\Delta\Omega/\Omega$ and $\Delta t$ as a function of $\Delta t_{\rm{sm}}$.  For $\Delta t_{\rm{sm}}< 5.0$, the two quantities are uncorrelated ($r_{\rm{p}}=-0.057$ and 0.050 for $\Delta t_{\rm{sm}}= 0.0$ and 1.0 respectively).  For $\Delta t_{\rm{sm}}\geq 5.0$, $r_{\rm{p}}$ differs significantly from zero (0.428 and 0.334 for $\Delta t_{\rm{sm}}=5.0$ and 10.0 respectively), suggesting that sizes and waiting times are correlated.  The interpretation of this correlation depends on the glitch definition.  Since the correlation emerges for smoothing scales that erase multi-peaked glitches and `jiggling', it is likely that this correlation exists if the glitch-finding algorithm does not distinguish between single- and multiple-vortex motion.  Simulations large enough to house several isolated regions of unpinning activity should not exhibit such a correlation \citep{Melatos:2008p204}.  

As observational duty cycles and measurement precision improve, e.g. with the advent of multibeam telescopes, astronomers will begin to resolve the spin-up phase of a glitch, multi-peaked glitches, and post-glitch oscillations \citep{Kramer:2010p12535}.  Our results suggest that improved measurement precision (equivalent to reducing $\Delta t_{\rm{sm}}$) should lead to a large increase in the number of observed glitches, without altering the shape of the size distribution.  For example, reducing $\Delta t_{\rm{sm}}$ five-fold doubles the number of glitches.  Conversely, the mean glitch rate increases.  More speculatively, we also predict that the low-$\Delta\Omega/\Omega$ end of the glitch size pdf is dominated by ongoing, non-collective vortex creep \citep[see][for example]{Alpar:1984p6781}, likely characterised by a Gaussian cut off.

\section{Pinning physics}\label{sec:pinning}

\begin{center}
\begin{table*}
\begin{tabular}{ | c | c | c | c | c | c | c |}
\hline
 $V_0$ 		&$N_{\rm{g}}$ 	&$10^4\langle\Delta\Omega/\Omega\rangle$&$\langle\Delta t\rangle$ 	&$A$ 	&$\gamma $($p_{\rm{KS}}$)	&$\lambda$($p_{\rm{KS}}$)\\
 \hline
 0.0	 &16	&13.00	&24.33	&1.72		&$-0.548$(0.809)	&0.042(0.895)\\
 8.3	 &48	&62.90	&18.01	&33.85		&$-0.818$(0.310)	&0.064(0.459)\\
 16.6	 &23	&187.3	&31.58	&28.51		&$-0.834$(0.762)	&0.029(0.761)\\
 33.3	 &26	&239.9	&21.62	&57.93		&$-0.827$(0.293)	&0.046(0.474)\\
 \hline
\end{tabular}
\caption{Glitch size and waiting-time statistics for different pinning potentials $V_0$ corresponding $\Omega(t)$ curves shown in Fig.~\ref{fig:ch5:om_V}.  $N_{\rm{g}}$ is the number of glitches detected, $\langle \Delta\Omega/\Omega\rangle$ is the mean glitch size, $\langle\Delta t\rangle$ is the mean waiting time, and $A$ is the activity parameter.  $\lambda$ is the mean glitch rate that parametrises the cumulative distribution of waiting times, $\gamma$ is the power-law index that parametrises the pdf of glitch sizes.  The K-S probability $p_{\rm{KS}}$ is given in parenthesis;  the fit can be rejected at the $1-p_{\rm{KS}}$ confidence level.  Simulation parameters:  $\eta = 1$, $\Delta V_i/V_0=0.0$, $R=12.5$, $\Delta x=0.15$, $\Delta t=5\times 10^{-4}$, $N_{\rm{EM}}/I_{\rm{c}}=10^{-3}$, $\Omega(t=0) = 0.8$, $n_{\rm{pin}}/n_{\rm{F}}=0.97$, $\Delta t_{\rm{sm}}=1.0$.}
\label{tab:ch5:pin_stat}
\end{table*}
\end{center}

Systematic differences in the pinning strength from pulsar to pulsar may result from different cooling histories \citep{Jones:2001p167}, which alter the morphology and abundance of crystal defects.  The role of the pinning strength $V_0$ in glitch physics is crucial.  If pinning is weak, vortices spread out smoothly and homologously, allowing the superfluid to decelerate in sympathy with the crust; differential rotation does not build up in stress reservoirs.  On the other hand, if pinning is strong, it may stiffen the vortex lattice to the point where the crust cracks before the vortices are unpinned by the Magnus force \citep{Horowitz:2009p10413}.  Intuitively, we expect that stronger pinning results in larger glitches; either vortices move further once they have unpinned, or more vortices unpin in order to reverse the crust-superfluid shear.  In this section we use simulations to test these intuitive relations between glitch distributions and pinning strength.  

\subsection{Equal pinning potentials}\label{subsec:pinning:strength}

We begin by examining how spin down depends on $V_0$, in the situation where all the pinning sites are of equal strength ($\Delta V_i=0$).  The four greyscale plots of superfluid density at the top of Fig.~\ref{fig:ch5:om_V} are the final states ($t=1025$) for simulations with $V_0=0$, 8.3, 16.6 and 33.3 (\emph{left} to \emph{right} respectively).  Vortices are easily identified as darker spots.  The lighter spots represent unoccupied pinning sites.  All vortices, except for one in the north-east quadrant of the second panel, are pinned. 

Without pinning, the vortex lattice expands smoothly and homologously, $\langle \hat{L}_z\rangle$ decreases smoothly, and the superfluid exerts a continuous spin-up torque on the crust via Eq.~(\ref{eq:ch5:feedback}).  The spin-up torque lessens the gradient of $\Omega$ (the \emph{black curve} in Fig.~\ref{fig:ch5:om_V}) when compared with no-feedback, linear spin down (the \emph{solid grey} curve). Small bumps in the \emph{black} curve in Fig.~\ref{fig:ch5:om_V} are numerical noise. 

We now switch on pinning, via an $11\times 11$ equidistant, equal-strength grid.  $\langle \hat{L}_z\rangle$ exhibits step-like decreases, accompanied by jumps in $\Omega$, in the \emph{dotted}, \emph{dashed} and \emph{dot-dashed}  curves in Fig.~\ref{fig:ch5:om_V}.   The first glitch (at $t_1$) occurs later when the pinning is stronger, viz. $t_1=200$ for $V_0=8.3$ compared to $t_1=440$ for $V_0=33.3$.  The delay can be understood by considering the velocity shear between pinned vortices and the superfluid in the absence of feedback:
\begin{equation}
 \Delta v(r,t) = v_{\rm{s}}-[\Omega(t=0)-(N_{\rm{EM}}/I_{\rm{c}})t]r~.
\end{equation}
Whilst vortices remain pinned, $v_{\rm{s}}$ is approximately constant.  Therefore, it takes longer for the Magnus force to unpin strongly pinned vortices.

Even in the strongest pinning case, before any unpinning has occurred, the spin-down curve is less steep than $\Omega(t) = tN_{\rm{EM}}/I_{\rm{c}}$ (the \emph{solid grey} curve in Fig.~\ref{fig:ch5:om_V}).  The reason is subtle; it is discussed in detail in \cite{Warszawski:2010lattice}.  While remaining pinned, vortices respond to increasing differential rotation by migrating smoothly to the outer edge of their pinning site.  Hence the crust experiences a gradual spin-up torque.  Put another way, pinning is not a binary condition; it is weakened gradually as the shear grows.

At the end of the simulation, $\Omega$ is highest ($\Omega=0.29$) for the weakest pinning and lowest ($\Omega=0.16$) for the strongest pinning, because stronger pinning potentials can sustain greater levels of stress. [Note that this is \emph{not} the reservoir effect \citep{Link:1999p2179,Wong:2001p192,Melatos:2008p204}, which predicts a correlation between glitch size and waiting time not found in observational data.] In other words, the areal density of remaining vortices after a fixed amount of time is proportional to the pinning strength.  

At this point we draw attention to vast discrepancies between typical pulsar parameters and the physical scales modelled by our simulations.  The extremely short simulation time ( $10^{-21}~\rm{s}$ in physical units), and the extremely high angular velocity ($\Omega=10^{24}~\rm{Hz}$ in pulsar units compared to the typical pulsar angular velocity $\Omega\sim 10~\rm{Hz}$) represent highly non-adiabatic conditions.  

\subsection{Statistics}\label{subsec:pinning:strength_stats}

A key goal of this paper is to understand the observed scale invariance of glitch sizes, which is believed to arise from collective vortex behaviour \citep{Melatos:2008p204}.  Global properties of the glitch distribution, such as the number of glitches, mean glitch size, and mean waiting time for each $V_0$ studied in Section~\ref{subsec:pinning:strength} are listed in Table~\ref{tab:ch5:pin_stat}; along with the activity parameter 
\begin{equation}
A=N_{\rm{g}}\sum \Delta\Omega_{\rm{g}}/\sum \Delta t~,
\end{equation} 
used by pulsar astronomers to characterise the total glitch-induced spin up of a pulsar \citep{Lyne:1998p12543}.  

In Fig.~\ref{fig:ch5:V_stats} we graph logarithmically binned size pdfs on log-log axes.  Our glitch-finding algorithm identifies 16, 48, 23 and 26 glitches for $V_0=0$, 8.3, 16.6 and 33.3 respectively.  The pdfs for simulations with pinning ($V_0>0$) exhibit a  bias towards larger glitches for larger $V_0$, which also shows up as a $\sim$four-fold increase in mean glitch size for $V_0=33.3$ compared to $V_0=8.3$.  Power-law fits to the pdfs yield similar indices (listed with the K-S probability in parenthesis in the penultimate column of Table~\ref{tab:ch5:pin_stat}) for $8.3\leq V_0\leq 33.3$.  However, the confidence level with which the fits can be rejected ranges from 23.8\% for $V_0=16.6$ to 70.7\% for $V_0=33.3$.  We remind the reader that glitches detected in the $V_0=0$ case arise from numerical noise, rather than vortex unpinning; this assertion is supported by the small mean glitch size ($10^4\langle \Delta\Omega/\Omega\rangle=13.00$). 

In the \emph{right} panel of Fig.~\ref{fig:ch5:V_stats} we graph cumulative distributions of waiting times for the simulations shown in Fig.~\ref{fig:ch5:om_V}.  Mean glitch rates, $\lambda$, derived from exponential fits, are listed in the final column of Table~\ref{tab:ch5:pin_stat}.  In \cite{Melatos:2008p204}, it was found that waiting times for seven of the nine pulsars that have glitched more than six times are well represented by exponential distributions.  Taking the results in Fig.~\ref{fig:ch5:V_stats} all together, we suggest that $\lambda$ does not depend monotonically on $V_0$ because there is another degree of freedom, $\langle \Delta\Omega/\Omega\rangle$, through which the additional stress can be released.

The activity parameter does not appear to scale monotonically with $V_0$.  However, $A$ almost doubles between $V_0=8.3$ and $V_0=33.3$.  We do not have a physical explanation for this behaviour at this time.  

\subsection{Pinning site abundance}\label{subsec:pinning:sites}
\begin{center}
\begin{table*}
\begin{tabular}{ | c| c| c| c| c| c| c|}
\hline
 $n_{\rm{pin}}/n_{\rm{F}}$ &$N_{\rm{g}}$ 	&$10^4\langle\Delta\Omega/\Omega\rangle$	&$\langle\Delta t\rangle$ 	&$A$	&$\gamma$($p_{\rm{KS}}$) &$\lambda$($p_{\rm{KS}}$)\\
 \hline
 0.11		 	&25	&14.84	&37.10	&2.02	&$-0.801$(0.911)&0.053(0.885)\\
 0.52		 	&27	&187.97	&27.93	&36.5	&$-0.846$(0.615)&0.035(0.450)\\
 0.97		 	&48	&92.41	&17.36	&51.5	&$-0.875$(0.432)&0.071(0.099)\\
 1.9		 	&49	&111.7	&15.95	&68.8	&$-0.889$(0.312)&0.068(0.362)\\
 \hline
\end{tabular}
\caption{Glitch size and waiting-time statistics for different $n_{\rm{pin}}/n_{\rm{F}}$ [corresponding $\Omega(t)$ curves shown in Fig.~\ref{fig:ch5:om_V}].  $N_{\rm{g}}$ is the number of glitches detected, $\langle \Delta\Omega/\Omega\rangle$ is the mean glitch size, $\langle\Delta t\rangle$ is the mean waiting time, and $A$ is the activity parameter.  $\lambda$ is the mean glitch rate that parametrises the cumulative distribution of waiting times, $\gamma$ is the power-law index that parametrises the pdf of glitch sizes.  The K-S probability $p_{\rm{KS}}$ is given in parenthesis;  the fit can be rejected at the $1-p_{\rm{KS}}$ confidence level.  Simulation parameters:  $V_i=16.6$, $\eta = 1$, $\Delta V_i/V_0=0.0$, $R=12.5$, $\Delta x=0.15$, $\Delta t=5\times 10^{-4}$, $N_{\rm{EM}}/I_{\rm{c}}=10^{-3}$, $\Omega(t=0) = 0.8$, $\Delta t_{\rm{sm}}=1.0$.}
\label{tab:ch5:density}
\end{table*}
\end{center}

If pinning occurs at individual nuclei, then $\sim 10^9$ pinning sites separates neighbouring vortices in a pulsar.  Alternatively, if pinning occurs at grain boundaries or along macroscopic faults in the crustal lattice, the ratio of pinning sites to vortices is not necessarily large, and pinning sites may not be evenly spaced.    In this section we explore the role of the pinning site density ($n_{\rm{pin}}$ sites per unit area) relative to vortex density ($n_{\rm{F}}=2\Omega/\kappa$  vortices per unit area).  We present results for simulations with:  (i) few or no pinning sites, (ii) $n_{\rm{pin}}\sim n_{\rm{F}}$, and (iii) many more pinning sites than vortices.  We parametrise the pinning site abundance in terms of the ratio $n_{\rm{pin}}/n_{\rm{F}}$ in the \emph{initial} state.

\begin{figure*}
\includegraphics[scale=0.45]{fig9a.epsi}
\includegraphics[scale=0.45]{fig9b.epsi}
\includegraphics[scale=0.45]{fig9c.epsi}
\includegraphics[scale=0.45]{fig9d.epsi}
\includegraphics[scale=1.0]{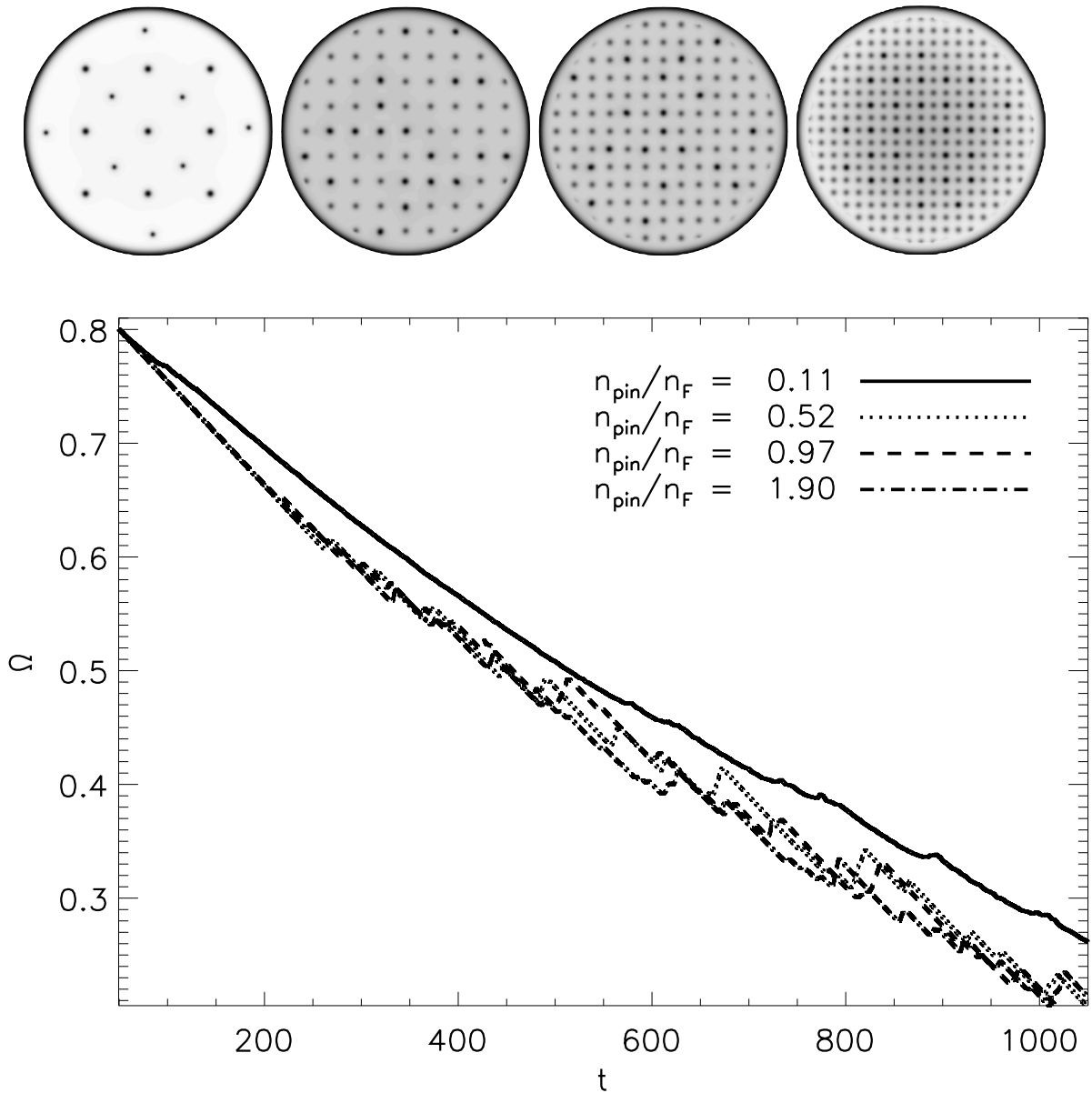}
\caption{Results from numerical experiments for different pinning abundances $n_{\rm{pin}}/n_{\rm{F}}$.  \emph{Top}:  Greyscale plots of the final superfluid density for $n_{\rm{pin}}/n_{\rm{F}}=0.11$, 0.52, 0.97 and 1.90 (\emph{left} to \emph{right} respectively).  The density scale runs from dark (low density) to light (high density).  The feint dots are unoccupied pinning sites; darker dots are pinned vortices.  \emph{Bottom}:  Angular velocity of the crust as a function of time $\Omega(t)$ for $n_{\rm{pin}}/n_{\rm{F}}=0.11$, 0.52, 0.97 and 1.90 (\emph{solid}, \emph{dotted}, \emph{dashed} and \emph{dot-dashed} respectively).  Simulation parameters: $V_i=16.6$, $\eta=1$, $\Delta V_i/V_0=0.0$,  $R=12.5$, $\Delta x=0.15$, $\Delta t=5\times 10^{-4}$, $N_{\rm{EM}}/I_{\rm{c}}=10^{-3}$, $\Omega(t=0) = 0.8$, $\Delta t_{\rm{sm}}=1.0$.}
\label{fig:ch5:om_sites}
\end{figure*}

Four pinning scenarios are illustrated in the top panels of Fig.~\ref{fig:ch5:om_sites} as greyscale plots of $|\psi|^2(\bm{x},t=1000)$ for $n_{\rm{pin}}/n_{\rm{F}}$ increasing from \emph{left} to \emph{right}.  The \emph{middle} panels both correspond to regime (ii).  In every panel except the leftmost one ($n_{\rm{pin}}/n_{\rm{F}}=0.11$), all vortices in the final state are pinned.  For  $n_{\rm{pin}}/n_{\rm{F}}=0.11$, even in the final state, there are only enough pinning sites to pin nine out of 17 vortices (\emph{i.e.} there are no unoccupied pinning sites).  Notably, for $n_{\rm{pin}}/n_{\rm{F}}=0.11$, the vortex lattice is regularly spaced, suggesting that pinning alters the orientation, but not the geometry, of the vortex lattice.  For $n_{\rm{pin}}/n_{\rm{F}}=0.52$ and $n_{\rm{pin}}/n_{\rm{F}}=0.97$, vortices are irregularly spaced, as they are forced to adhere to the geometry of the pinning grid.  When pinning sites significantly outnumber vortices (\emph{right} panel, $n_{\rm{pin}}/n_{\rm{F}}=1.90$), the vortex lattice is once again evenly spaced, because enough pinning sites are available for a vortex to `choose' a site near its equilibrium (Abrikosov) position.  It should be noted that this equilibrium position changes as the crust spins down.  

The \emph{bottom} panel of Fig.~\ref{fig:ch5:om_sites} graphs the spin-down curves for the four pinning abundances illustrated in the top panels.  The example with few pinning sites ($n_{\rm{pin}}/n_{\rm{F}}=0.11$, \emph{solid} curve) is relatively smooth.  Although our algorithm detects 25 glitches, they are small ($\langle\Delta\Omega/\Omega\rangle=14.84\times 10^{-4}$; \emph{cf.} $\langle\Delta\Omega/\Omega\rangle=13.00\times 10^{-4}$ for $V_0=0$) and well separated ($\langle\Delta t\rangle=37.10$), suggesting that they arise in the rare event that an outwardly moving vortex runs over a pinning site.  By contrast, for $n_{\rm{pin}}\sim n_{\rm{F}}$ (\emph{dotted}, \emph{dashed} and \emph{dash-dotted} curves), the spin down is roughly linear until vortices begin to unpin at $t\sim 275$, whereupon vortices move towards the edge in spasmodic jumps between pinning sites (vortices do not stop at \emph{all} pinning sites in their path).  For all $n_{\rm{pin}}/n_{\rm{F}}>0.11$, the gradual outward migration of vortices across their pinning sites reduces $\dot{\Omega}$, even before vortices have unpinned.  For $n_{\rm{pin}}/n_{\rm{F}}=0.11$, $\dot{\Omega}$ is considerably less than $N_{\rm{EM}}/I_{\rm{c}}$ for $t\gtrsim 50$, as the unpinned vortices move smoothly outward almost as soon as the spin-down torque is switched on.

It is physically interesting that the total spin down of the crust over long time periods is almost identical for  $n_{\rm{pin}}/n_{\rm{F}}> 1$ and $n_{\rm{pin}}/n_{\rm{F}}\sim 1$.  The aggregate effect of glitches is insensitive to increases in $n_{\rm{pin}}/n_{\rm{F}}$ over and above unity.  One interpretation is to say that the average distance travelled by a vortex once it unpins is of order $n_{\rm{F}}^{-1/2}$, rather than $n_{\rm{pin}}^{-1/2}$, so that vortices maintain an approximate Abrikosov lattice, without developing the large-scale spatial inhomogeneities required in self-organised critical avalanche models \citep{Warszawski:2008p4510}.  However, this hypothesis is challenged by three trends.  First, the number of glitches and activity parameter increase for $n_{\rm{pin}}/n_{\rm{F}}>0.11$; these qualities scale with $n_{\rm{pin}}^{-1/2}$, even though $n_{\rm{F}}^{-1/2}$ is approximately fixed.  Second, the mean waiting time is inversely proportional to $n_{\rm{pin}}/n_{\rm{F}}$; see Table~\ref{tab:ch5:density} for details and the \emph{right} panel of Fig.~\ref{fig:ch5:om_sites} for cumulative distributions of $\Delta t$.  Third,     the mean glitch size does not change monotonically with $n_{\rm{pin}}/n_{\rm{F}}$.  Taken together, these results suggest that the typical distance travelled by a vortex in a glitch is a play off between the number of vortices that unpin, the distance to the next available pinning site, and the distance the vortex must move to reverse the shear that unpinned it.  

\subsection{Maximum size}

\begin{figure*}
\includegraphics[scale=0.365,angle=90]{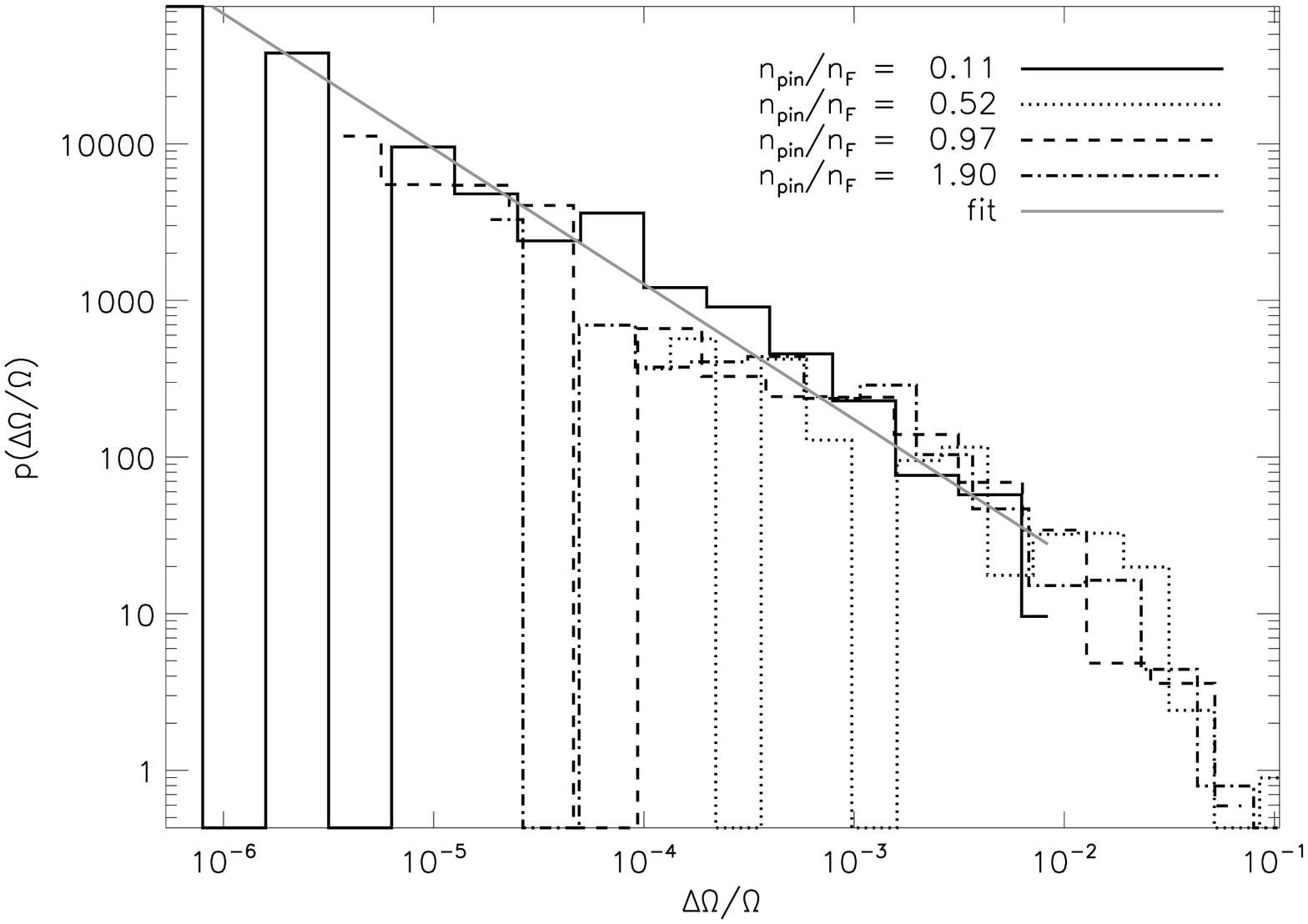}
\includegraphics[scale=0.365,angle=90]{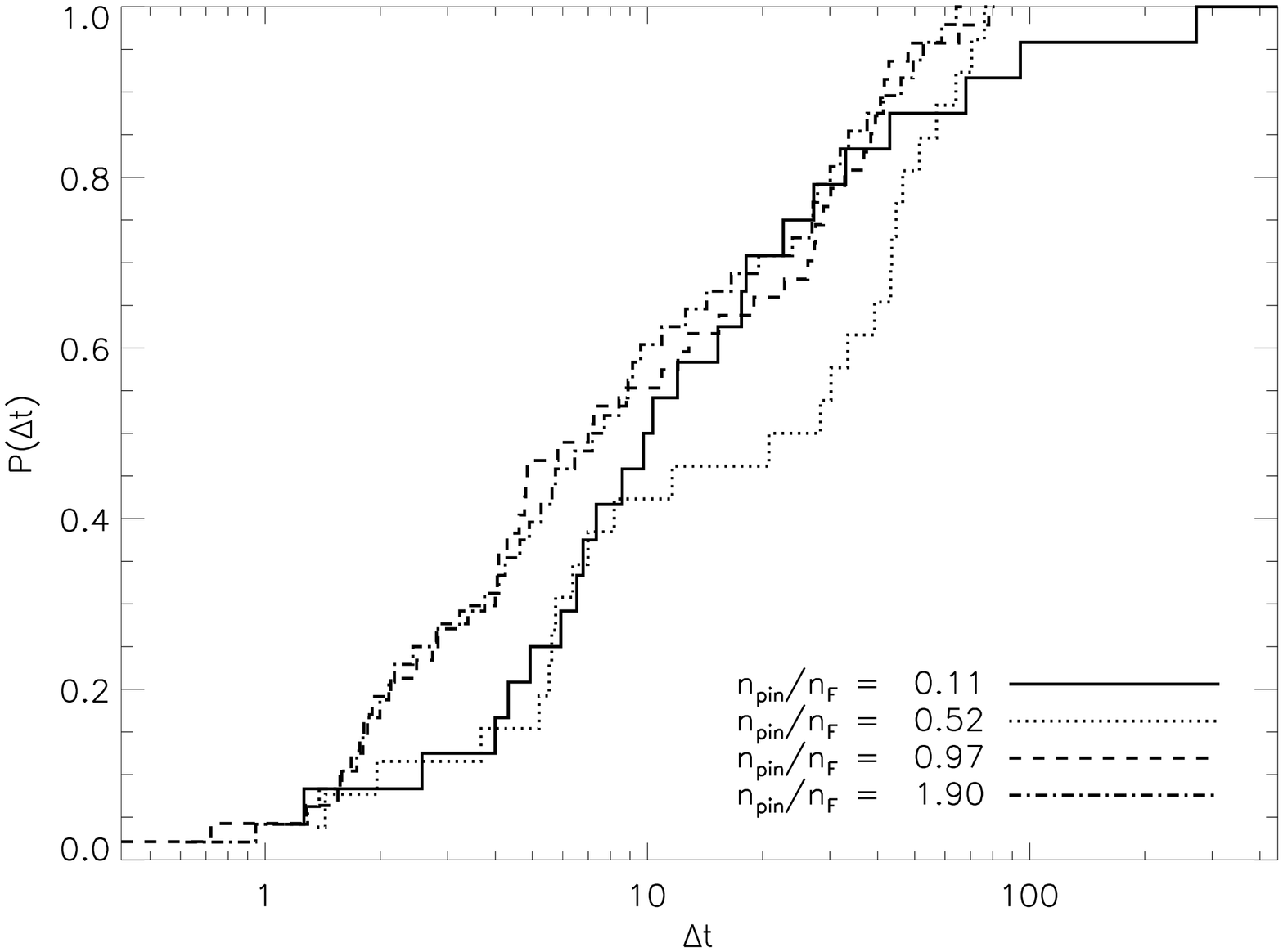}
\caption{Glitch distributions for spin-down experiments with different pinning abundances $n_{\rm{pin}}/n_{\rm{F}}$ for the $\Omega$ curves graphed in Fig.~\ref{fig:ch5:om_sites}.  \emph{Left}:  Probability density function (pdf) of fractional glitch sizes $p(\Delta\Omega/\Omega)$  for $n_{\rm{pin}}/n_{\rm{F}}=0.11$, 0.52, 0.97 and 1.90 (\emph{solid}, \emph{dotted}, \emph{dashed} and \emph{dot-dashed} respectively) logarithmically binned and graphed on log-log axes.  The \emph{solid grey} curve is a power-law least squares fit to the \emph{solid} histogram with index $-0.801$.  \emph{Right}:  Cumulative probability of glitch waiting times $p(\Delta t)$ graphed on log-linear axes for $n_{\rm{pin}}/n_{\rm{F}}=0.11$, 0.52, 0.97 and 1.90 (\emph{solid}, \emph{dotted}, \emph{dashed} and \emph{dot-dashed} respectively). Simulation parameters:  $V_0=16.6$, $\eta=1$, $\Delta V_i/V_0=0.0$, $R=12.5$, $\Delta x=0.15$, $\Delta t=5\times 10^{-4}$, $N_{\rm{EM}}/I_{\rm{c}}=10^{-3}$, $\Omega(t=0) = 0.8$, $\Delta t_{\rm{sm}}=1.0$.}
\label{fig:ch5:sites_stats}
\end{figure*}

Glitch size is governed by the total change in $\langle \hat{L}_z\rangle$ and hence depends on the number of vortices that unpin, the distance that each vortex travels before repinning, and the location from which it unpinned.  From the pdfs of $\Delta \Omega/\Omega$ in Fig.~\ref{fig:ch5:sites_stats}, we see that the largest glitch is approximately the same for all $n_{\rm{pin}}\gtrsim n_{\rm{F}}$.  If vortex motion is determined by the first encounter with a  pinning site, rather than the imperative to reduce shear and maintain equal spacing, then we expect the mean glitch size to scale inversely with the density of pinning sites.  This is not observed.  By contrast, maximum glitch size does scale monotonically with $V_0$ [max($10^4\Delta\Omega/\Omega$) = 36, 538, 1057, and 1641 for $V_0=0$, 8.3, 16.6 and 33.3 respectively], because vortices need to move further to reduce the shear sustained by stronger pinning.

\subsection{Unequal pinning potentials}\label{subsec:pinning:delta}

\begin{center}
\begin{table*}
\begin{tabular}{ | c| c| c| c| c| c| c|}
\hline
 $\Delta V_i/V_0$ &$N_{\rm{g}}$ &$10^4\langle\Delta\Omega/\Omega\rangle$&$\langle\Delta t\rangle$ 	&$A$ &$\gamma$($p_{\rm{KS}}$) &$\lambda$($p_{\rm{KS}}$)\\
 \hline
 0.0		 &28		&231.2	&21.72			&59.82	&$-0.818$(0.212)&$0.048$(0.360)\\
 0.25		 &28		&173.3	&22.13			&44.13	&$-0.676$(0.596)&$0.043$(0.712)\\
 0.5		 &27		&231.0	&26.74			&47.19	&$-0.805$(0.504)&$0.038$(0.232)\\
 0.75		 &24		&261.8	&33.42			&37.80	&$-0.776$(0.671)&$0.051$(0.008)\\
 1.0		 &30		&206.2	&26.45			&47.10	&$-0.901$(0.253)&$0.070$(0.055)\\
 \hline
\end{tabular}
\caption{Glitch size and waiting-time statistics for different ranges of pinning strengths $\Delta V_i/V_0$ [corresponding $\Omega(t)$ curves shown in Fig.~\ref{fig:ch5:om_flat}].  $N_{\rm{g}}$ is the number of glitches detected, $\langle \Delta\Omega/\Omega\rangle$ is the mean glitch size, $\langle\Delta t\rangle$ is the mean waiting time, and $A$ is the activity parameter.  $\lambda$ is the mean glitch rate that parametrises the cumulative distribution of waiting times, $\gamma$ is the power-law index that parametrises the pdf of glitch sizes.  The K-S probability $p_{\rm{KS}}$ is given in parenthesis;  the fit can be rejected at the $1-p_{\rm{KS}}$ confidence level.  Simulation parameters:  $V_0=16.6$, $\eta = 1$, $n_{\rm{pin}}/n_{\rm{F}}$, $R=12.5$, $\Delta x=0.15$, $\Delta t=5\times 10^{-4}$, $N_{\rm{EM}}/I_{\rm{c}}=10^{-3}$, $\Omega(t=0) = 0.8$, $\Delta t_{\rm{sm}}=1.0$.}
\label{tab:ch5:delta}
\end{table*}
\end{center}
Pinned vortices at a given radius experience an equal Magnus force.  If every pinning site is of equal strength, then vortices at a given radius should unpin simultaneously, yielding periodic glitches of equal size, contrary to observations \citep{Melatos:2008p204,Melatos:2009p4511}.  In this section, we investigate the effect of varying the width  of the pinning strength distribution $\Delta V_i/V_0$. The glitch statistics from these simulations are summarised in Table~\ref{tab:ch5:delta}.

\begin{figure*}
\includegraphics[scale=1.0]{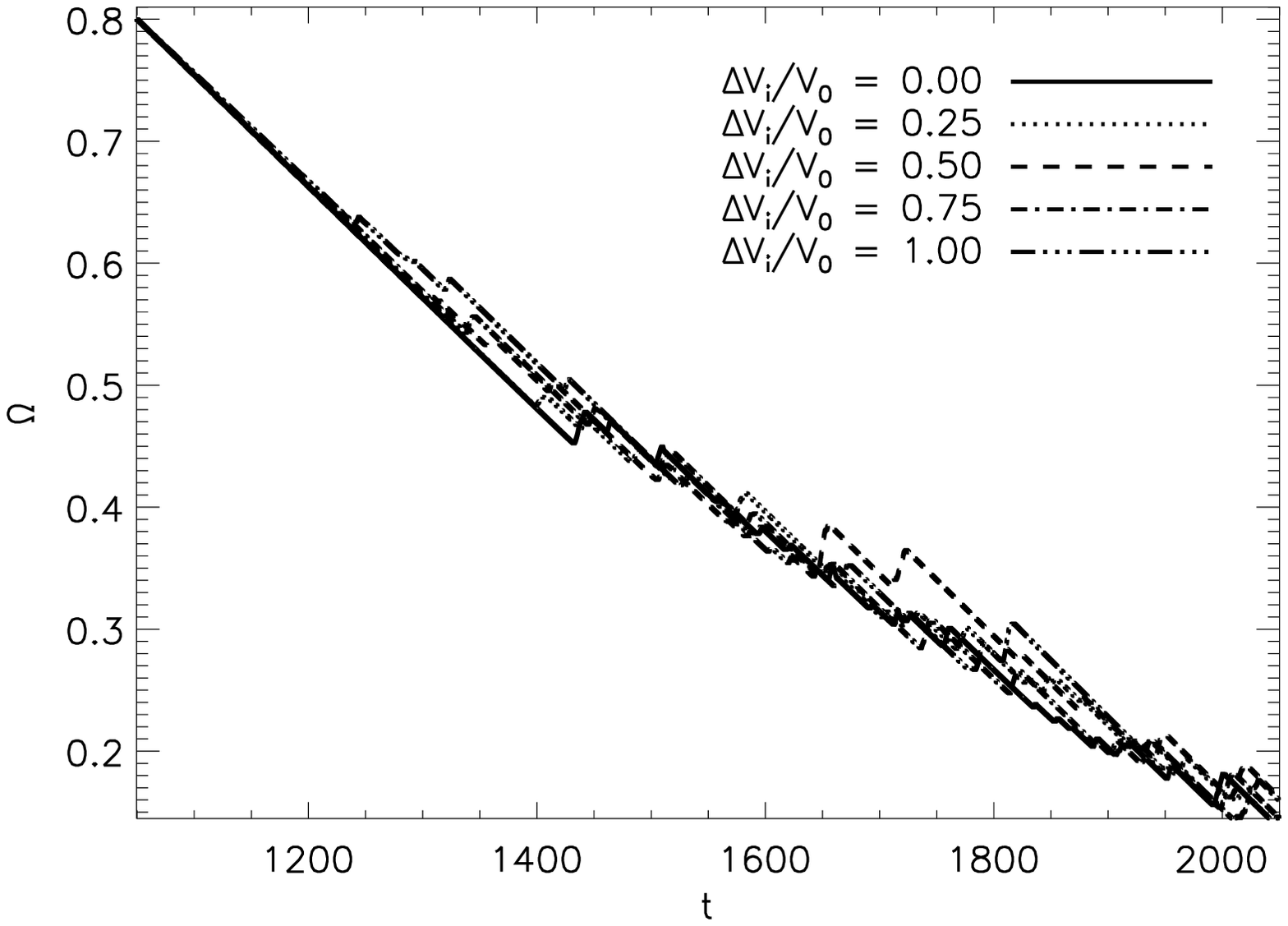}
\caption{Angular velocity of the crust as a function of time $\Omega(t)$ for spin-down experiments with $\Delta V_i/V_0=0.0$, 0.25, 0.5, 0.75 and 1.0 (\emph{solid}, \emph{dotted}, \emph{dashed}, \emph{dot-dashed} and \emph{triple-dot-dashed} respectively).  Simulation parameters:  $V_0=16.6$, $\eta=1$, $\Delta V_i/V_0=0.0$, $R=12.5$, $\Delta x=0.15$, $\Delta t=5\times 10^{-4}$, $N_{\rm{EM}}/I_{\rm{c}}=10^{-3}$, $\Omega(t=0) = 0.8$, $\Delta t_{\rm{sm}}=1.0$.}
\label{fig:ch5:om_flat}
\end{figure*}

\begin{figure*}
\includegraphics[scale=0.365,angle=90]{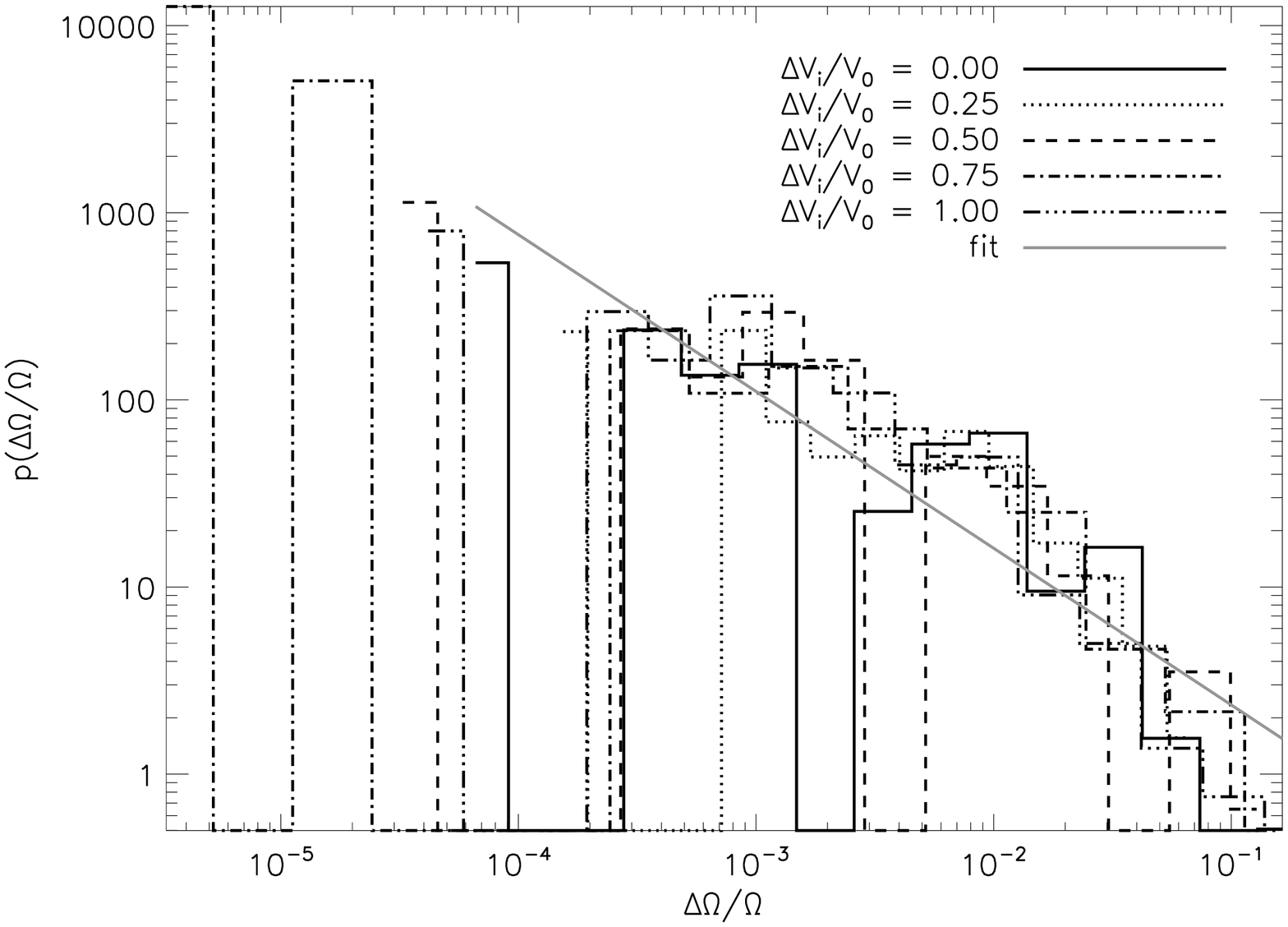}
\includegraphics[scale=0.365,angle=90]{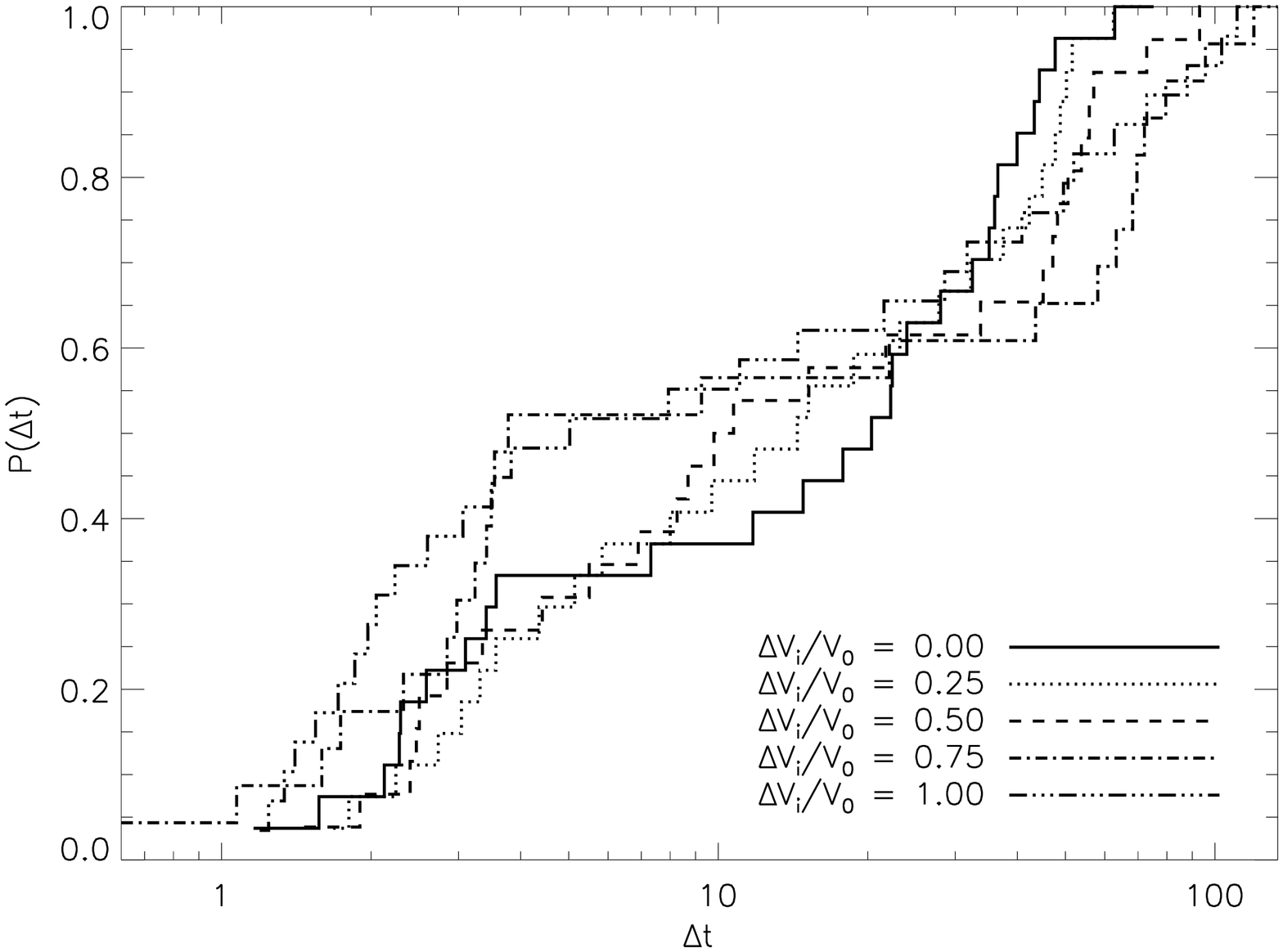}
\caption{Glitch distributions for spin-down experiments with different $\Delta V_i/V_0$ for the $\Omega(t)$ curves graphed in Fig.~\ref{fig:ch5:om_flat}.  \emph{Left}:  Probability density function (pdf) of fractional glitch sizes ($\Delta\Omega/\Omega$) for $\Delta V_i/V_0=0.0$, 0.25, 0.5, 0.75 and 1.0 (\emph{solid}, \emph{dotted}, \emph{dashed}, \emph{dot-dashed} and \emph{triple-dot-dashed} respectively) logarithmically binned and graphed on log-log axes.  The \emph{solid grey} curve is a power-law least squares fit to the \emph{solid} histogram with index $-0.818$.  \emph{Right}:  Cumulative probability of glitch waiting times $p(\Delta t)$ graphed on log-linear axes for $\Delta V_i/V_0=0.0$, 0.25, 0.5, 0.75 and 1.0 (\emph{solid}, \emph{dotted}, \emph{dashed}, \emph{dot-dashed} and \emph{triple-dot-dashed} respectively). Simulation parameters:  $V_0=16.6$, $\eta=1$, $R=12.5$, $\Delta x=0.15$, $\Delta t=5\times 10^{-4}$, $N_{\rm{EM}}/I_{\rm{c}}=10^{-3}$, $\Omega(t=0) = 0.8$, $\Delta t_{\rm{sm}}=1.0$.}
\label{fig:ch5:flat_stats}
\end{figure*}

The spin-down curves in Fig.~\ref{fig:ch5:om_flat} demonstrate that varying $\Delta V_i/V_0$ makes little difference to the total spin down over the time scale of the simulation; $\Omega(t=1000)$ is approximately the same for all $0\leq\Delta V_i/V_0\leq 1$.  
However, the timing of the first departure from linear spin-down scales as $t_{\rm{g}1}\approx 1250+200(1-\Delta V_i/V_0)$.  In addition, the glitch statistics change systematically with $\Delta V_i/V_0$.  The pdfs of $\Delta\Omega/\Omega$ graphed in the \emph{left} panel of Fig.~\ref{fig:ch5:flat_stats} confirm that the minimum glitch size depends on $\Delta V_i/V_0$, via the minimum pinning strength.  The correlation is not strict because the number of draws from the pinning strength distribution is relatively small ($11\times11=121$), so the minimum $V_i$ does not monotonically scale with $\Delta V_i/V_0$.  We detect no systematic change in the exponent of power-law fits to the size pdf.  However, we note that the fits can be rejected with high confidence for $\Delta V_i/V_0=0$ (78.8\%) and $\Delta V_i/V_0=1.0$ (74.7\%).

Cumulative waiting time distributions are graphed in the \emph{right} panel Fig.~\ref{fig:ch5:flat_stats}.  Exponential fits reveal a tendency for the mean glitch rate to increase with $\Delta V_i/V_0$ ($\lambda=0.07$ for $\Delta V_i/V_0=1.0$ and $0.043$ for $\Delta V_i/V_0=0.0$).  This finding disagrees with the uniform pinning scenarios discussed in Section~\ref{subsec:pinning:strength} and Section~\ref{subsec:pinning:strength_stats}, in which $\lambda$ does not scale inversely with $V_0$.

To interpret the above findings, we suggest that, even for $\Delta V_i/V_0=0$, the effective potential at each pinning site is a function of its departure from an equilibrium position in the Abrikosov lattice.  If a vortex is slightly inside or outside (radially) its Feynman position, the effective pinning energy is decreased or increased respectively, and hence the distribution of pinning strengths can be treated as heterogeneous, even when $\Delta V_i=0$. In a pulsar ($n_{\rm{pin}}/n_{\rm{F}}\sim 10^{14}$ if all nuclear sites can pin; see Table~\ref{tab:ch5:pulsar}), the pinned vortices resemble an Abrikosov array.  Hence we expect that $\Delta V_i/V_0$ is more influential in a pulsar than in the simulations presented here.


\section{Stellar parameters}\label{sec:structure}
\subsection{Crust moment of inertia}\label{subsec:structure:eta}
\begin{center}
\begin{table*}
\begin{tabular}{ | c| c| c| c| c| c| c|}
\hline
 $\eta$ &$N_{\rm{g}}$ &$10^4\langle\Delta\Omega/\Omega\rangle$&$\langle\Delta t\rangle$ 	&$A$ &$\gamma$($p_{\rm{KS}}$) &$\lambda$($p_{\rm{KS}}$)\\
 \hline
 1.0		 &27		&188.0		&27.93		&36.45 	&$-0.846$(0.615)&0.035 (0.450)\\
 2.0		 &37		&74.73		&25.19		&22.00 	&$-0.760$(0.916)&0.065 (0.152)\\
 \hline
\end{tabular}
\caption{Glitch size and waiting-time statistics for different crustal moments of inertia $\eta$ [corresponding $\Omega(t)$ curves shown in Fig.~\ref{fig:ch5:om_eps}].  $N_{\rm{g}}$ is the number of glitches detected, $\langle \Delta\Omega/\Omega\rangle$ is the mean glitch size, $\langle\Delta t\rangle$ is the mean waiting time, and $A$ is the activity parameter.  $\lambda$ is the mean glitch rate that parametrises the cumulative distribution of waiting times, $\gamma$ is the power-law index that parametrises the pdf of glitch sizes.  The K-S probability $p_{\rm{KS}}$ is given in parenthesis;  the fit can be rejected at the $1-p_{\rm{KS}}$ confidence level.  Simulation parameters:  $V_0=16.6$, $\Delta V_i/V_0=0.0$, $n_{\rm{pin}}/n_{\rm{F}}$, $R=12.5$, $\Delta x=0.15$, $\Delta t=5\times 10^{-4}$, $N_{\rm{EM}}/I_{\rm{c}}=10^{-3}$, $\Omega(t=0) = 0.8$, $\Delta t_{\rm{sm}}=1.0$.}
\label{tab:ch5:eta}
\end{table*}
\end{center}

\begin{figure*}
\includegraphics[scale=1.0]{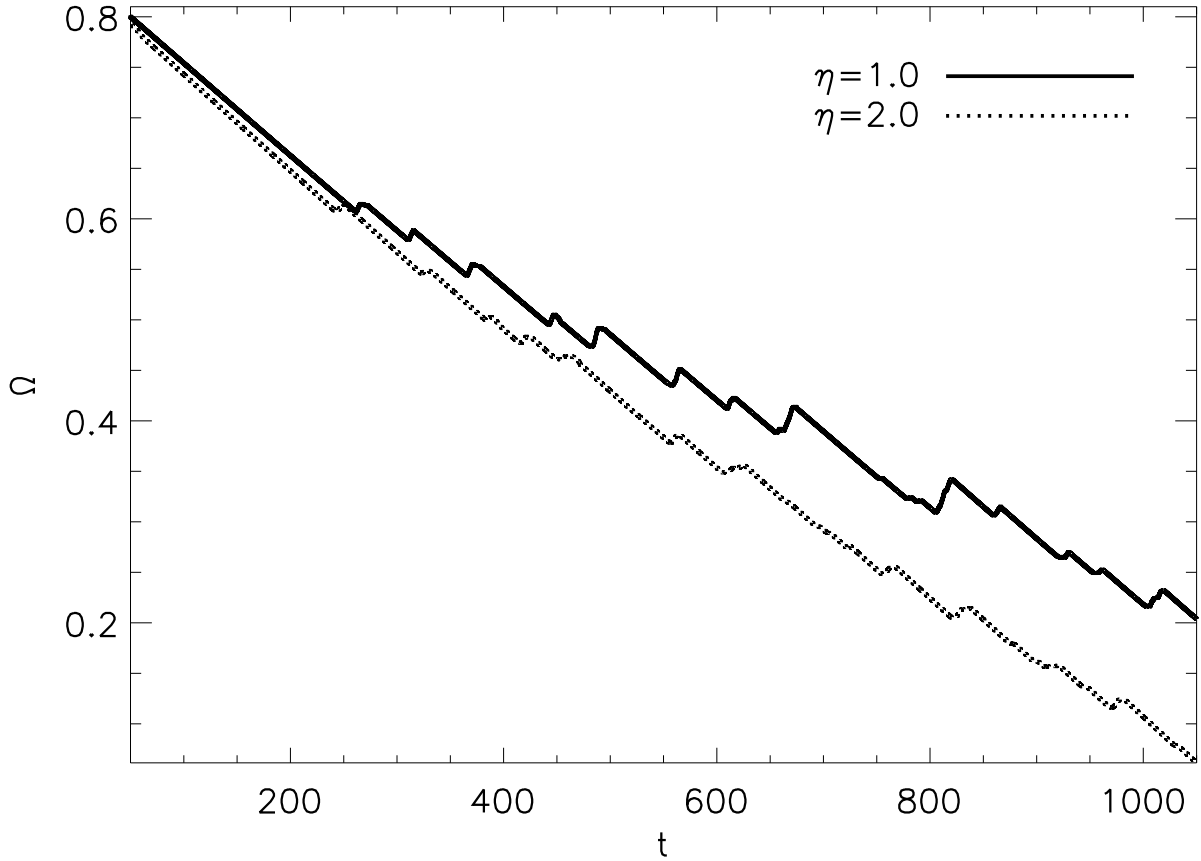}
\caption{Angular velocity of the crust as a function of time $\Omega(t)$ for spin-down experiments with lighter ($\eta =1$) and heavier ($\eta=2$) crusts (\emph{solid} and \emph{dotted} curves respectively). Simulation parameters:  $V_0=16.6$, $\Delta V_i/V_0=0.0$, $R=12.5$, $\Delta x=0.15$, $\Delta t=5\times 10^{-4}$, $N_{\rm{EM}}/I_{\rm{c}}=10^{-3}$, $\Omega(t=0) = 0.8$, $\Delta t_{\rm{sm}}=1.0$.}
\label{fig:ch5:om_eps}
\end{figure*}
\begin{figure*}
\includegraphics[scale=0.365,angle=90]{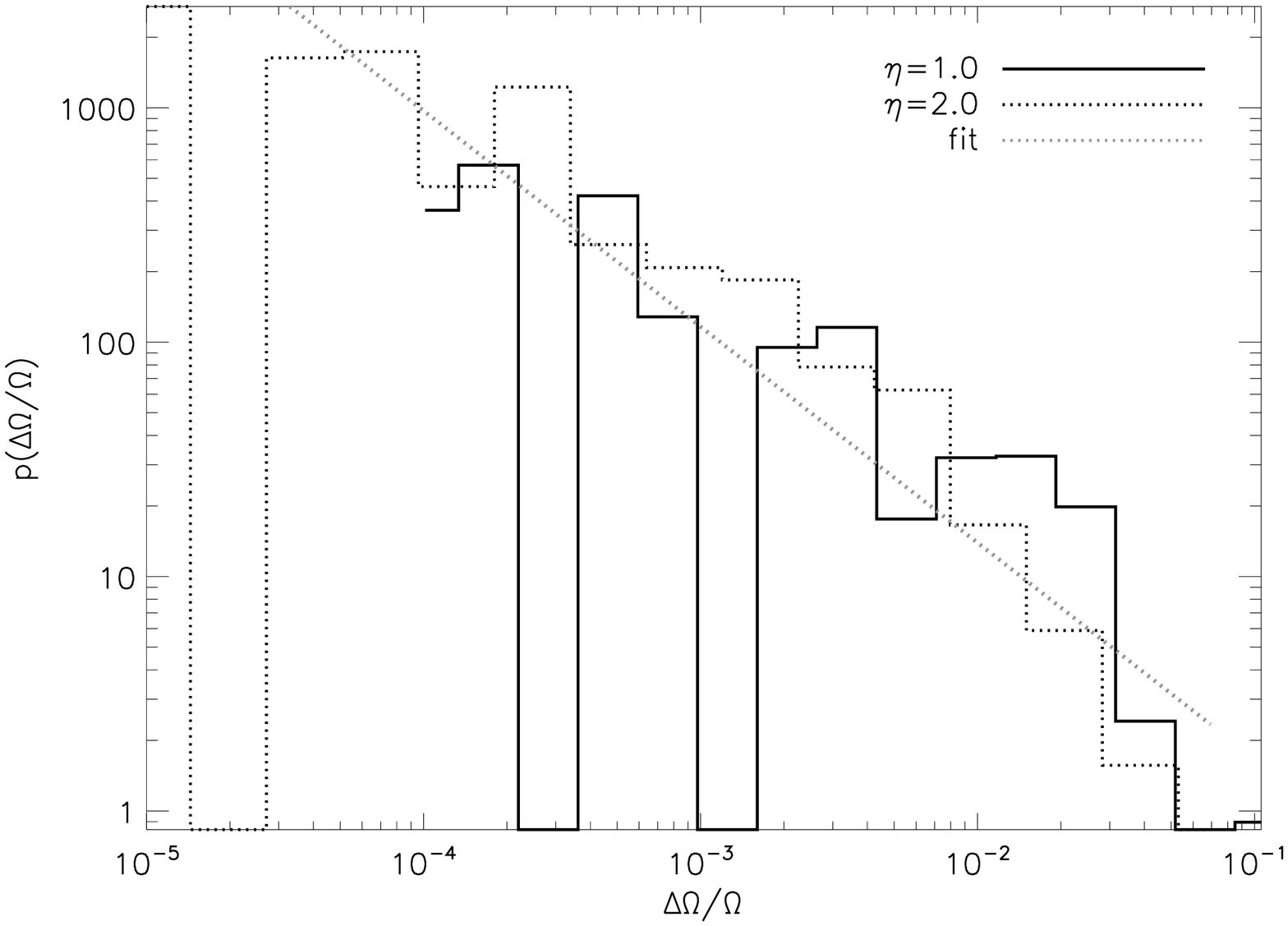}
\includegraphics[scale=0.365,angle=90]{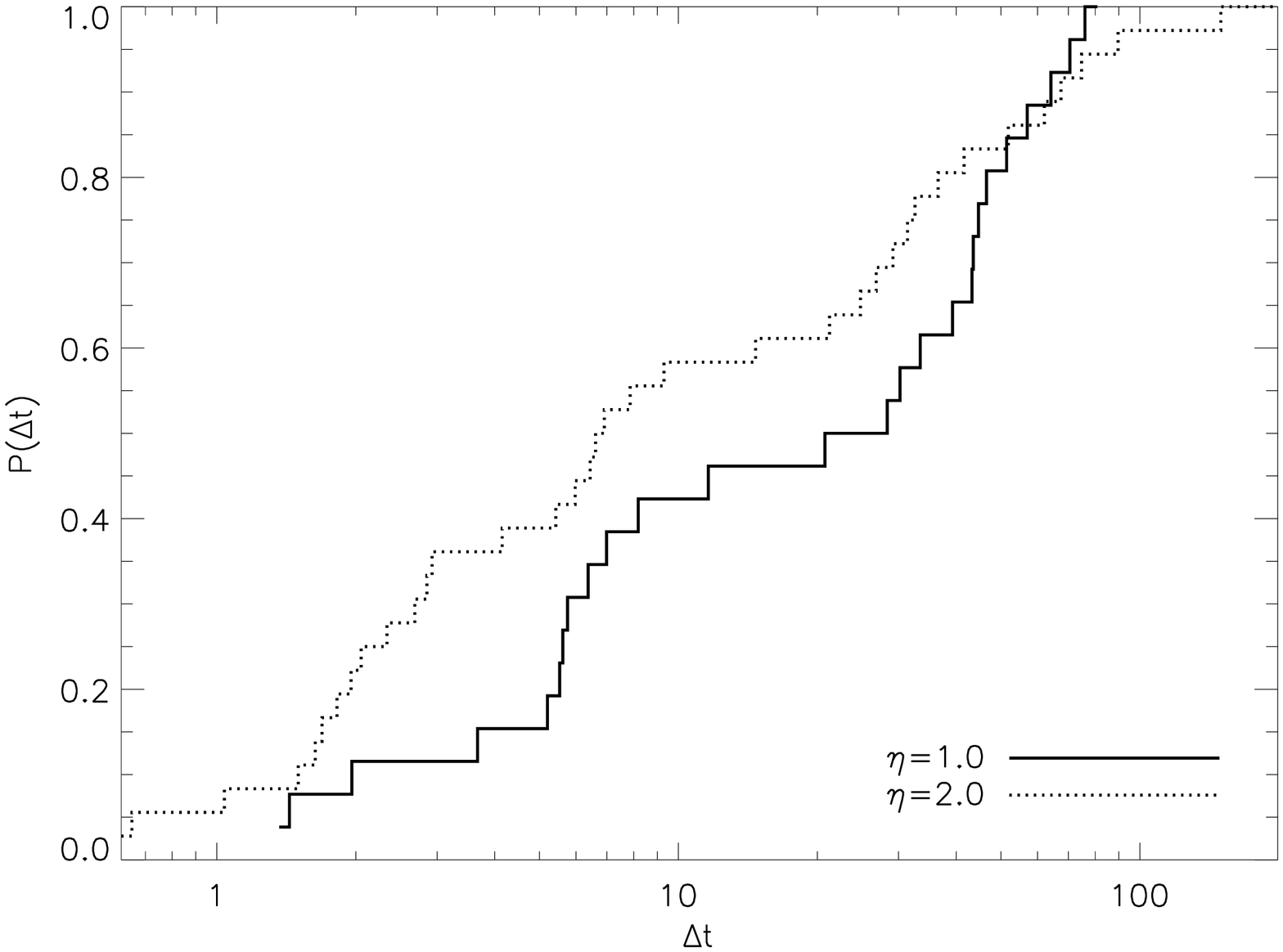}
\caption{Glitch distributions for spin-down experiments with different $\Delta V_i/V_0$ for the $\Omega(t)$ curves graphed in Fig.~\ref{fig:ch5:om_eps}.  \emph{Left}:  Probability density function (pdf) of fractional glitch sizes ($\Delta\Omega/\Omega$) for $\eta=1$ and 2 (\emph{solid} and \emph{dotted} curves respectively) logarithmically binned and graphed on log-log axes.  The \emph{dotted grey} curve is a power-law least squares fit to the \emph{dotted} histogram with index $-0.846$.  \emph{Right}:  Cumulative pdf of glitch waiting times $p(\Delta t)$ graphed on log-linear axes for $\eta=1$ and 2 (\emph{solid} and \emph{dotted} curves respectively). Simulation parameters:  $V_i=16.6$, $\Delta V_i/V_0=0.0$, $R=12.5$, $\Delta x=0.15$, $\Delta t=5\times 10^{-4}$, $N_{\rm{EM}}/I_{\rm{c}}=10^{-3}$, $\Omega(t=0) = 0.8$, $\Delta t_{\rm{sm}}=1.0$.}
\label{fig:ch5:eps_stats}
\end{figure*}

The inertia of the pulsar crust, relative to the inertia of the pinned superfluid, controls the crust's response to changes in $\langle \hat{L}_z\rangle$.  Let $\eta=I_{\rm{c}}/I_{\rm{S}}$ be the ratio of the crust's moment of inertia, $I_{\rm{c}}$, to the moment of inertia that the superfluid would have if it were a rigid body.  If the crust is light (small $\eta$), small decreases in $\langle \hat{L}_z\rangle$ produce large crustal spin-up events.  For sufficiently small $\eta$, even vortex creep \citep{Alpar:1984p6781,Link:1993p47} should produce small but observable glitches.  For larger $\eta$, on the other hand, crustal spin up is smaller.  In what follows, we adopt the view proffered by \cite{Link:1999p2179} that the `crust' is composed of the rigid nuclear lattice, all charged fluid components that are coupled electromagnetically to the rigid lattice, and the unpinned superfluid (which may be a large fraction of the total).  Hydrodynamic models of glitch recovery suggest $\eta\sim 1$ \citep{VanEysden:2010}.  Results for the numerical experiments presented in this section are summarised in Table~\ref{tab:ch5:eta}.

As $\eta$ increases, the mean glitch size diminishes; a given change in $\langle\hat{L}_z \rangle$ results in a smaller change in $\Omega$.  For example, from the spin-down curve in Fig.~\ref{fig:ch5:om_eps}, we see that doubling $\eta$ halves $\Omega(t=1000)$.  Put another way, as $\eta$ increases, unpinning and outward vortex motion do progressively less to reduce the crust-superfluid shear.  Hence, the timing of the unpinning events changes with $\eta$.  Cumulative waiting time distributions are graphed in the \emph{right} panel of Fig.~\ref{fig:ch5:eps_stats}.  They show that the mean waiting time decreases as $\eta$ increases.  The waiting-time distribution is weighted to the low  $\Delta t$ end for $\eta=2.0$ compared to $\eta =1.0$ (e.g. $\langle \Delta t\rangle= 27.93,~25.19$ and $\lambda=0.035,~0.065$ for $\eta = 1.0,~2.0$).  

The resistance of a heavy crust to spin up also produces smaller glitches.  This result is confirmed by considering the pdf of $\Delta\Omega/\Omega$ for two cases, $\eta=1.0$ and 2.0. Looking at the \emph{left} panel in Fig.~\ref{fig:ch5:eps_stats}, we find that the heavier crust is accompanied by smaller glitches, e.g $10^4\langle\Delta\Omega/\Omega\rangle= 188$, 74.73 for $\eta = 1.0$, 2.0. 

The vertical distance between the curves in Fig.~\ref{fig:ch5:om_eps} increases with time, demonstrating that the increased glitch rate is not sufficient to compensate for reduced glitch sizes.  Over longer time periods than those simulated here, $\dot{\Omega}$ self-regulates so that the crust-superfluid shear exactly equals the value required to sustain the average unpinning rate leading to $\dot{\Omega}$ (given $N_{\rm{EM}}$ and $\eta$).  The activity parameter $A$ provides a means of comparing the relative changes in response to the spin-down torque in $\langle \Delta t\rangle$ and $\langle\Delta\Omega/\Omega\rangle$; $A$ decreases by $\sim33\%$ as $\eta$ doubles from $\eta=1.0$ to 2.0, demonstrating again that the increased glitch rate does not compensate for the large decrease in glitch size. 

\subsection{Electromagnetic spin-down torque}\label{subsec:structure:omdot}

\begin{center}
\begin{table*}
\begin{tabular}{| c| c| c| c| c| c| c| c|}
\hline
 $N_{\rm{EM}}/I_{\rm{c}}$ &$N_{\rm{g}}$ 	&$10^{4}\langle\Delta\Omega/\Omega\rangle$	&$\langle\Delta t\rangle$ 	&$A$ 	&$I_{\rm{c}}\langle\Delta\Omega_{\rm{g}}\rangle/(\langle\Delta t\rangle N_{\rm{EM}})$&$\gamma$($p_{\rm{KS}}$) &$\lambda$($p_{\rm{KS}}$)\\\\
 \hline
 $10^{-4.0}$		 &3	&84.28	&703.4		&0.0958	&0.119 & $-1.082$(0.929) & 0.333 (0.810)\\
 $10^{-3.5}$		 &39	&43.68	&51.42		&9.52	&0.269 & $-0.720$(0.514) & 0.0289 (0.002)\\
 $10^{-3.0}$		 &27	&68.50	&27.93		&13.28 	&0.245 & $-0.819$(0.913) & 0.0403 (0.450)\\
 $10^{-2.5}$		 &84	&45.39	&10.44		&73.98 	&0.137 & $-0.984$(0.0013) & 0.112 (0.058)\\
 \hline
\end{tabular}
\caption{Glitch size and waiting-time statistics for different $N_{\rm{EM}}/I_{\rm{c}}$ for the $\Omega(t)$ curves shown in Fig.~\ref{fig:ch5:om_omdot}.  $N_{\rm{g}}$ is the number of glitches detected, $\langle \Delta\Omega/\Omega\rangle$ is the mean glitch size, $\langle\Delta t\rangle$ is the mean waiting time, and $A$ is the activity parameter.  $\lambda$ is the mean glitch rate that parametrises the cumulative distribution of waiting times, $\gamma$ is the power-law index that parametrises the pdf of glitch sizes.  The K-S probability $p_{\rm{KS}}$ is given in parenthesis;  the fit can be rejected at the $1-p_{\rm{KS}}$ confidence level.    Simulation parameters:  $V_0=16.6$, $\eta=1$, $\Delta V_i/V_0=0.0$, $n_{\rm{pin}}/n_{\rm{F}}$, $R=12.5$, $\Delta x=0.15$, $\Delta t=5\times 10^{-4}$, $N_{\rm{EM}}/I_{\rm{c}}=10^{-3}$, $\Omega(t=0) = 0.8$, $\Delta t_{\rm{sm}}=1.0$.}
\label{tab:ch5:omdot}
\end{table*}
\end{center}

\begin{figure*}
\includegraphics[scale=1.0]{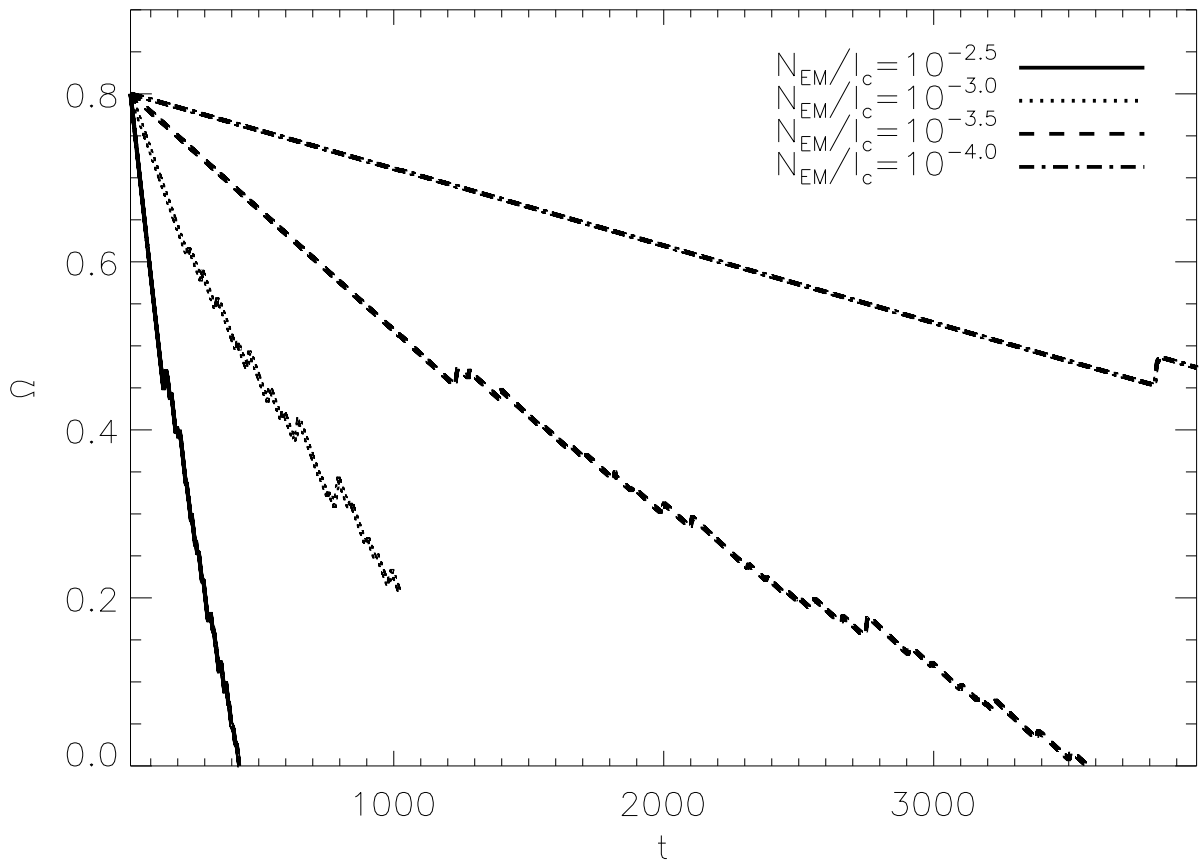}
\caption{Angular velocity of the crust as a function of time $\Omega(t)$ for spin-down experiments with different electromagnetic spin-down torques $N_{\rm{EM}}/I_{\rm{c}}=-10^{-2.5}$, $-10^{-3.0}$, $-10^{-3.5}$ and $-10^{-4.0}$ (\emph{solid}, \emph{dotted}, \emph{dashed} and \emph{dot-dashed} curves respectively). Simulation parameters:  $V_0=16.6$, $\eta=1$, $\Delta V_i/V_0=0.0$, $R=12.5$, $\Delta x=0.15$, $\Delta t=5\times 10^{-4}$, $\Omega(t=0) = 0.8$, $\Delta t_{\rm{sm}}=1.0$.}
\label{fig:ch5:om_omdot}
\end{figure*}

\begin{figure*}
\includegraphics[scale=0.365,angle=90]{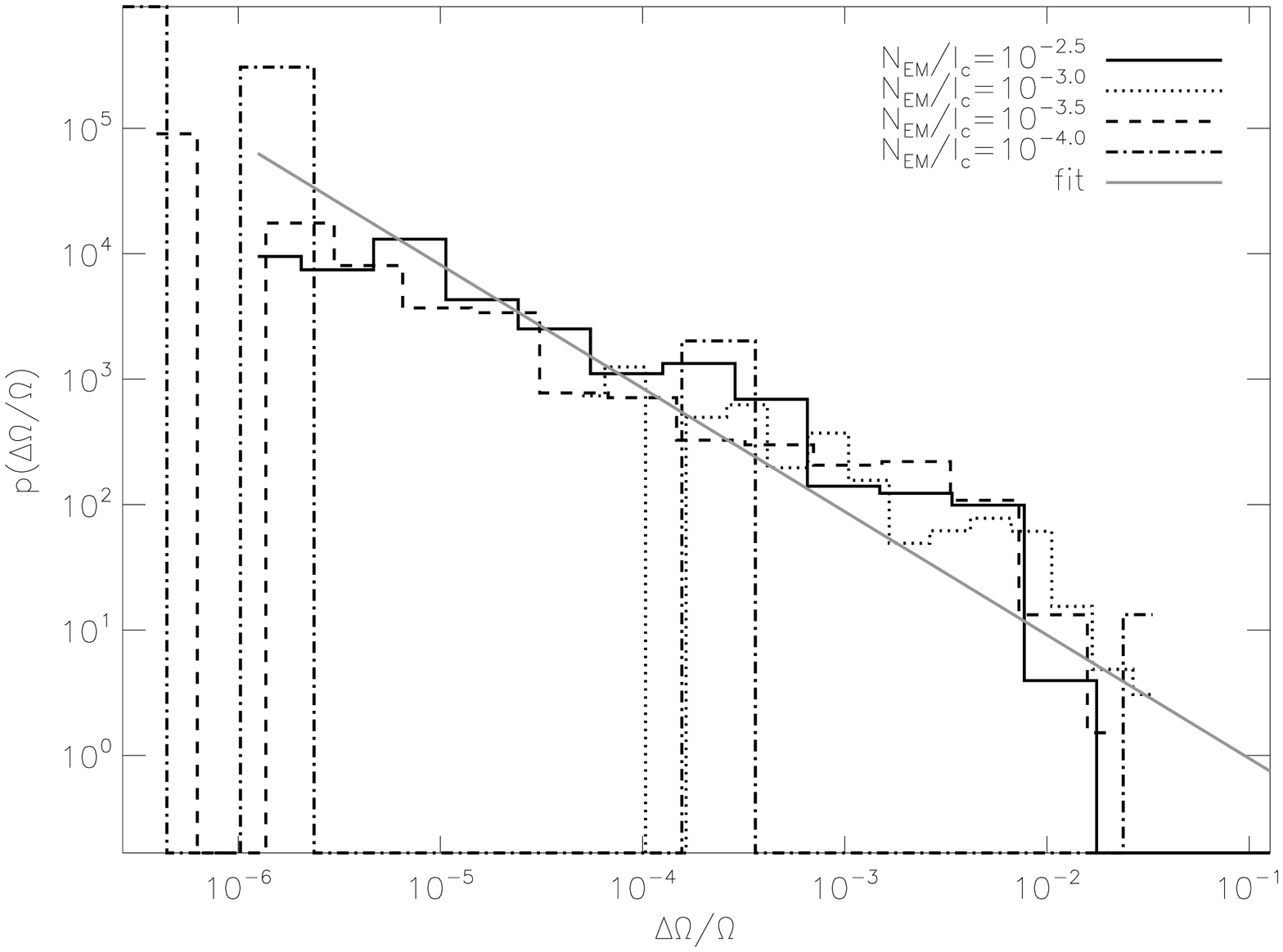}
\includegraphics[scale=0.365,angle=90]{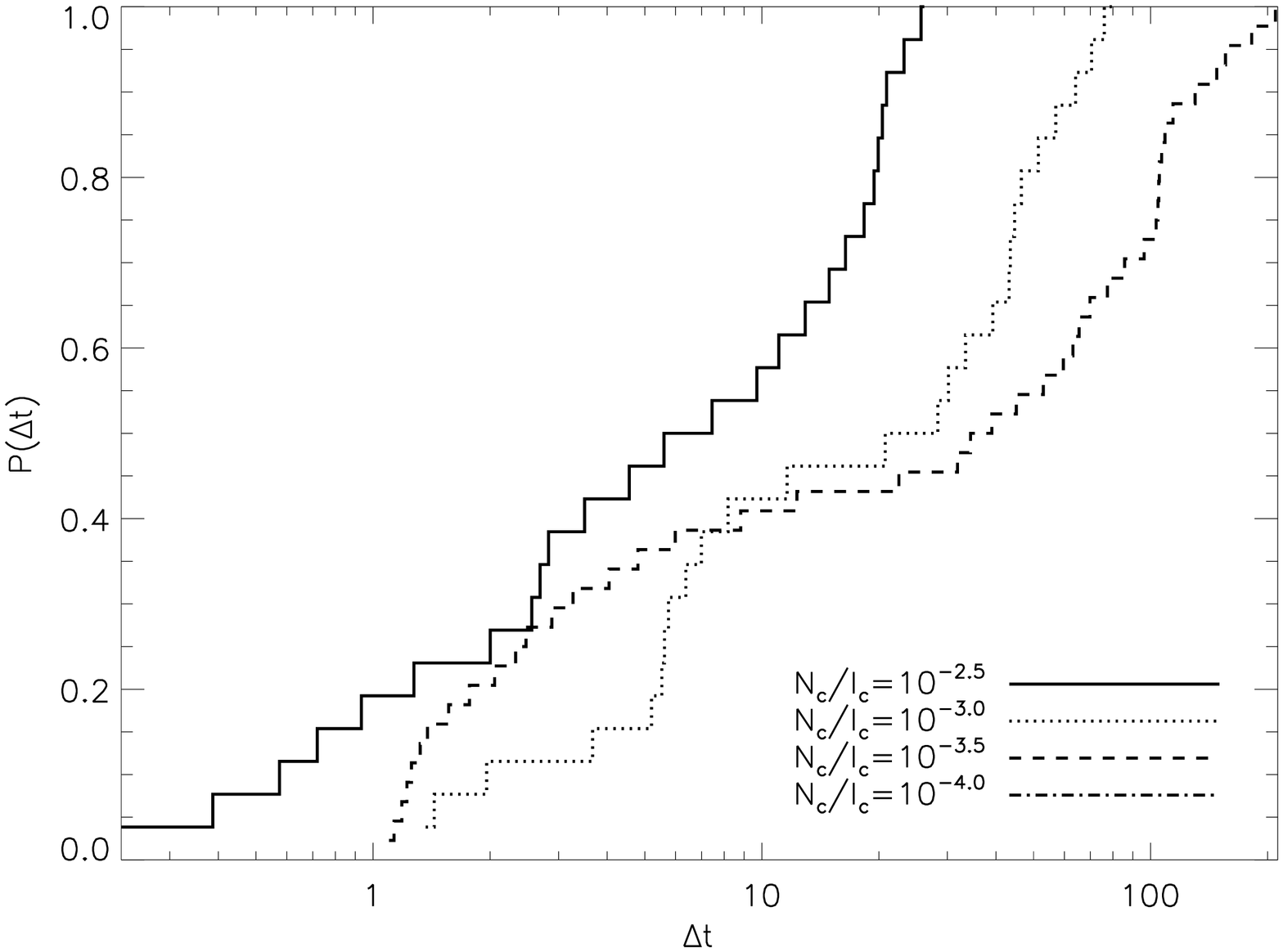}
\caption{Glitch distributions for spin-down experiments with different electromagnetic spin-down torques  $N_{\rm{EM}}/I_{\rm{c}}$ for the $\Omega(t)$ curves graphed in Fig.~\ref{fig:ch5:om_omdot}.  \emph{Left}:  Probability density function (pdf) of fractional glitch sizes ($\Delta\Omega/\Omega$) for $N_{\rm{EM}}/I_{\rm{c}}=-10^{-2.5}$, $-10^{-3.0}$, $-10^{-3.5}$ and $-10^{-4.0}$ (\emph{solid}, \emph{dotted}, \emph{dashed} and \emph{dot-dashed} curves respectively) logarithmically binned and graphed on log-log axes.  The \emph{solid grey} curve is a power-law least squares fit to the \emph{solid} histogram with index $-0.984$.  \emph{Right}:  Cumulative probability of glitch waiting times $p(\Delta t)$ graphed on log-linear axes for $N_{\rm{EM}}/I_{\rm{c}}=-10^{-2.5}$, $-10^{-3.0}$, $-10^{-3.5}$ and $-10^{-4.0}$ (\emph{solid}, \emph{dotted}, \emph{dashed} and \emph{dot-dashed} curves respectively). Simulation parameters:  $V_0=16.6$, $\eta=1$, $\Delta V_i/V_0=0.0$, $R=12.5$, $\Delta x=0.15$, $\Delta t=5\times 10^{-4}$,  $\Omega(t=0) = 0.8$, $\Delta t_{\rm{sm}}=1.0$.}
\label{fig:ch5:omdot_stats}
\end{figure*}

The magnetic dipole torque acting on a pulsar ultimately drives the crust-superfluid shear essential to the (un)pinning model of pulsar glitches.  Here, we quantify the torque in terms of an electromagnetic spin-down rate, $N_{\rm{EM}}/I_{\rm{c}}$, in the absence of feedback.  The larger the torque, the more quickly Magnus stresses build up in the pinned vortex lattice.   Results from simulations in which $N_{\rm{EM}}/I_{\rm{c}}$ is varied are summarised in Table~\ref{tab:ch5:omdot}.

In Fig.~\ref{fig:ch5:om_omdot} we graph spin-down curves for $N_{\rm{EM}}/I_{\rm{c}}=10^{-2.5}$ (\emph{solid}), $10^{-3}$ (\emph{dashed}), $10^{-3.5}$ (\emph{dotted}) and $10^{-4}$ (\emph{dot-dash}).  For $N_{\rm{EM}}/I_{\rm{c}}=10^{-4.0}$, we detect few glitches ($N_{\rm{g}}=3$), owing to the small shear that accumulates over the total lifetime of the simulation  [$t_{\rm{total}}N_{\rm{EM}}/I_{\rm{c}}=0.4$].  Importantly, the first glitch does not appear at the same value of $\Omega$ for different spin-down torques; $\Omega(t_1)=0.45$, 0.46, 0.61 and 0.44 for $N_{\rm{EM}}/I_{\rm{c}}=10^{-4.0}$, $10^{-3.5}$, $10^{-3}$ and $10^{-2.5}$ respectively.  This result reinforces the findings of hysteresis reported in \cite{Jackson:2006p3196} and \cite{Warszawski:2010lattice}: the spin-down history, not merely the current $\Omega$, determines the number and position of vortices.  

For a crust-superfluid system with $\eta = 1.0$, the observed crust spin-down rate is $N_{\rm{EM}}/(2I_{\rm{c}})$; see Eq.~(\ref{eq:ch5:feedback}) with $dL_{\rm{z}}/dt= I_{\rm{c}}d\Omega/dt$.  For $N_{\rm{g}}\gg 1$, the results in Fig.~\ref{fig:ch5:om_omdot} demonstrate that the fraction of the torque-driven spin down $t_{\rm{total}}N_{\rm{EM}}/I_{\rm{c}}$  that is reversed by glitches decreases as $N_{\rm{EM}}/I_{\rm{c}}$ increases.  Quantitatively, we find $I_{\rm{c}}\langle\Delta\Omega_{\rm{g}}\rangle/(\langle\Delta t\rangle N_{\rm{EM}})=0.269$, 0.245 and 0.137 for $N_{\rm{EM}}/I_{\rm{c}}=10^{-3.5}$, $10^{-3.0}$ and $10^{-4.0}$ respectively.  We interpret this to mean that weaker torques drive the system adiabatically, allowing the superfluid to match the crust's deceleration.  In a pulsar, the slow spin down ($\dot{\Omega}/\Omega\lesssim -10^{-10}~\rm{s}^{-1}$) corresponds to the low-$N_{\rm{EM}}$ regime.  Hence, we can expect on large time scales the superfluid deceleration matches that of the crust, thus avoiding unfettered growth of differential motion.

Size and waiting-time distributions for $N_{\rm{EM}}/I_{\rm{c}}>10^{-4.0}$ are plotted as logarithmically binned pdfs and cumulative probability distributions in the \emph{left} and \emph{right} panels of Fig.~\ref{fig:ch5:omdot_stats} respectively.  As $N_{\rm{EM}}/I_{\rm{c}}$ increases, one obtains more frequent, smaller glitches.  The results reported in Table~\ref{tab:ch5:omdot} show that the slopes of power-law fits to the size pdf also increase monotonically with $N_{\rm{EM}}/I_{\rm{c}}$, as does the rate extracted from exponential fits to waiting-time pdfs (from $\lambda=0.0289$ to 0.112 for $N_{\rm{EM}}/I_{\rm{c}}=10^{-3.5}$ and $N_{\rm{EM}}/I_{\rm{c}}=10^{-2.5}$ respectively; since the $N_{\rm{EM}}/I_{\rm{c}}=10^{-4}$ curve contains only 3 glitches, we do not trust the fitted $\lambda$).  Systematic increases in the activity parameter $A$ (from $A=0.0958$ to 73.98 for $N_{\rm{EM}}/I_{\rm{c}}=10^{-4.0}$ and $N_{\rm{EM}}/I_{\rm{c}}=10^{-2.5}$ respectively) confirm that increasing $N_{\rm{EM}}/I_{\rm{c}}$ does not trigger the same vortex movements as for smaller $N_{\rm{EM}}/I_{\rm{c}}$.

\section{Large-scale simulations}\label{sec:ch5:big}

\begin{center}
\begin{table*}
\begin{tabular}{ | c| c| c| c| c| c| c| c|}
\hline
 $\Delta t_{\rm{sm}}$ &$\Delta V_i/V_0$ &$N_{\rm{g}}$ &$10^4\langle\Delta\Omega/\Omega\rangle$&$\langle\Delta t\rangle$ 	&$A$ &$\gamma$($p_{\rm{KS}}$) &$\lambda$($p_{\rm{KS}}$)\\
 \hline
 0.0	&0.0		 &4639		&0.208	&0.192		&1007 	&-1.26($0.0000$)&1.99($0.000$)\\
 0.0	&0.8		 &196		&0.500	&6.25		&3.14 	&-1.08($0.0007$)&0.337($0.000$)\\
 0.1	&0.0		 &491		&0.752	&1.55		&47.65 	&-1.23($0.1480$)&1.53($0.000$)\\
 0.1	&0.8		 &139		&0.591	&8.81		&1.87 	&-1.05($0.0214$)&0.192($0.0002$)\\
 \hline
\end{tabular}
\caption{Glitch size and waiting-time statistics for different $\Delta t_{\rm{sm}}$ and $\Delta V_i/V_0$ [corresponding $\Omega(t)$ curves shown in Fig.~\ref{fig:ch5:om_big}] for large-scale simulations containing $\sim 200$ vortices.  $N_{\rm{g}}$ is the number of glitches detected, $\langle \Delta\Omega/\Omega\rangle$ is the mean glitch size, $\langle\Delta t\rangle$ is the mean waiting time, and $A$ is the activity parameter.  $\lambda$ is the mean glitch rate that parametrises the cumulative distribution of waiting times, $\gamma$ is the power-law index that parametrises the pdf of glitch sizes.  The K-S probability $p_{\rm{KS}}$ is given in parenthesis;  the fit can be rejected at the $1-p_{\rm{KS}}$ confidence level.  Simulation parameters:  $V_0=36$, $\eta=1$, $\Delta V_i/V_0=0.0$, $n_{\rm{pin}}/n_{\rm{F}}$, $R=38$, $\Delta x=0.15$, $\Delta t=5\times 10^{-4}$, $N_{\rm{EM}}/I_{\rm{c}}=10^{-3}$, $\Omega(t=0) = 0.8$.}
\label{tab:ch5:big}
\end{table*}
\end{center}

\begin{figure*}
\includegraphics[scale=0.7,angle=90]{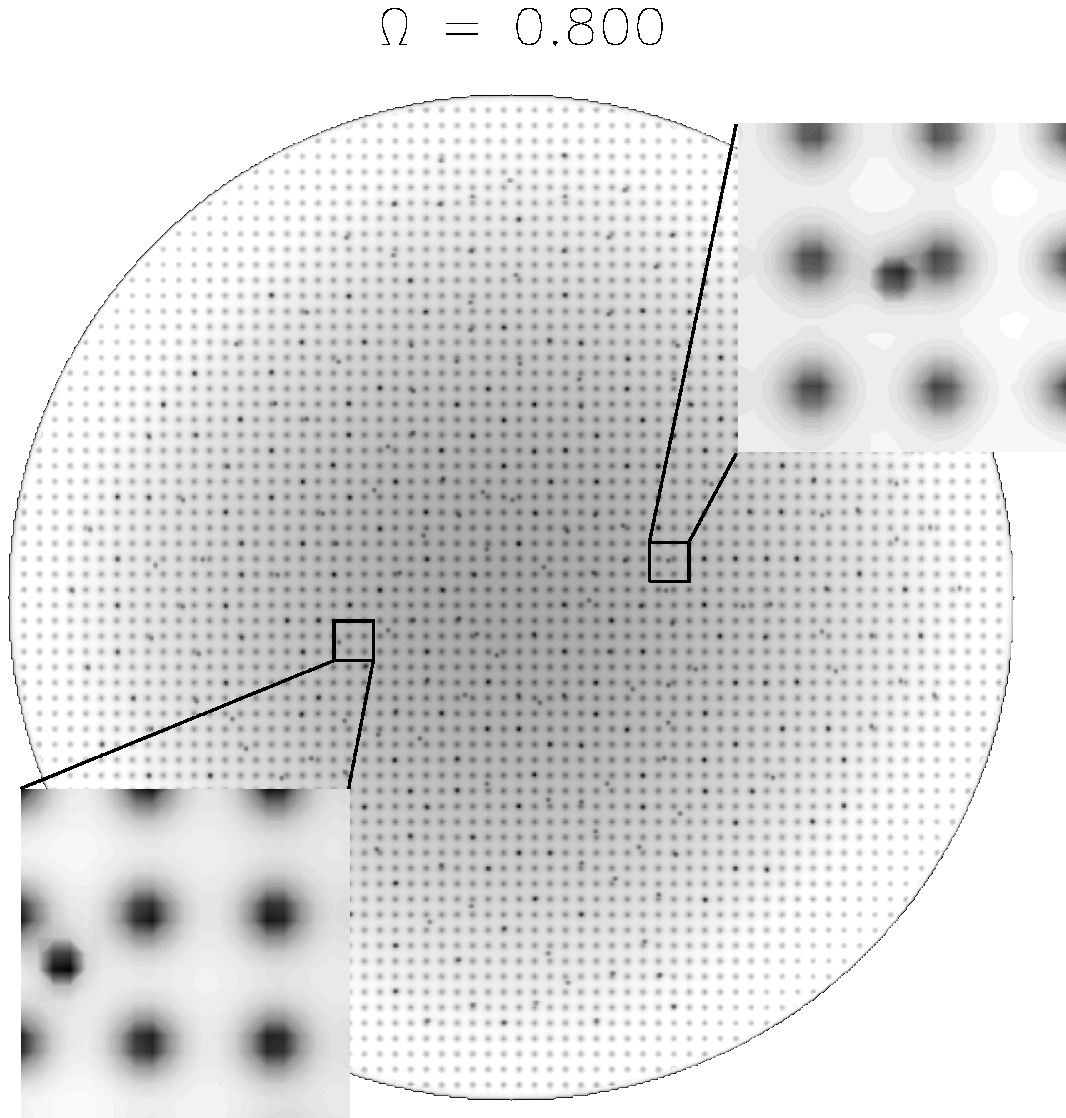}
\hspace{1cm}\includegraphics[scale=0.7,angle=90]{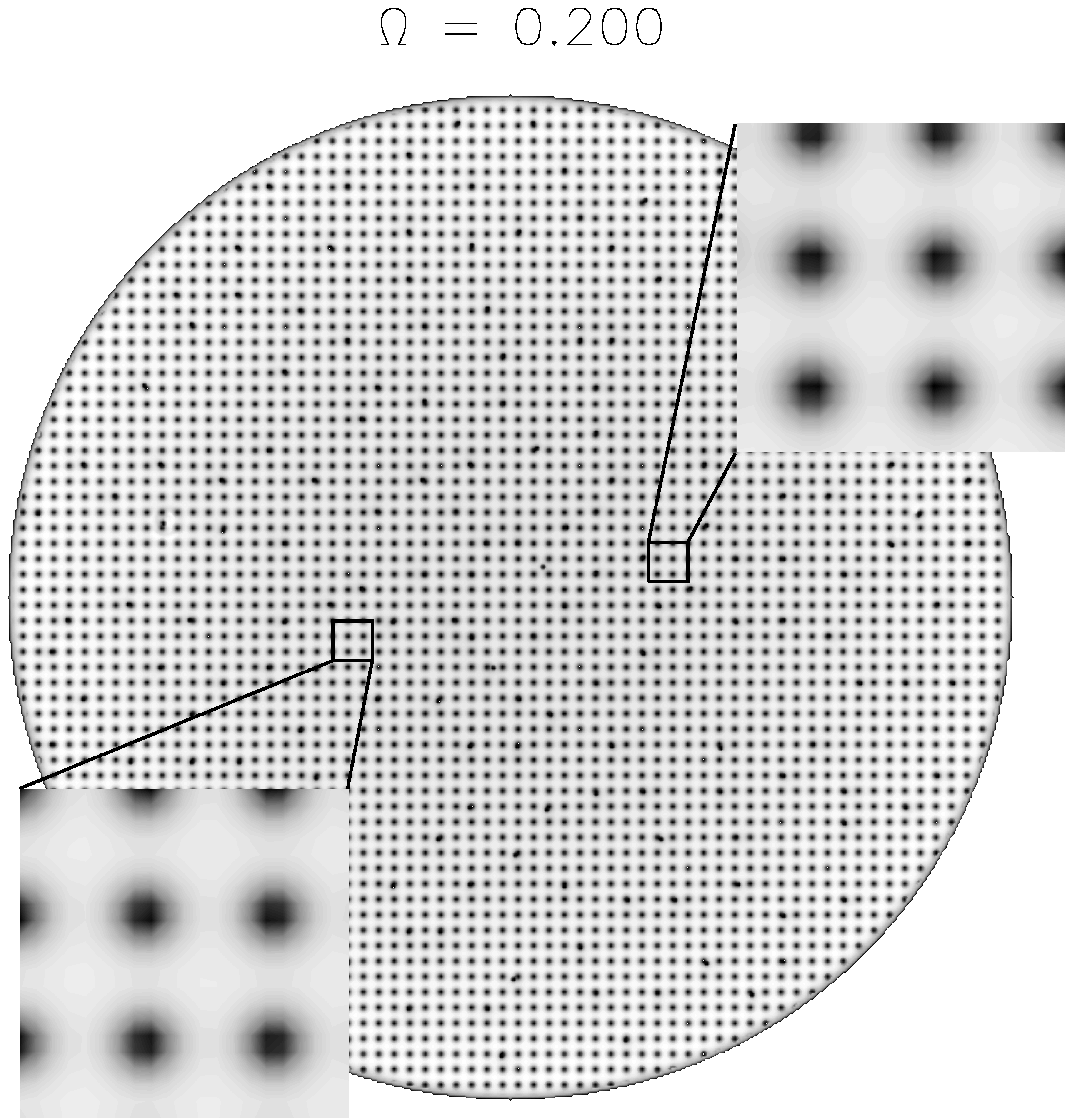}
\caption{Greyscale plots of the initial ($\Omega=0.8$, \emph{left}) and final ($\Omega=0.2$, \emph{right}) superfluid density $|\psi|^2$ for a $600\times600$ simulation with $R=38$ and $V_i=36$.  The insets zoom in on regions with dimensions $3\times3$.  Vortices are recognisable as the darker dots, initially clustered towards the centre of the crust and evenly spaced at the end of the simulation.  Unpinned vortices can be seen in the initial but not the final insets. Simulation parameters:   $V_0=36$, $\eta=1$, $N_{\rm{EM}}/I_{\rm{c}}=10^{-3}$, $R=38$, $\Delta x=0.15$, $\Delta t=5\times 10^{-4}$,  $\Omega(t=0) = 0.8$, $\Delta t_{\rm{sm}}=0.1$.}
\label{fig:ch5:big_image}
\end{figure*}
\begin{figure*}
\includegraphics[scale=1.0]{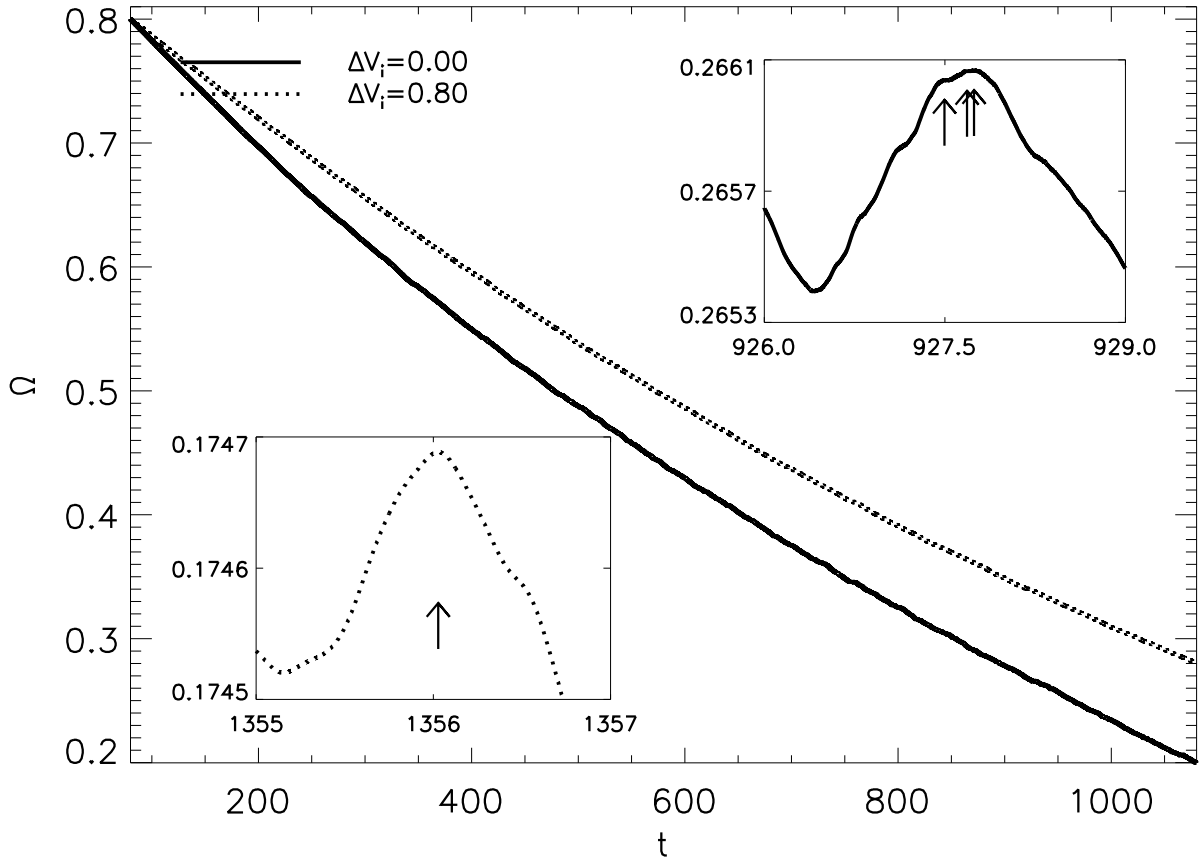}
\caption{Angular velocity of the crust as a function of time $\Omega(t)$ for large-scale spin-down experiments with $\Delta V_i/V_0=0.0$ and 0.8 (\emph{solid} and \emph{dotted} curves respectively). Greyscale plots of the initial and final superfluid density $|\psi|^2$ for $\Delta V_i/V_0=0.0$ are shown in Fig.~\ref{fig:ch5:big_image}.  \emph{Left inset}: Close-up of glitch at $t=804.3$ from $\Delta V_i/V_0=0.8$ curve. \emph{Right inset}:  Close-up of glitch at $t=806.1$ from $\Delta V_i/V_0=0.0$ curve.  Glitch positions are marked as \emph{arrows}.  Simulation parameters:   $V_0=36$, $\eta=1$, $N_{\rm{EM}}/I_{\rm{c}}=10^{-3}$, $R=38$, $\Delta x=0.15$, $\Delta t=5\times 10^{-4}$,  $\Omega(t=0) = 0.8$, $\Delta t_{\rm{sm}}=0.1$.}
\label{fig:ch5:om_big}
\end{figure*}
\begin{figure*}
\includegraphics[scale=0.365,angle=90]{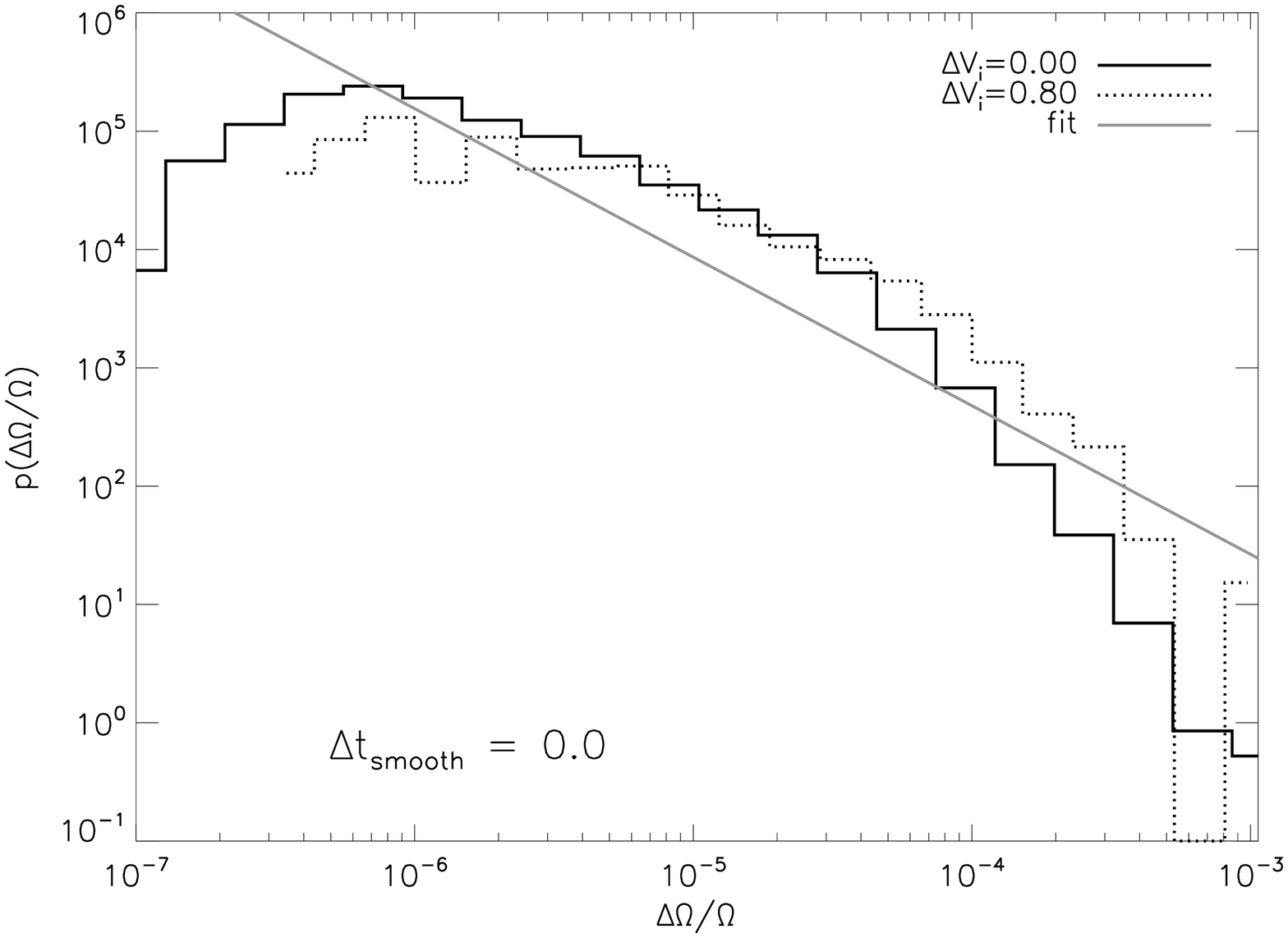}
\includegraphics[scale=0.365,angle=90]{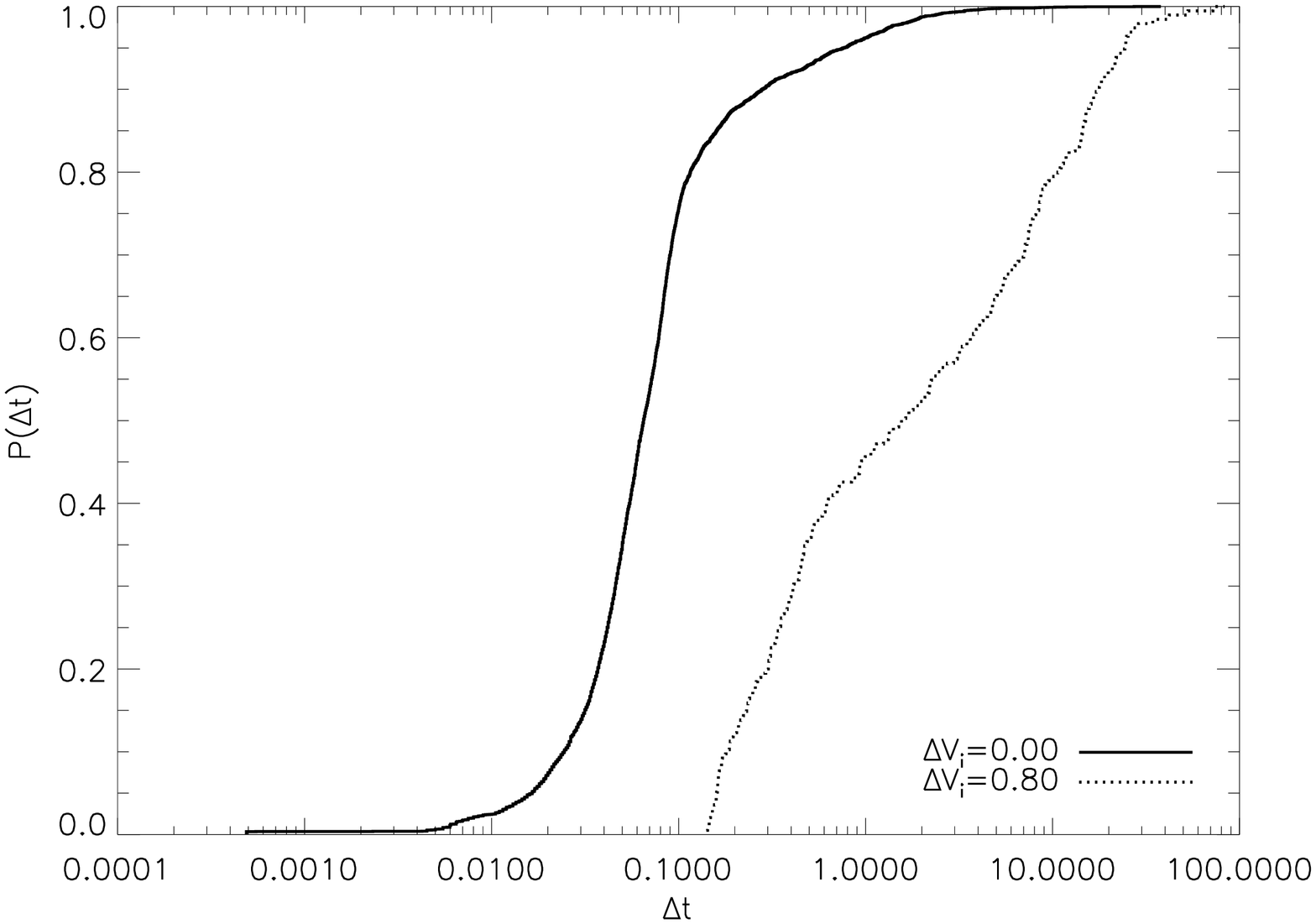}
\includegraphics[scale=0.365,angle=90]{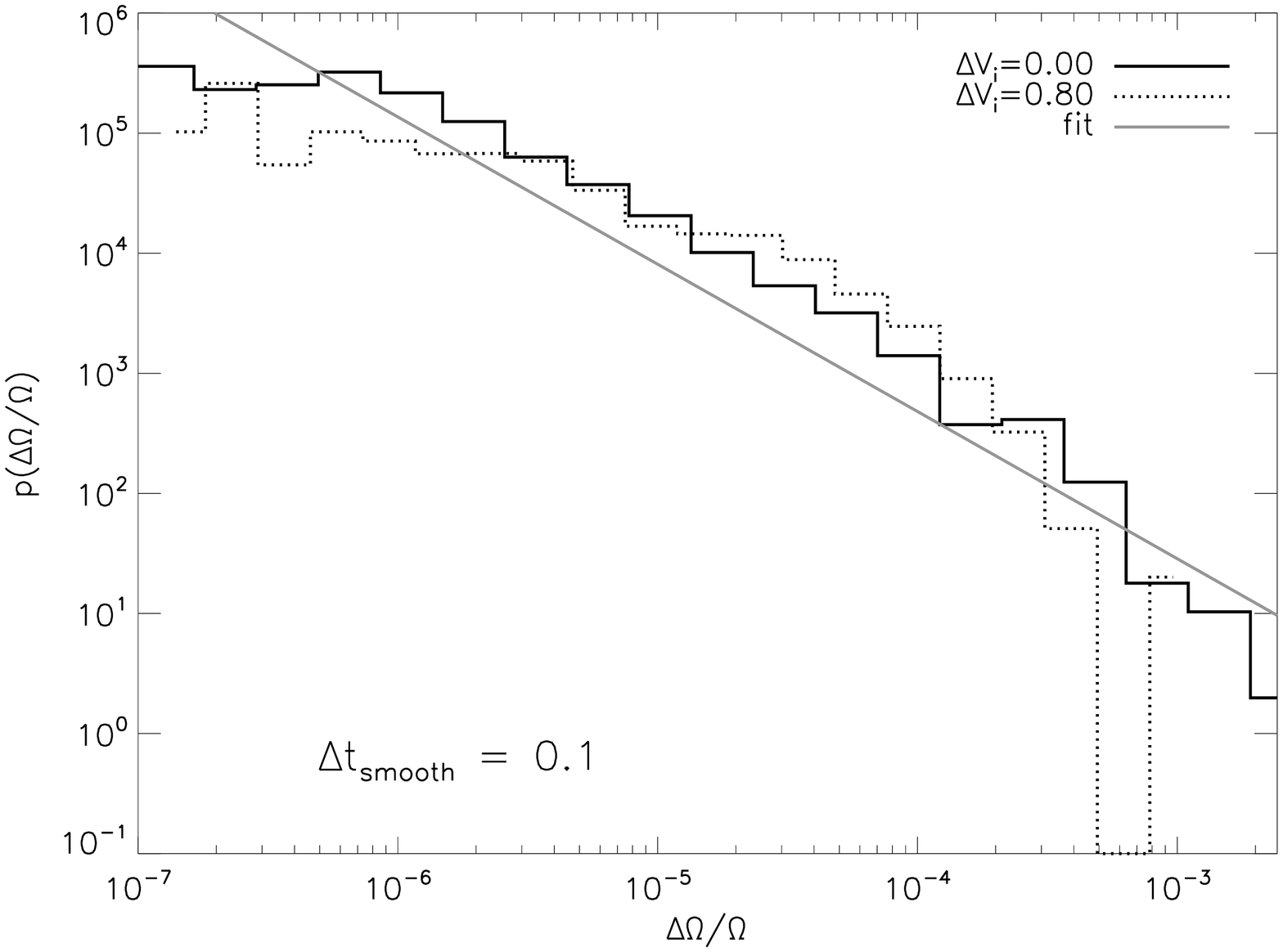}
\includegraphics[scale=0.365,angle=90]{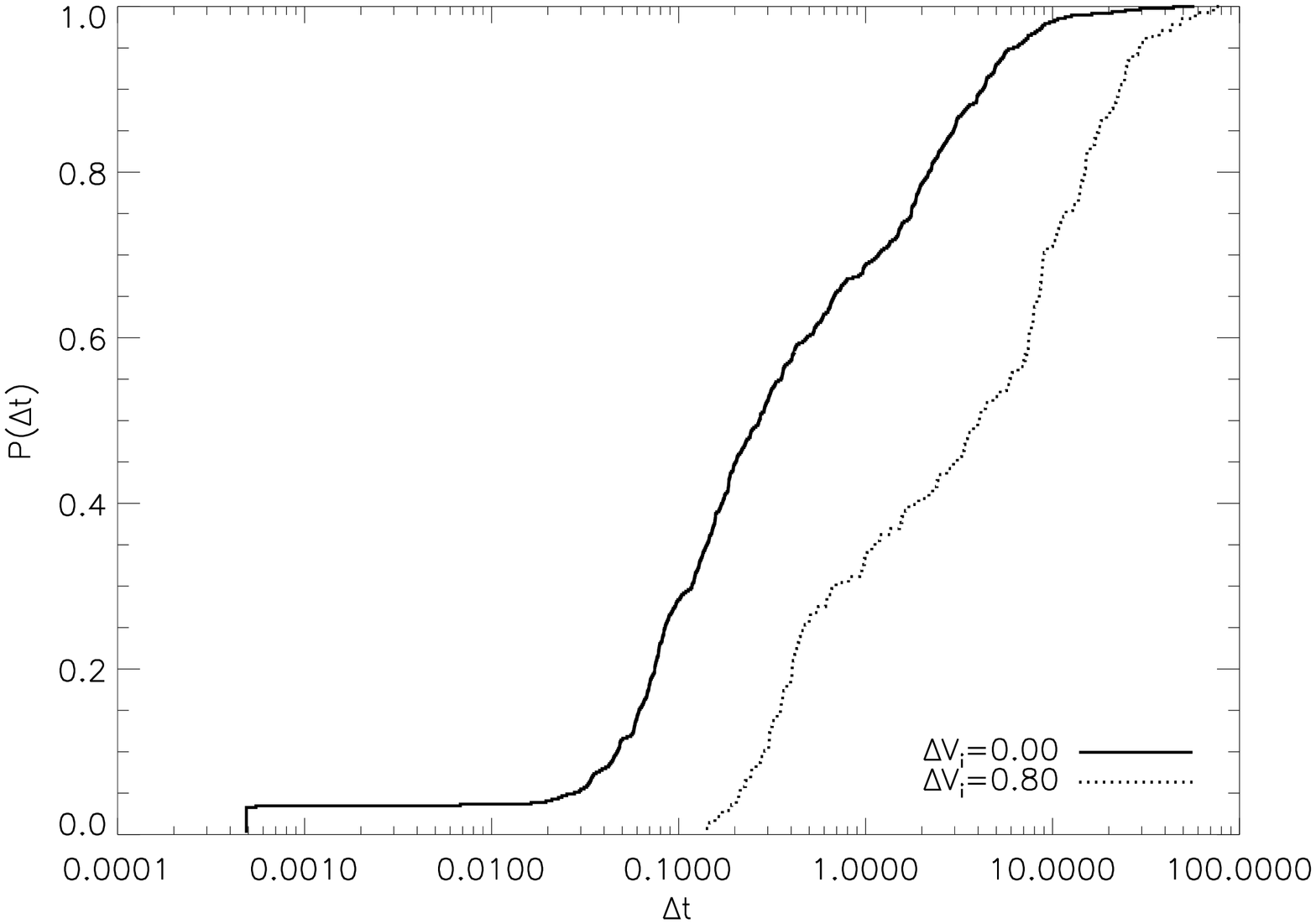}
\caption{\emph{Left}:  Probability density function (pdf) of fractional glitch sizes ($\Delta\Omega/\Omega$) for the $\Omega(t)$ curves graphed in Fig.~\ref{fig:ch5:om_big} for $\Delta V_i/V_0=0.0$ and 0.8 (\emph{solid} and \emph{dotted} curves respectively) unsmoothed (\emph{top}) and smoothed with top-hat window function of width $\Delta t_{\rm{sm}}=0.1$ (\emph{bottom}), logarithmically binned and graphed on log-log axes.  The \emph{solid grey} curves are power-law least squares fits to the \emph{solid} histograms for $\Delta\Omega/\Omega\geq 7\times 10^{-7}$, with indices $-1.26$ and $-1.23$ respectively.  \emph{Right}:  Cumulative pdf of glitch waiting times $p(\Delta t)$ graphed on log-linear axes with $\Delta V_i/V_0=0.0$ and 0.8 (\emph{solid} and \emph{dotted} curves respectively) unsmoothed (\emph{top}) and smoothed with top-hat window function of width $\Delta t_{\rm{sm}}=0.1$ (\emph{bottom}). Simulation parameters:  $V_0=36$, $\eta=1$, $N_{\rm{EM}}/I_{\rm{c}}=10^{-3}$, $R=38$, $\Delta x=0.15$, $\Delta t=5\times 10^{-4}$,  $\Omega(t=0) = 0.8$, $\Delta t_{\rm{sm}}=0.1$.}
\label{fig:ch5:big_stats}
\end{figure*}

The scale-invariant statistics of pulsar glitches are expected to emerge most clearly in the many-vortex limit.  In this section we increase the system size, and hence the vortex number, by an order of magnitude.  Prevented by computational expense from undertaking a full parameter study, we present examples of two systems with uniform and variable pinning strengths  ($\Delta V_i/V_0=0.0$ and 0.8 respectively), each with $\sim200$ vortices in total.  The simulations ran for $\sim 2~\rm{weeks}$ on a single 2.5~GHz processor.   Statistics from the simulations are summarised in Table~\ref{tab:ch5:big}.

To maintain the same spatial resolution as in the simulation, we increase the crust radius to $R=38$ and impose a $66\times66$ pinning grid (with $V_0=36$).  We also increase the number of particles in the system 10-fold (via $n_0$), to provide sufficient contrast to distinguish vortices from the background density.  We initialise the experiment by instantaneously accelerating the crust from rest to $\Omega = 0.8$ and letting the system reach a steady state.  Once a steady state is achieved, the crust is subjected to a spin-down torque $N_{\rm{EM}}=10^{-3}I_{\rm{c}}$.  

The greyscale plot in the \emph{left} panel of  Fig.~\ref{fig:ch5:big_image} depicts the initial superfluid density for uniform pinning ($\Delta V_i/V_0=0.0$).  The two insets zoom in on regions where vortices sit interstitially.  The smooth increase in $|\psi|^2$ with radius, indicated by the radial colour gradient, represents the standard parabolic density profile, analogous to the meniscus of a rotating bucket of water \citep{Warszawski:2010lattice}.  Initially the vortices (dark dots in Fig.~\ref{fig:ch5:big_image}; light dots are unoccupied pinning sites) are surrounded by a vortex-free corona in the region $4\leq r\leq 38$, highlighting the distortion and hysteresis wrought by the pinning grid \citep[see Section IVB of][]{Warszawski:2010lattice}.  By the end of the simulation, the vortices are approximately evenly distributed across the pinning grid; see the greyscale plot in the \emph{right} panel of Fig.~\ref{fig:ch5:big_image}.  The radial density gradient is shallower because $\Omega$ is lower than its initial value.  We can see that vortices have moved from their initial positions; vortices present in the insets in the \emph{left} panel of Fig.~\ref{fig:ch5:big_image} are absent from the \emph{right} panel.

The spin-down curves for $\Delta V_i/V_0=0.0$ and 0.8 (\emph{solid} and \emph{dotted} curves in Fig.~\ref{fig:ch5:om_big} respectively), smoothed with a top-hat window function of width $\Delta t_{\rm{sm}}=0.1$, reveal only small glitches ($10^4\langle \Delta\Omega_{\rm{g}}/\Omega{\rm{g}}\rangle=0.752$ and 0.591 for $\Delta V_i/V_0=0.0$ and 0.8 respectively).  This result is unsurprising.  In larger systems, each vortex corresponds to a smaller fraction of the total angular momentum.  To achieve glitches as large as those produced by a $R=12.5$ system, $\sim 10$ times more vortices must unpin simultaneously.  Dissipation, which suppresses sound waves, renders knock-on events less likely; hence such large-scale collective unpinning is rare in the systems simulated here.  The insets show close-ups of the largest glitch (\emph{top right} and \emph{bottom left} for $\Delta V_i/V_0=0.0$ and $\Delta V_i/V_0=0.8$ respectively).  The spin-up phase of the largest glitch in the $\Delta V_i/V_0=0.0$ curve is bumpy. In fact, the vertical distance between the bumps is of order the height of the largest glitch in the $\Delta V_i/V_0=0.8$ curve, which does not exhibit the same `bumpiness'.  

Pdfs of $\Delta\Omega/\Omega$ and cumulative distributions of $\Delta t$ are graphed in the \emph{left} and \emph{right} columns of Fig.~\ref{fig:ch5:big_stats} respectively; the solid curves are for $\Delta V_i/V_0=0.0$ and the \emph{dotted} curves are for $\Delta V_i/V_0=0.8$. Since there is an obvious turnover in the size distribution at $\Delta\Omega/\Omega\approx 7\times 10^{-7}$, power-law fits are to $p(\Delta\Omega/\Omega\geq 7\times 10^{-7})$.  The \emph{top} panels refer to glitches detected in the unsmoothed $\Omega$ curve, whereas the \emph{bottom} panels are smoothed with $\Delta t_{\rm{sm}}=0.1$.  For both values of $\Delta V_i/V_0$, smoothing flattens the size distribution.  Comparison with Fig.~\ref{fig:ch5:V_stats} reveals that the smallest glitch in the larger system is six times smaller than the smallest glitches for the $R=12.5$ systems discussed in previous sections, which agrees with the $\sim$six$-$fold increase in vortex number $N_{\rm{v}}$.  If a single vortex is responsible for the smallest glitch, then the size of the smallest glitch should scale inversely with $N_{\rm{v}}$.  For $\Delta t_{\rm{sm}}=0.1$, the fitted power law is marginally shallower for $\Delta V_i/V_0=0.8$ ($\gamma=-1.05$) than for  $\Delta V_i/V_0=0.0$ ($\gamma=-1.23$).  

The waiting times tend to lengthen as $\Delta V_i/V_0$ increases, as observed in smaller systems (see Fig.~\ref{fig:ch5:flat_stats}).  For $\Delta V_i/V_0=0.0$, smoothing broadens the waiting-time distribution significantly.  The \emph{top right} panel of Fig.~\ref{fig:ch5:flat_stats} reveals a strong periodicity at $\Delta t\approx 0.05$ in the waiting-time distribution for $\Delta t_{\rm{sm}}=0.0$.  This may be evidence of the periodic glitches expected from a system in which stress increases over time by the same amount on vortices pinned with equal strength.   Despite the overall slower spin down for $\Delta V_i/V_0=0.8$ (cf. Fig.~\ref{fig:ch5:om_big}), the longer waiting times are not matched by larger glitches. This is likely due to one of two reasons:  (i) much of the crust-superfluid shear is reversed by smooth vortex motion, unhindered by weak pinning; or (ii) glitches are too small to cause a bump (positive change) in $\Omega$, and hence are not detected by our glitch-finding algorithm.  

Two significant differences between this experiment and those described in Section~\ref{sec:canonical}---Section~\ref{sec:structure} prevent us from attributing the altered spin-down behaviour (smaller glitches) entirely to system size. First, the ratio of pinning sites to vortices is significantly higher ($n_{\rm{pin}}/n_{\rm{F}}\approx 5$) in this experiment than in the smaller systems, whilst the spacing of the pinning sites, relative to the vortex core size, is smaller. It may be that such a closely spaced pinning grid presents the vortices with a nearly homogeneous, elevated potential, effectively weakening the pinning strength.  Second, since the Magnus force on pinned vortices is proportional to $|\bm{v}-\bm{v}_{\rm{s}}|\approx r(\langle L_z\rangle/I_{\rm{s}}-\Omega)$ at the pinning site, the larger radii of many of the pinned vortices in the $R=38$ system make for larger Magnus forces for smaller values of $(\langle L_z\rangle/I_{\rm{s}}-\Omega)$.  

\section{Discussion}\label{sec:ch5:conc_GPE}
\begin{center}
\begin{table*}
\begin{tabular}{ | c| c| c| c| c| c| c|}
\hline
quantity & $\Delta t_{\rm{sm}}$ &$V_0$ & $\Delta V_i/V_0$ & $n_{\rm{pin}}/n_{\rm{F}}$ & $\eta$ & $N_{\rm{EM}}$\\
\hline
$\langle\Delta\Omega/\Omega\rangle$	&$+$ &$+$	&.		&.	&$-$		&. \\
$\langle\Delta t\rangle$						&$+$ &.		&$+$	&$-$	&$-$	&$-$ \\
$A$														&. &$+$		&.		&$+$	&$-$		&$+$\\
 \hline
\end{tabular}
\caption{Summary of how simulated glitch statistics depend on the physical levers of the model.  A plus (minus) sign indicates that the quantity in the left column increases (decreases) when the parameter in the column heading increases.  A dot indicates no clear correlation.}
\label{tab:ch5:summary}
\end{table*}
\end{center}

The numerical experiments described in this paper demonstrate from first principles that a superfluid coupled to its container via vortex pinning spins down spasmodically.  The experiments are designed to explore how the distributions of pulsar glitch sizes and waiting times depend on pinning and stellar parameters.  Computational limitations prevent us from probing the parameter regime most appropriate to the pulsar glitch problem.  However, we do our best to mimic astrophysical conditions by adopting the same ordering of dimensionless parameters as in nature.  Qualitatively, and with respect to the canonical parameters listed in Table~\ref{tab:ch5:pulsar}, the pulsar problem involves weak (but still effective) pinning, slow electromagnetic spin down, a huge number of vortices, an even larger number of pinning sites, and a heavy crust. 
 
In order to speed up the process of compiling glitch statistics and minimise its subjective element, we devise an automated glitch finder.  Since the glitch-finding algorithm identifies all positive jumps in $\Delta\Omega$, it is necessary to smooth $\Omega$ so as to filter out post-glitch oscillations or numerical noise.  In the future, improvements in timing and observational duty cycle should allow real pulsar spin-down curves to be analysed in the same way.

Our simulations lend new insights into aspects of the physics of pulsar glitches. 
\begin{enumerate}
 \item Stronger pinning causes larger glitches, with $\langle \Delta\Omega_{\rm{g}}/\Omega\rangle\approx 7.6V_i$.  This is not as obvious as it may sound.  Prima facie it is equally likely that strong pinning leads to a large average shear with small, frequent fluctuations.  In fact, this does not happen.
 \item When pinning sites and vortices are equally numerous, vortices conform to the geometry of the pinning grid, producing larger glitches than if pinning sites are more abundant than vortices.  Waiting times are also longer for $n_{\rm{pin}}\sim n_{\rm{F}}$ than for $n_{\rm{pin}}> n_{\rm{F}}$.
 \item  Glitch sizes are insensitive to increases in the pinning site abundance beyond parity (\emph{i.e.} $n_{\rm{F}}\sim n_{\rm{pin}}$), because vortices move across many pinning sites to negate the local shear, rather than stopping at an adjacent pinning site.  The distance travelled by an unpinned vortex is a key input to heuristic glitch models \citep[][for example]{Warszawski:2008p4510,Melatos:2009p4511}.  
 \item A broader range of $V_i$ results in a broader range of glitch sizes and more frequent glitches. 
 \item A heavier crust produces more large glitches and longer waiting times than does a lighter crust.
 \item A strong spin-down torque catalyses smaller, more frequent glitches than a weak torque.
 \item Pulsar astronomers derive the strength of the pulsar magnetic field from the spin-down rate of the crust without allowing for superfluid feedback.  The results in this paper demonstrate that this is inaccurate.  Even when no vortices have yet unpinned (e.g. $t\leq 430$ in the \emph{dot-dashed} curve in Fig.~\ref{fig:ch5:om_V}), $\dot{\Omega}$ is a function of both the external torque and the superfluid dynamics [via Eq.~(\ref{eq:ch5:feedback})]. As vortices migrate towards the edge of their pinning sites, the smooth, internal spin-up torque effectively reduces the crust's spin-down rate. But this effect is much smaller in a real pulsar than in our simulations.  For example, if all vortices move radially outward by the width of a nucleus, $\Delta\langle\hat{L}_z\rangle/\langle\hat{L}_z\rangle\approx 10^{-18}$.  A larger, smooth spin-up torque results from homogeneous outward vortex creep.
\end{enumerate}

Table~\ref{tab:ch5:summary} summarises how measures of glitch activity, like the mean size and waiting time and the glitch activity parameter, depend on the underlying physical levers, like the pinning strength/abundance and the electromagnetic torque.  A dot indicates where there is no/weak dependence; a plus (minus) sign indicates that the measure of activity increases (decreases) as the physical variable increases. Intuitively, we expect that a positive correlation with mean glitch size is accompanied by a positive correlation with mean waiting time.  For $V_0$, for example, this is not the case; mean glitch size increases with increasing $V_0$, whereas mean waiting time does not change monotonically with $V_0$.  One possible explanation is that the simulation has not run long enough to establish a critical velocity shear that ensures a net outward flux of vortices that matches the superfluid spin down to that of the crust \citep{Alpar:1984p6781,Melatos:2009p4511}.  The increasing vertical distance between the spin-down curves in Figure~\ref{fig:ch5:om_V} confirms that this critical shear has not yet been established.

If pinning is not strong at the densities where only a neutron superfluid is present [\cite{Donati:2003p97} show that $V_0$ depends on density], then glitches may arise from the physics of the outer pulsar core, where the neutron superfluid coexists with a superconducting proton fluid.  Magnetic flux tubes in the superconducting fluid complicate the superfluid vortex dynamics, rendering the simple Magnus-force dynamics in this paper invalid.  Hence, simulations presented in this paper are relevant only to the pulsar inner crust, whose density lies in the range $10^{11}~\rm{g~cm}^{-3}\lesssim\rho\lesssim 10^{14}~\rm{g~cm}^{-3}$, where the majority of protons are confined within nuclei \citep{Pethick:2000}.

The discrete $\Omega$ jumps in the spin-down curves graphed throughout this paper are analogous to the step changes in magnetic field seen in Fig.~1 of \cite{Field:1995p155}, caused by magnetic flux tube avalanches in type II superconductors.  Superconductor experiments exhibit power-law size distributions over two decades, with exponents ranging from $-1.4$ to $-2.2$.  Several authors have successfully modelled the superconducting system as a self-organised critical system fluctuating about a Bean state, governed by collective unpinning and motion of flux tubes \citep{Bassler:1998p8}.  A natural theoretical framework for a self-organised critical system with knock-on mechanisms is a branching process.  The characteristic power-law index of such systems is $-1.5$ \citep{Carreras:2002p8220}, consistently steeper than the size distributions reported in this paper. We therefore conclude that either:  (i) we are not simulating a self-organised critical system, or (ii) our definition of event size and/or identification algorithm is not correct.

Although our glitch size distributions are not inconsistent with power laws, we do not find conclusive evidence of collective vortex behaviour, e.g. many-generation cascades of unpinning events in movies of $|\psi|^2$.  This is not surprising.  \cite{Warszawski:2010individual} found that sound waves from moving vortices unpin nearby vortices only when dissipation is low.  In this paper's experiments, for computational expediency, we use $\gamma =0.05$, whereas \cite{Warszawski:2010individual}  required $\gamma = 0.0025$ to facilitate an acoustic knock-on event given the same characteristic pinning strengths as studied in this paper.  In contrast, the proximity knock-on mechanism identified by \cite{Warszawski:2010individual} does occur in the simulations presented here.  Hence collective dynamics are not excluded completely.

Collective effects emerge most fully in large systems.  We therefore present preliminary results from large, computationally expensive simulations with $\sim 200$ vortices. The large simulations yield a broader range of glitch sizes than smaller systems.  Power-law fits to the entire size distribution can be rejected with high confidence.  Imposing a minimum cut-off, to create a subset of larger glitches, yields a distribution that can be fitted by a power law that is rejected with less confidence.  Physically, this may be equivalent to distinguishing between creep-like vortex motion and avalanche events. 

In addition to requiring larger system sizes, reliable pulsar glitch simulations must also address the role of pinning along the length of a vortex.  When a three-dimensional vortex moves radially outward, it either `unzips' from pinning sites along its length or moves in small vertical segments that behave effectively as individual vortices.  Three dimensions also introduce the possibility of turbulent superflow, i.e. a polarised vortex tangle \citep{Schwarz:1988p60,Barenghi,Mongiovi:2007p198,Kobayashi:2007p156}.  Tangled vortices can form bundles, which behave as multiply-quantised vortices.  A full understanding requires further investigation using 3D simulations. 

Finally, as well as expanding GPE simulations to larger systems and higher resolution, we recommend that progress in understanding pulsar glitches can be achieved via cellular automata \citep{Morley:1996p2128,Bassler:1998p8,Warszawski:2008p4510}.  The generic behaviour observed in GPE simulations can be used to devise microscopic rules for such automata.

\section*{Acknowledgements}
We thank Dr Andrew Martin and Dr Natalia Berloff for enlightening discussions on many of the topics addressed in this paper.  LW acknowledges the support of an Australian Postgraduate Award.

\bibliography{ms_revised}
\bibliographystyle{mn2e}

\end{document}